%
\expandafter\ifx\csname phyzzx\endcsname\relax
 \message{It is better to use PHYZZX format than to
          \string\input\space PHYZZX}\else
 \wlog{PHYZZX macros are already loaded and are not
          \string\input\space again}%
 \endinput \fi
\catcode`\@=11 
\let\rel@x=\relax
\let\n@expand=\relax
\def\pr@tect{\let\n@expand=\noexpand}
\let\protect=\pr@tect
\let\gl@bal=\global
%
%
%
\newfam\cpfam
\newdimen\b@gheight             \b@gheight=12pt
\newcount\f@ntkey               \f@ntkey=0
\def\f@m{\afterassignment\samef@nt\f@ntkey=}
\def\samef@nt{\fam=\f@ntkey \the\textfont\f@ntkey\rel@x}
\def\setstr@t{\setbox\strutbox=\hbox{\vrule height 0.85\b@gheight
                                depth 0.35\b@gheight width\z@ }}
\input phyzzx.fonts
%
\def\rm{\n@expand\f@m0 }
\def\mit{\n@expand\f@m1 }         
\def\cal{\n@expand\f@m2 }
\def\it{\n@expand\f@m\itfam}
\def\sl{\n@expand\f@m\slfam}
\def\bf{\n@expand\f@m\bffam}
\def\tt{\n@expand\f@m\ttfam}
\def\caps{\n@expand\f@m\cpfam}    
\def\em@{\rel@x\ifnum\f@ntkey=0 \it \else
        \ifnum\f@ntkey=\bffam \it \else \rm \fi \fi }
\def\em{\n@expand\em@}
\def\fourteenpoint{\fourteenf@nts \samef@nt \b@gheight=14pt \setstr@t }
\def\twelvepoint{\twelvef@nts \samef@nt \b@gheight=12pt \setstr@t }
\def\tenpoint{\tenf@nts \samef@nt \b@gheight=10pt \setstr@t }
\normalbaselineskip = 20pt plus 0.2pt minus 0.1pt
\normallineskip = 1.5pt plus 0.1pt minus 0.1pt
\normallineskiplimit = 1.5pt
\newskip\normaldisplayskip
\normaldisplayskip = 20pt plus 5pt minus 10pt
\newskip\normaldispshortskip
\normaldispshortskip = 6pt plus 5pt
\newskip\normalparskip
\normalparskip = 6pt plus 2pt minus 1pt
\newskip\skipregister
\skipregister = 5pt plus 2pt minus 1.5pt
\newif\ifsingl@
\newif\ifdoubl@
\newif\iftwelv@  \twelv@true
\def\singlespace{\singl@true\doubl@false\spaces@t}
\def\doublespace{\singl@false\doubl@true\spaces@t}
\def\normalspace{\singl@false\doubl@false\spaces@t}
\def\Tenpoint{\tenpoint\twelv@false\spaces@t}
\def\Twelvepoint{\twelvepoint\twelv@true\spaces@t}
\def\spaces@t{\rel@x
      \iftwelv@ \ifsingl@\subspaces@t3:4;\else\subspaces@t1:1;\fi
       \else \ifsingl@\subspaces@t3:5;\else\subspaces@t4:5;\fi \fi
      \ifdoubl@ \multiply\baselineskip by 5
         \divide\baselineskip by 4 \fi }
\def\subspaces@t#1:#2;{
      \baselineskip = \normalbaselineskip
      \multiply\baselineskip by #1 \divide\baselineskip by #2
      \lineskip = \normallineskip
      \multiply\lineskip by #1 \divide\lineskip by #2
      \lineskiplimit = \normallineskiplimit
      \multiply\lineskiplimit by #1 \divide\lineskiplimit by #2
      \parskip = \normalparskip
      \multiply\parskip by #1 \divide\parskip by #2
      \abovedisplayskip = \normaldisplayskip
      \multiply\abovedisplayskip by #1 \divide\abovedisplayskip by #2
      \belowdisplayskip = \abovedisplayskip
      \abovedisplayshortskip = \normaldispshortskip
      \multiply\abovedisplayshortskip by #1
        \divide\abovedisplayshortskip by #2
      \belowdisplayshortskip = \abovedisplayshortskip
      \advance\belowdisplayshortskip by \belowdisplayskip
      \divide\belowdisplayshortskip by 2
      \smallskipamount = \skipregister
      \multiply\smallskipamount by #1 \divide\smallskipamount by #2
      \medskipamount = \smallskipamount \multiply\medskipamount by 2
      \bigskipamount = \smallskipamount \multiply\bigskipamount by 4 }
\def\normalbaselines{ \baselineskip=\normalbaselineskip
   \lineskip=\normallineskip \lineskiplimit=\normallineskip
   \iftwelv@\else \multiply\baselineskip by 4 \divide\baselineskip by 5
     \multiply\lineskiplimit by 4 \divide\lineskiplimit by 5
     \multiply\lineskip by 4 \divide\lineskip by 5 \fi }
\Twelvepoint  
\interlinepenalty=50
\interfootnotelinepenalty=5000
\predisplaypenalty=9000
\postdisplaypenalty=500
\hfuzz=1pt
\vfuzz=0.2pt
\newdimen\HOFFSET  \HOFFSET=0pt
\newdimen\VOFFSET  \VOFFSET=0pt
\newdimen\HSWING   \HSWING=0pt
\dimen\footins=8in
%
%
%
\newskip\pagebottomfiller
\pagebottomfiller=\z@ plus \z@ minus \z@
\def\pagecontents{
   \ifvoid\topins\else\unvbox\topins\vskip\skip\topins\fi
   \dimen@ = \dp255 \unvbox255
   \vskip\pagebottomfiller
   \ifvoid\footins\else\vskip\skip\footins\footrule\unvbox\footins\fi
   \ifr@ggedbottom \kern-\dimen@ \vfil \fi }
\def\makeheadline{\vbox to 0pt{ \skip@=\topskip
      \advance\skip@ by -12pt \advance\skip@ by -2\normalbaselineskip
      \vskip\skip@ \line{\vbox to 12pt{}\the\headline} \vss
      }\nointerlineskip}
\def\makefootline{\baselineskip = 1.5\normalbaselineskip
                 \line{\the\footline}}
\newif\iffrontpage
\newif\ifp@genum
\def\nopagenumbers{\p@genumfalse}
\def\pagenumbers{\p@genumtrue}
\pagenumbers
\newtoks\paperheadline
\newtoks\paperfootline
\newtoks\letterheadline
\newtoks\letterfootline
\newtoks\letterinfo
\newtoks\date
\paperheadline={\hfil}
\paperfootline={\hss\iffrontpage\else\ifp@genum\tenrm\folio\hss\fi\fi}
\letterheadline{\iffrontpage \hfil \else
    \rm \ifp@genum page~~\folio\fi \hfil\the\date \fi}
\letterfootline={\iffrontpage\the\letterinfo\else\hfil\fi}
\letterinfo={\hfil}
\def\monthname{\rel@x\ifcase\month 0/\or January\or February\or
   March\or April\or May\or June\or July\or August\or September\or
   October\or November\or December\else\number\month/\fi}
\def\today{\monthname~\number\day, \number\year}
\date={\today}
\headline=\paperheadline 
\footline=\paperfootline 
\countdef\pageno=1      \countdef\pagen@=0
\countdef\pagenumber=1  \pagenumber=1
\def\advancepageno{\gl@bal\advance\pagen@ by 1
   \ifnum\pagenumber<0 \gl@bal\advance\pagenumber by -1
    \else\gl@bal\advance\pagenumber by 1 \fi
    \gl@bal\frontpagefalse  \swing@ }
\def\folio{\ifnum\pagenumber<0 \romannumeral-\pagenumber
           \else \number\pagenumber \fi }
\def\swing@{\ifodd\pagenumber \gl@bal\advance\hoffset by -\HSWING
             \else \gl@bal\advance\hoffset by \HSWING \fi }
\def\footrule{\dimen@=\prevdepth\nointerlineskip
   \vbox to 0pt{\vskip -0.25\baselineskip \hrule width 0.35\hsize \vss}
   \prevdepth=\dimen@ }
\let\footnotespecial=\rel@x
\newdimen\footindent
\footindent=24pt
\def\Textindent#1{\noindent\llap{#1\enspace}\ignorespaces}
\def\Vfootnote#1{\insert\footins\bgroup
   \interlinepenalty=\interfootnotelinepenalty \floatingpenalty=20000
   \singl@true\doubl@false\Tenpoint
   \splittopskip=\ht\strutbox \boxmaxdepth=\dp\strutbox
   \leftskip=\footindent \rightskip=\z@skip
   \parindent=0.5\footindent \parfillskip=0pt plus 1fil
   \spaceskip=\z@skip \xspaceskip=\z@skip \footnotespecial
   \Textindent{#1}\footstrut\futurelet\next\fo@t}

\def\vfootnote#1{\Vfootnote{${#1}$}}
\def\footnote#1{\attach{#1}\vfootnote{#1}}

\def\foot{\attach\footsymbolgen\vfootnote{\footsymbol}}
\let\footsymbol=\star
\newcount\lastf@@t           \lastf@@t=-1
\newcount\footsymbolcount    \footsymbolcount=0
\newif\ifPhysRev
\def\footsymbolgen{\bumpfootsymbolcount \generatefootsymbol \footsymbol }
\def\bumpfootsymbolcount{\rel@x
   \iffrontpage \bumpfootsymbolpos \else \advance\lastf@@t by 1
     \ifPhysRev \bumpfootsymbolneg \else \bumpfootsymbolpos \fi \fi
   \gl@bal\lastf@@t=\pagen@ }
\def\bumpfootsymbolpos{\ifnum\footsymbolcount <0
                            \gl@bal\footsymbolcount =0 \fi
    \ifnum\lastf@@t<\pagen@ \gl@bal\footsymbolcount=0
     \else \gl@bal\advance\footsymbolcount by 1 \fi }
\def\bumpfootsymbolneg{\ifnum\footsymbolcount >0
             \gl@bal\footsymbolcount =0 \fi
         \gl@bal\advance\footsymbolcount by -1 }
\def\fd@f#1 {\xdef\footsymbol{\mathchar"#1 }}
\def\generatefootsymbol{\ifcase\footsymbolcount \fd@f 13F \or \fd@f 279
        \or \fd@f 27A \or \fd@f 278 \or \fd@f 27B \else
        \ifnum\footsymbolcount <0 \fd@f{023 \number-\footsymbolcount }
         \else \fd@f 203 {\loop \ifnum\footsymbolcount >5
                \fd@f{203 \footsymbol } \advance\footsymbolcount by -1
                \repeat }\fi \fi }

\def\nonfrenchspacing{\sfcode`\.=3001 \sfcode`\!=3000 \sfcode`\?=3000
        \sfcode`\:=2000 \sfcode`\;=1500 \sfcode`\,=1251 }
\nonfrenchspacing
\newdimen\d@twidth
{\setbox0=\hbox{s.} \gl@bal\d@twidth=\wd0 \setbox0=\hbox{s}
        \gl@bal\advance\d@twidth by -\wd0 }
\def\removehglue{\loop \unskip \ifdim\lastskip >\z@ \repeat }
\def\roll@ver#1{\removehglue \nobreak \count255 =\spacefactor \dimen@=\z@
        \ifnum\count255 =3001 \dimen@=\d@twidth \fi
        \ifnum\count255 =1251 \dimen@=\d@twidth \fi
    \iftwelv@ \kern-\dimen@ \else \kern-0.83\dimen@ \fi
   #1\spacefactor=\count255 }
\def\step@ver#1{\rel@x \ifmmode #1\else \ifhmode
        \roll@ver{${}#1$}\else {\setbox0=\hbox{${}#1$}}\fi\fi }
\def\attach#1{\step@ver{\strut^{\mkern 2mu #1} }}
%
%
%
\newcount\chapternumber      \chapternumber=0
\newcount\sectionnumber      \sectionnumber=0
\newcount\equanumber         \equanumber=0
\let\chapterlabel=\rel@x
\let\sectionlabel=\rel@x
\newtoks\chapterstyle        \chapterstyle={\Number}
\newtoks\sectionstyle        \sectionstyle={\Number}
\newskip\chapterskip         \chapterskip=\bigskipamount
\newskip\sectionskip         \sectionskip=\medskipamount
\newskip\headskip            \headskip=8pt plus 3pt minus 3pt
\newdimen\chapterminspace    \chapterminspace=15pc
\newdimen\sectionminspace    \sectionminspace=10pc
\newdimen\referenceminspace  \referenceminspace=15pc
\newif\ifcn@                 \cn@true
\newif\ifcn@@                \cn@@false
\def\numberedchapters{\cn@true}
\def\unnumberedchapters{\cn@false\sequentialequations}
\def\chapterreset{\gl@bal\advance\chapternumber by 1
   \ifnum\equanumber<0 \else\gl@bal\equanumber=0\fi
   \sectionnumber=0 \let\sectionlabel=\rel@x
   \ifcn@ \gl@bal\cn@@true {\pr@tect
       \xdef\chapterlabel{\the\chapterstyle{\the\chapternumber}}}%
    \else \gl@bal\cn@@false \gdef\chapterlabel{\rel@x}\fi }
\def\@alpha#1{\count255='140 \advance\count255 by #1\char\count255}
 \def\alphabetic{\n@expand\@alpha}
\def\@Alpha#1{\count255='100 \advance\count255 by #1\char\count255}
 \def\Alphabetic{\n@expand\@Alpha}
\def\@Roman#1{\uppercase\expandafter{\romannumeral #1}}
 \def\Roman{\n@expand\@Roman}
\def\@roman#1{\romannumeral #1}    \def\roman{\n@expand\@roman}
\def\@number#1{\number #1}         \def\Number{\n@expand\@number}
\def\BLANK#1{\rel@x}               
\def\titleparagraphs{\interlinepenalty=9999
     \leftskip=0.03\hsize plus 0.22\hsize minus 0.03\hsize
     \rightskip=\leftskip \parfillskip=0pt
     \hyphenpenalty=9000 \exhyphenpenalty=9000
     \tolerance=9999 \pretolerance=9000
     \spaceskip=0.333em \xspaceskip=0.5em }
\def\titlestyle#1{\par\begingroup \titleparagraphs
     \iftwelv@\fourteenpoint\else\twelvepoint\fi
   \noindent #1\par\endgroup }
\def\spacecheck#1{\dimen@=\pagegoal\advance\dimen@ by -\pagetotal
   \ifdim\dimen@<#1 \ifdim\dimen@>0pt \vfil\break \fi\fi}
\def\chapter#1{\par \penalty-300 \vskip\chapterskip
   \spacecheck\chapterminspace
   \chapterreset \titlestyle{\ifcn@@\chapterlabel.~\fi #1}
   \nobreak\vskip\headskip \penalty 30000
   {\pr@tect\wlog{\string\chapter\space \chapterlabel}} }

\def\section#1{\par \ifnum\lastpenalty=30000\else
   \penalty-200\vskip\sectionskip \spacecheck\sectionminspace\fi
   \gl@bal\advance\sectionnumber by 1
   {\pr@tect
   \xdef\sectionlabel{\ifcn@@ \chapterlabel.\fi
       \the\sectionstyle{\the\sectionnumber}}%
   \wlog{\string\section\space \sectionlabel}}%
   \noindent {\caps\enspace\sectionlabel.~~#1}\par
   \nobreak\vskip\headskip \penalty 30000 }
\def\subsection#1{\par
   \ifnum\the\lastpenalty=30000\else \penalty-100\smallskip \fi
   \noindent\undertext{#1}\enspace \vadjust{\penalty5000}}

\def\undertext#1{\vtop{\hbox{#1}\kern 1pt \hrule}}

\def\ack{\subsection{Acknowledgements:}}
\def\APPENDIX#1#2{\par\penalty-300\vskip\chapterskip
   \spacecheck\chapterminspace \chapterreset \xdef\chapterlabel{#1}
   \titlestyle{APPENDIX #2} \nobreak\vskip\headskip \penalty 30000
   \wlog{\string\Appendix~\chapterlabel} }
\def\Appendix#1{\APPENDIX{#1}{#1}}
\def\appendix{\APPENDIX{A}{}}
%
%
%
\def\eqname#1{\rel@x {\pr@tect
  \ifnum\equanumber<0 \xdef#1{{\rm(\number-\equanumber)}}%
     \gl@bal\advance\equanumber by -1
  \else \gl@bal\advance\equanumber by 1
   \xdef#1{{\rm(\ifcn@@ \chapterlabel.\fi \number\equanumber)}}\fi
  }#1}
\def\eq{\eqname\?}
\def\eqn{\eqno\eqname}

\def\eqinsert#1{\noalign{\dimen@=\prevdepth \nointerlineskip
   \setbox0=\hbox to\displaywidth{\hfil #1}
   \vbox to 0pt{\kern 0.5\baselineskip\hbox{$\!\box0\!$}\vss}
   \prevdepth=\dimen@}}
%

%
%
\def\GENITEM#1;#2{\par \hangafter=0 \hangindent=#1
    \Textindent{$ #2 $}\ignorespaces}
\outer\def\newitem#1=#2;{\gdef#1{\GENITEM #2;}}

\newdimen\itemsize                \itemsize=30pt
\newitem\item=1\itemsize;
\newitem\sitem=1.75\itemsize;     
\newitem\ssitem=2.5\itemsize;     
\outer\def\newlist#1=#2&#3&#4;{\toks0={#2}\toks1={#3}%
   \count255=\escapechar \escapechar=-1
   \alloc@0\list\countdef\insc@unt\listcount     \listcount=0
   \edef#1{\par
      \countdef\listcount=\the\allocationnumber
      \advance\listcount by 1
      \hangafter=0 \hangindent=#4
      \Textindent{\the\toks0{\listcount}\the\toks1}}
   \expandafter\expandafter\expandafter
    \edef\c@t#1{begin}{\par
      \countdef\listcount=\the\allocationnumber \listcount=1
      \hangafter=0 \hangindent=#4
      \Textindent{\the\toks0{\listcount}\the\toks1}}
   \expandafter\expandafter\expandafter
    \edef\c@t#1{con}{\par \hangafter=0 \hangindent=#4 \noindent}
   \escapechar=\count255}
\def\c@t#1#2{\csname\string#1#2\endcsname}
\newlist\point=\Number&.&1.0\itemsize;
\newlist\subpoint=(\alphabetic&)&1.75\itemsize;
\newlist\subsubpoint=(\roman&)&2.5\itemsize;
%

%
%
%
%
\newcount\referencecount     \referencecount=0
\newcount\lastrefsbegincount \lastrefsbegincount=0
\newif\ifreferenceopen       \newwrite\referencewrite
\newdimen\refindent          \refindent=30pt
\def\normalrefmark#1{\attach{\scriptscriptstyle [ #1 ] }}
\let\PRrefmark=\attach
\def\NPrefmark#1{\step@ver{{\;[#1]}}}
\def\refmark#1{\rel@x\ifPhysRev\PRrefmark{#1}\else\normalrefmark{#1}\fi}
\def\refend@{\refmark{\number\referencecount}}
\def\refend{\refend@{}\space }
\def\refsend{\refmark{\count255=\referencecount
   \advance\count255 by-\lastrefsbegincount
   \ifcase\count255 \number\referencecount
   \or \number\lastrefsbegincount,\number\referencecount
   \else \number\lastrefsbegincount-\number\referencecount \fi}\space }
\def\REFNUM#1{\rel@x \gl@bal\advance\referencecount by 1
    \xdef#1{\the\referencecount }}
\def\Refnum#1{\REFNUM #1\refend@ } 
\def\REF#1{\REFNUM #1\R@FWRITE\ignorespaces}
\def\Ref#1{\Refnum #1\REFWRITE }
\def\ref{\Ref\?}
\def\REFS#1{\REFNUM #1\gl@bal\lastrefsbegincount=\referencecount
    \REFWRITE }

\def\r@fitem#1{\par \hangafter=0 \hangindent=\refindent \Textindent{#1}}
\def\refitem#1{\r@fitem{#1.}}
\def\NPrefitem#1{\r@fitem{[#1]}}
\def\NPrefs{\let\refmark=\NPrefmark \let\refitem=NPrefitem}
\def\REFWRITE{\R@FWRITE\rel@x }
\def\R@FWRITE#1{\ifreferenceopen \else \gl@bal\referenceopentrue
     \immediate\openout\referencewrite=\jobname.refs
     \toks@={\begingroup \refoutspecials \catcode`\^^M=10 }%
     \immediate\write\referencewrite{\the\toks@}\fi
    \immediate\write\referencewrite{\noexpand\refitem %
                                    {\the\referencecount}}%
    \p@rse@ndwrite \referencewrite #1}
\begingroup
 \catcode`\^^M=\active \let^^M=\relax %
 \gdef\p@rse@ndwrite#1#2{\begingroup \catcode`\^^M=12 \newlinechar=`\^^M%
         \chardef\rw@write=#1\sc@nlines#2}%
 \gdef\sc@nlines#1#2{\sc@n@line \g@rbage #2^^M\endsc@n \endgroup #1}%
 \gdef\sc@n@line#1^^M{\expandafter\toks@\expandafter{\deg@rbage #1}%
         \immediate\write\rw@write{\the\toks@}%
         \futurelet\n@xt \sc@ntest }%
\endgroup
\def\sc@ntest{\ifx\n@xt\endsc@n \let\n@xt=\rel@x
       \else \let\n@xt=\sc@n@notherline \fi \n@xt }
\def\sc@n@notherline{\sc@n@line \g@rbage }
\def\deg@rbage#1{}
\let\g@rbage=\relax    \let\endsc@n=\relax
\def\refout{\par\penalty-400\vskip\chapterskip
   \spacecheck\referenceminspace
   \ifreferenceopen \Closeout\referencewrite \referenceopenfalse \fi
   \line{\fourteenrm\hfil REFERENCES\hfil}\vskip\headskip
   \input \jobname.refs
   }
\def\refoutspecials{\sfcode`\.=1000 \interlinepenalty=1000
         \rightskip=\z@ plus 1em minus \z@ }
\def\Closeout#1{\toks0={\par\endgroup}\immediate\write#1{\the\toks0}%
   \immediate\closeout#1}
%
%
\newcount\figurecount     \figurecount=0
\newcount\tablecount      \tablecount=0
\newif\iffigureopen       \newwrite\figurewrite
\newif\iftableopen        \newwrite\tablewrite
\def\FIGNUM#1{\rel@x \gl@bal\advance\figurecount by 1
    \xdef#1{\the\figurecount}}
\def\FIGURE#1{\FIGNUM #1\F@GWRITE\ignorespaces }

\def\figitem#1{\r@fitem{#1)}}
\def\FIGWRITE{\F@GWRITE\rel@x }
\def\TABNUM#1{\rel@x \gl@bal\advance\tablecount by 1
    \xdef#1{\the\tablecount}}
\def\TABLE#1{\TABNUM #1\T@BWRITE\ignorespaces }

\def\tabitem#1{\r@fitem{#1:}}
\def\TABWRITE{\T@BWRITE\rel@x }
\def\F@GWRITE#1{\iffigureopen \else \gl@bal\figureopentrue
     \immediate\openout\figurewrite=\jobname.figs
     \toks@={\begingroup \catcode`\^^M=10 }%
     \immediate\write\figurewrite{\the\toks@}\fi
    \immediate\write\figurewrite{\noexpand\figitem %
                                 {\the\figurecount}}%
    \p@rse@ndwrite \figurewrite #1}
\def\T@BWRITE#1{\iftableopen \else \gl@bal\tableopentrue
     \immediate\openout\tablewrite=\jobname.tabs
     \toks@={\begingroup \catcode`\^^M=10 }%
     \immediate\write\tablewrite{\the\toks@}\fi
    \immediate\write\tablewrite{\noexpand\tabitem %
                                 {\the\tablecount}}%
    \p@rse@ndwrite \tablewrite #1}
\def\figout{\par\penalty-400
   \vskip\chapterskip\spacecheck\referenceminspace
   \iffigureopen \Closeout\figurewrite \figureopenfalse \fi
   \line{\fourteenrm\hfil FIGURE CAPTIONS\hfil}\vskip\headskip
   \input \jobname.figs
   }
\def\tabout{\par\penalty-400
   \vskip\chapterskip\spacecheck\referenceminspace
   \iftableopen \Closeout\tablewrite \tableopenfalse \fi
   \line{\fourteenrm\hfil TABLE CAPTIONS\hfil}\vskip\headskip
   \input \jobname.tabs
   }
%
%
%
\newbox\picturebox
\def\p@cht{\ht\picturebox }
\def\p@cwd{\wd\picturebox }
\def\p@cdp{\dp\picturebox }
\newdimen\xshift
\newdimen\yshift
\newdimen\captionwidth
\newskip\captionskip
\captionskip=15pt plus 5pt minus 3pt
\def\fullwidth{\captionwidth=\hsize }
\newtoks\Caption
\newif\ifcaptioned
\newif\ifselfcaptioned
\def\caption{\captionedtrue \Caption }
\newcount\linesabove
\newif\iffileexists
\newtoks\picfilename
\def\fil@#1 {\fileexiststrue \picfilename={#1}}
\def\file#1{\if=#1\let\n@xt=\fil@ \else \def\n@xt{\fil@ #1}\fi \n@xt }
\def\pl@t{\begingroup \pr@tect
    \setbox\picturebox=\hbox{}\fileexistsfalse
    \let\height=\p@cht \let\width=\p@cwd \let\depth=\p@cdp
    \xshift=\z@ \yshift=\z@ \captionwidth=\z@
    \Caption={}\captionedfalse
    \linesabove =0 \picturedefault }
\def\plot{\pl@t \selfcaptionedfalse }
\def\Picture#1{\gl@bal\advance\figurecount by 1
    \xdef#1{\the\figurecount}\pl@t \selfcaptionedtrue }

\def\s@vepicture{\iffileexists \parsefilename \redopicturebox \fi
   \ifdim\captionwidth>\z@ \else \captionwidth=\p@cwd \fi
   \xdef\lastpicture{\iffileexists
        \setbox0=\hbox{\raise\the\yshift \vbox{%
              \moveright\the\xshift\hbox{\picturedefinition}}}%
        \else \setbox0=\hbox{}\fi
         \ht0=\the\p@cht \wd0=\the\p@cwd \dp0=\the\p@cdp
         \vbox{\hsize=\the\captionwidth \line{\hss\box0 \hss }%
              \ifcaptioned \vskip\the\captionskip \noexpand\Tenpoint
                \ifselfcaptioned Figure~\the\figurecount.\enspace \fi
                \the\Caption \fi }}%
    \endgroup }
\let\endpicture=\s@vepicture
\def\savepicture#1{\s@vepicture \global\let#1=\lastpicture }
\def\displaypicture{\fullwidth \s@vepicture $$\lastpicture $${}}
\def\toppicture{\fullwidth \s@vepicture \topinsert
    \lastpicture \medskip \endinsert }
\def\midpicture{\fullwidth \s@vepicture \midinsert
    \lastpicture \endinsert }
%
%
\def\leftpicture{\pres@tpicture
    \dimen@i=\hsize \advance\dimen@i by -\dimen@ii
    \setbox\picturebox=\hbox to \hsize {\box0 \hss }%
    \wr@paround }
\def\rightpicture{\pres@tpicture
    \dimen@i=\z@
    \setbox\picturebox=\hbox to \hsize {\hss \box0 }%
    \wr@paround }
\def\pres@tpicture{\gl@bal\linesabove=\linesabove
    \s@vepicture \setbox\picturebox=\vbox{
         \kern \linesabove\baselineskip \kern 0.3\baselineskip
         \lastpicture \kern 0.3\baselineskip }%
    \dimen@=\p@cht \dimen@i=\dimen@
    \advance\dimen@i by \pagetotal
    \par \ifdim\dimen@i>\pagegoal \vfil\break \fi
    \dimen@ii=\hsize
    \advance\dimen@ii by -\parindent \advance\dimen@ii by -\p@cwd
    \setbox0=\vbox to\z@{\kern-\baselineskip \unvbox\picturebox \vss }}
\def\wr@paround{\Caption={}\count255=1
    \loop \ifnum \linesabove >0
         \advance\linesabove by -1 \advance\count255 by 1
         \advance\dimen@ by -\baselineskip
         \expandafter\Caption \expandafter{\the\Caption \z@ \hsize }%
      \repeat
    \loop \ifdim \dimen@ >\z@
         \advance\count255 by 1 \advance\dimen@ by -\baselineskip
         \expandafter\Caption \expandafter{%
             \the\Caption \dimen@i \dimen@ii }%
      \repeat
    \edef\n@xt{\parshape=\the\count255 \the\Caption \z@ \hsize }%
    \par\noindent \n@xt \strut \vadjust{\box\picturebox }}
\let\picturedefault=\relax
\let\parsefilename=\relax
\def\redopicturebox{\let\picturedefinition=\rel@x
   \errhelp=\disabledpictures
   \errmessage{This version of TeX cannot handle pictures.  Sorry.}}
\newhelp\disabledpictures
     {You will get a blank box in place of your picture.}
%
%
%
%
%
%
%
%
%
%
\def\FRONTPAGE{\ifvoid255\else\vfill\penalty-20000\fi
   \gl@bal\pagenumber=1     \gl@bal\chapternumber=0
   \gl@bal\equanumber=0     \gl@bal\sectionnumber=0
   \gl@bal\referencecount=0 \gl@bal\figurecount=0
   \gl@bal\tablecount=0     \gl@bal\frontpagetrue
   \gl@bal\lastf@@t=0       \gl@bal\footsymbolcount=0
   \gl@bal\cn@@false }

\def\papers{\papersize\headline=\paperheadline\footline=\paperfootline}
\def\papersize{\hsize=35pc \vsize=50pc \hoffset=0pc \voffset=1pc
   \advance\hoffset by\HOFFSET \advance\voffset by\VOFFSET
   \pagebottomfiller=0pc
   \skip\footins=\bigskipamount \normalspace }
\papers  
%
%
\newskip\lettertopskip       \lettertopskip=20pt plus 50pt
\newskip\letterbottomskip    \letterbottomskip=\z@ plus 100pt
\newskip\signatureskip       \signatureskip=40pt plus 3pt
\def\lettersize{\hsize=6.5in \vsize=8.5in \hoffset=0in \voffset=0.5in
   \advance\hoffset by\HOFFSET \advance\voffset by\VOFFSET
   \pagebottomfiller=\letterbottomskip
   \skip\footins=\smallskipamount \multiply\skip\footins by 3
   \singlespace }
\def\MEMO{\lettersize \headline=\letterheadline \footline={\hfil }%
   \let\rule=\memorule \FRONTPAGE \memohead }

\def\memodate{\afterassignment\MEMO \date }
\def\memit@m#1{\smallskip \hangafter=0 \hangindent=1in
    \Textindent{\caps #1}}
\def\subject{\memit@m{Subject:}}
\def\topic{\memit@m{Topic:}}
\def\from{\memit@m{From:}}
\def\to{\rel@x \ifmmode \rightarrow \else \memit@m{To:}\fi }
\def\memorule{\medskip\hrule height 1pt\bigskip}  
\def\memohead{\centerline{\fourteenrm MEMORANDUM}}
\newwrite\labelswrite
\newtoks\rw@toks
\def\letters{\lettersize
   \headline=\letterheadline \footline=\letterfootline
   \immediate\openout\labelswrite=\jobname.lab}

\let\letterhead=\rel@x
\def\addressee#1{\medskip\line{\hskip 0.75\hsize plus\z@ minus 0.25\hsize
                               \the\date \hfil }%
   \vskip \lettertopskip
   \ialign to\hsize{\strut ##\hfil\tabskip 0pt plus \hsize \crcr #1\crcr}
   \writelabel{#1}\medskip \noindent\hskip -\spaceskip \ignorespaces }
\def\rwl@begin#1\cr{\rw@toks={#1\crcr}\rel@x
   \immediate\write\labelswrite{\the\rw@toks}\futurelet\n@xt\rwl@next}
\def\rwl@next{\ifx\n@xt\rwl@end \let\n@xt=\rel@x
      \else \let\n@xt=\rwl@begin \fi \n@xt}
\let\rwl@end=\rel@x
\def\writelabel#1{\immediate\write\labelswrite{\noexpand\labelbegin}
     \rwl@begin #1\cr\rwl@end
     \immediate\write\labelswrite{\noexpand\labelend}}
\newtoks\FromAddress         \FromAddress={}
\newtoks\sendername          \sendername={}
\newbox\FromLabelBox
\newdimen\labelwidth          \labelwidth=6in
\def\makelabels{\afterassignment\Makelabels \sendername=}
\def\Makelabels{\FRONTPAGE \letterinfo={\hfil } \MakeFromBox
     \immediate\closeout\labelswrite  \input \jobname.lab\vfil\eject}
\let\labelend=\rel@x
\def\labelbegin#1\labelend{\setbox0=\vbox{\ialign{##\hfil\cr #1\crcr}}
     \MakeALabel }
\def\MakeFromBox{\gl@bal\setbox\FromLabelBox=\vbox{\Tenpoint
     \ialign{##\hfil\cr \the\sendername \the\FromAddress \crcr }}}
\def\MakeALabel{\vskip 1pt \hbox{\vrule \vbox{
        \hsize=\labelwidth \hrule\bigskip
        \leftline{\hskip 1\parindent \copy\FromLabelBox}\bigskip
        \centerline{\hfil \box0 } \bigskip \hrule
        }\vrule } \vskip 1pt plus 1fil }
\def\signed#1{\par \nobreak \bigskip \dt@pfalse \begingroup
  \everycr={\noalign{\nobreak
            \ifdt@p\vskip\signatureskip\gl@bal\dt@pfalse\fi }}%
  \tabskip=0.5\hsize plus \z@ minus 0.5\hsize
  \halign to\hsize {\strut ##\hfil\tabskip=\z@ plus 1fil minus \z@\crcr
          \noalign{\gl@bal\dt@ptrue}#1\crcr }%
  \endgroup \bigskip }
\newbox\letterb@x
\def\lettertext{\par \vskip\parskip \unvcopy\letterb@x \par }
\def\multiletter{\setbox\letterb@x=\vbox\bgroup
      \everypar{\vrule height 1\baselineskip depth 0pt width 0pt }
      \singlespace \topskip=\baselineskip }
\def\letterend{\par\egroup}
%
%
%
\newskip\frontpageskip
\newtoks\Pubnum   
\newtoks\Pubtype  \let\pubtype=\Pubtype
\newif\ifp@bblock  \p@bblocktrue
\def\PH@SR@V{\doubl@true \baselineskip=24.1pt plus 0.2pt minus 0.1pt
             \parskip= 3pt plus 2pt minus 1pt }
\def\PHYSREV{\papers\PhysRevtrue\PH@SR@V}

\def\titlepage{\FRONTPAGE\papers\ifPhysRev\PH@SR@V\fi
   \ifp@bblock\p@bblock \else\hrule height\z@ \rel@x \fi }
\def\nopubblock{\p@bblockfalse}
\def\endpage{\vfil\break}
\frontpageskip=12pt plus .5fil minus 2pt
\Pubtype={}
\Pubnum={}
\def\p@bblock{\begingroup \tabskip=\hsize minus \hsize
   \baselineskip=1.5\ht\strutbox \topspace-2\baselineskip
   \halign to\hsize{\strut ##\hfil\tabskip=0pt\crcr
       \the\Pubnum\crcr\the\date\crcr\the\pubtype\crcr}\endgroup}
\def\title#1{\vskip\frontpageskip \titlestyle{#1} \vskip\headskip }
\def\author#1{\vskip\frontpageskip\titlestyle{\twelvecp #1}\nobreak}
\def\andauthor{\vskip\frontpageskip\centerline{and}\author}

\def\address#1{\par\kern 5pt\titlestyle{\twelvepoint\it #1}}
\def\andaddress{\par\kern 5pt \centerline{\sl and} \address}

\def\JHL{\address{Joseph Henry Laboratories\break
      Princeton University\break Princeton, New Jersey 08544}}
\def\abstract{\par\dimen@=\prevdepth \hrule height\z@ \prevdepth=\dimen@
   \vskip\frontpageskip\centerline{\fourteenrm ABSTRACT}\vskip\headskip }

%
%
%
\def\ie{\hbox{\it i.e.}}       \def\etc{\hbox{\it etc.}}

\def\\{\rel@x \ifmmode \backslash \else {\tt\char`\\}\fi }
\def\sequentialequations{\rel@x \if\equanumber<0 \else
  \gl@bal\equanumber=-\equanumber \gl@bal\advance\equanumber by -1 \fi }
\def\journal#1&#2(#3){\begingroup \let\journal=\dummyj@urnal
    \unskip, \sl #1\unskip~\bf\ignorespaces #2\rm
    (\afterassignment\j@ur \count255=#3), \endgroup\ignorespaces }
\def\j@ur{\ifnum\count255<100 \advance\count255 by 1900 \fi
          \number\count255 }
\def\dummyj@urnal{%
    \toks@={Reference foul up: nested \journal macros}%
    \errhelp={Your forgot & or ( ) after the last \journal}%
    \errmessage{\the\toks@ }}

\def\topspace{\hrule height 0pt depth 0pt \vskip}
\def\coeff#1#2{{\textstyle{#1\over #2}}}
\def\half{\coeff12 }
\def\partder#1#2{{\partial #1\over\partial #2}}
\def\Buildrel#1\under#2{\mathrel{\mathop{#2}\limits_{#1}}}
\def\becomes#1{\mathchoice{\becomes@\scriptstyle{#1}}
   {\becomes@\scriptstyle{#1}} {\becomes@\scriptscriptstyle{#1}}
   {\becomes@\scriptscriptstyle{#1}}}
\def\becomes@#1#2{\mathrel{\setbox0=\hbox{$\m@th #1{\,#2\,}$}%
        \mathop{\hbox to \wd0 {\rightarrowfill}}\limits_{#2}}}
\def\bra#1{\left\langle #1\right|}
\def\ket#1{\left| #1\right\rangle}

\def\VEV#1{\left\langle #1\right\rangle}

\def\Tr{\mathop{\rm Tr}\nolimits}

\let\int=\intop         \let\oint=\ointop
\def\lsim{\mathrel{\mathpalette\@versim<}}
\def\gsim{\mathrel{\mathpalette\@versim>}}
\def\@versim#1#2{\vcenter{\offinterlineskip
        \ialign{$\m@th#1\hfil##\hfil$\crcr#2\crcr\sim\crcr } }}
\def\big#1{{\hbox{$\left#1\vbox to 0.85\b@gheight{}\right.\n@space$}}}
\def\Big#1{{\hbox{$\left#1\vbox to 1.15\b@gheight{}\right.\n@space$}}}
\def\bigg#1{{\hbox{$\left#1\vbox to 1.45\b@gheight{}\right.\n@space$}}}
\def\Bigg#1{{\hbox{$\left#1\vbox to 1.75\b@gheight{}\right.\n@space$}}}
\def\){\mskip 2mu\nobreak }
%
%
%
\let\sec@nt=\sec
\def\sec{\rel@x\ifmmode\let\n@xt=\sec@nt\else\let\n@xt\section\fi\n@xt}
\def\obsolete#1{\message{Macro \string #1 is obsolete.}}
\def\firstsec#1{\obsolete\firstsec \section{#1}}
\def\firstsubsec#1{\obsolete\firstsubsec \subsection{#1}}
\def\thispage#1{\obsolete\thispage \gl@bal\pagenumber=#1\frontpagefalse}
\def\thischapter#1{\obsolete\thischapter \gl@bal\chapternumber=#1}
\def\splitout{\obsolete\splitout\rel@x}
\def\prop{\obsolete\prop \propto }
\def\nextequation#1{\obsolete\nextequation \gl@bal\equanumber=#1
   \ifnum\the\equanumber>0 \gl@bal\advance\equanumber by 1 \fi}
\def\BOXITEM{\afterassigment\B@XITEM\setbox0=}
\def\B@XITEM{\par\hangindent\wd0 \noindent\box0 }
%
%
%
\def\phyzzx{PHY\setbox0=\hbox{Z}\copy0 \kern-0.5\wd0 \box0 X}
        
\everyjob{\xdef\today{\monthname~\number\day, \number\year}
        \input myphyx.tex }
\message{ by V.K.}
\input phyzzx.local
\catcode`\@=12 
%

%
%
%
%
\font\seventeenrm =cmr17
\font\fourteenrm  =cmr12 scaled\magstep1
\font\twelverm    =cmr12
\font\ninerm      =cmr9
\font\sixrm       =cmr6

\font\fourteenbf  =cmbx12 scaled\magstep1
\font\twelvebf    =cmbx12
\font\ninebf      =cmbx9
\font\sixbf       =cmbx6
\font\seventeeni  =cmmi12 scaled\magstep2    \skewchar\seventeeni='177
\font\fourteeni   =cmmi12 scaled\magstep1     \skewchar\fourteeni='177
\font\twelvei     =cmmi12                       \skewchar\twelvei='177
\font\ninei       =cmmi9                          \skewchar\ninei='177
\font\sixi        =cmmi6                           \skewchar\sixi='177
\font\seventeensy =cmsy10 scaled\magstep3    \skewchar\seventeensy='60
\font\fourteensy  =cmsy10 scaled\magstep2     \skewchar\fourteensy='60
\font\twelvesy    =cmsy10 scaled\magstep1       \skewchar\twelvesy='60
\font\ninesy      =cmsy9                          \skewchar\ninesy='60
\font\sixsy       =cmsy6                           \skewchar\sixsy='60

\font\fourteenex  =cmex10 scaled\magstep2
\font\twelveex    =cmex10 scaled\magstep1

\font\fourteensl  =cmsl12 scaled\magstep1
\font\twelvesl    =cmsl12
\font\ninesl      =cmsl9

\font\fourteenit  =cmti12 scaled\magstep1
\font\twelveit    =cmti12
\font\nineit      =cmti9
\font\fourteentt  =cmtt12 scaled\magstep1
\font\twelvett    =cmtt12
\font\fourteencp  =cmcsc10 scaled\magstep2
\font\twelvecp    =cmcsc10 scaled\magstep1
\font\tencp       =cmcsc10
%
%
\def\fourteenf@nts{\relax
    \textfont0=\fourteenrm          \scriptfont0=\tenrm
      \scriptscriptfont0=\sevenrm
    \textfont1=\fourteeni           \scriptfont1=\teni
      \scriptscriptfont1=\seveni
    \textfont2=\fourteensy          \scriptfont2=\tensy
      \scriptscriptfont2=\sevensy
    \textfont3=\fourteenex          \scriptfont3=\twelveex
      \scriptscriptfont3=\tenex
    \textfont\itfam=\fourteenit     \scriptfont\itfam=\tenit
    \textfont\slfam=\fourteensl     \scriptfont\slfam=\tensl
    \textfont\bffam=\fourteenbf     \scriptfont\bffam=\tenbf
      \scriptscriptfont\bffam=\sevenbf
    \textfont\ttfam=\fourteentt
    \textfont\cpfam=\fourteencp }
\def\twelvef@nts{\relax
    \textfont0=\twelverm          \scriptfont0=\ninerm
      \scriptscriptfont0=\sixrm
    \textfont1=\twelvei           \scriptfont1=\ninei
      \scriptscriptfont1=\sixi
    \textfont2=\twelvesy           \scriptfont2=\ninesy
      \scriptscriptfont2=\sixsy
    \textfont3=\twelveex          \scriptfont3=\tenex
      \scriptscriptfont3=\tenex
    \textfont\itfam=\twelveit     \scriptfont\itfam=\nineit
    \textfont\slfam=\twelvesl     \scriptfont\slfam=\ninesl
    \textfont\bffam=\twelvebf     \scriptfont\bffam=\ninebf
      \scriptscriptfont\bffam=\sixbf
    \textfont\ttfam=\twelvett
    \textfont\cpfam=\twelvecp }
\def\tenf@nts{\relax
    \textfont0=\tenrm          \scriptfont0=\sevenrm
      \scriptscriptfont0=\fiverm
    \textfont1=\teni           \scriptfont1=\seveni
      \scriptscriptfont1=\fivei
    \textfont2=\tensy          \scriptfont2=\sevensy
      \scriptscriptfont2=\fivesy
    \textfont3=\tenex          \scriptfont3=\tenex
      \scriptscriptfont3=\tenex
    \textfont\itfam=\tenit     \scriptfont\itfam=\seveni  
    \textfont\slfam=\tensl     \scriptfont\slfam=\sevenrm 
    \textfont\bffam=\tenbf     \scriptfont\bffam=\sevenbf
      \scriptscriptfont\bffam=\fivebf
    \textfont\ttfam=\tentt
    \textfont\cpfam=\tencp }
%
%
%
%
%
%
%
%
%
%
%
%
%
%

\Pubnum={}
\pubtype={}

\def\memohead{\line{\fourteenrm Princeton University:\ \twelverm
     Department of Physics.\hfil\twelveit \the\date}\medskip}

\def\JHLhead{\leftline{\vbox{\baselineskip=17pt
    \def\\{\kern 1pt} \ialign{&\sixbf ##\hfil\cr
     \fourteenrm Princeton University\quad &
     \sevenbf DEPARTMENT OF PHYSICS\\: JOSEPH HENRY LABORATORIES\cr
     & JADWIN HALL\cr & POST OFFICE BOX \teni\\7\\0\\8\cr
     & PRINCETON, NEW JERSEY \teni\\0\\8\\5\\4\\4\cr } }}\medskip }
\let\letterhead=\JHLhead

\FromAddress={\crcr Joseph Henry Laboratories\cr Princeton University\cr
    Princeton, New Jersey 08544\crcr}

\letterinfo={\ninerm \hfil Telephone: (609) 258--\phoneext\qquad
                      Telex: (609) 258--6360\hfil }
\def\phoneext{4400}

\edef\memorule{\medskip\hrule\kern 2pt\hrule \noindent
      \llap{\vbox to 0pt{ \vskip 1in\normalbaselines \tabskip=0pt plus 1fil
        \halign to 0.99in{\seventeenrm\hfil ##\hfil\cr
          M\cr E\cr M\cr O\cr R\cr A\cr N\cr D\cr U\cr M\cr}
        \vss }}\par\medskip}

%
%

%
%

\let\rule=\memorule

\def\pri{^{\, \prime }}

\def\boxit#1{\vbox{\hrule\hbox{\vrule\kern3pt
\vbox{\kern3pt#1\kern3pt}\kern3pt\vrule}\hrule}}

%
%

\input epsf

\def\tablist#1{\singlespace\halign{\tabskip0pt
\vtop{\hsize2.5in\noindent##\vfil}\tabskip20pt &
\vtop{\hsize2.5in\noindent##\vfil}\tabskip0pt\cr#1}}

\hsize=6.0in
\sequentialequations

\def\rk#1{$^{#1}$}                        
\overfullrule=0pt
\catcode`\@=11

\def\dsl{double-scaling limit}
\def \PH{\hat \psi}
\def \PE{\psi^\epsilon}
\def \E{\epsilon}

\def \L{\lambda}

\def \T{\theta}
\def \P{\psi}
\def \D{\delta}
\def \DX{\Delta x}
\def \W{\omega}

\def \sh{\; {\rm sh} \;}
\def \ch{\; {\rm ch} \;}
\def \th{\; {\rm th} \;}
\def \cth{\; {\rm cth} \;}
\def \DM{ {\partial \over {\partial \mu}}}
\def \O{{\cal O}}
\def \CO{{\cal O}}
\def \G{\Gamma}
\def \tT{{\tilde T}}
\def \tq{{\tilde q}}

\def\CO{{\cal O}}
\def\e{\epsilon}
\def\KT{Kosterlitz-Thouless}
\def\eqaligntwo#1{\null\,\vcenter{\openup\jot\m@th
\ialign{\strut\hfil
$\displaystyle{##}$&$\displaystyle{{}##}$&$\displaystyle{{}##}$\hfil
\crcr#1\crcr}}\,}
\catcode`\@=12
\REF\RS{V.~Kazakov, \PL {\bf 150}, 282 (1985);
J.~Ambj\o rn, B.~Durhuus, and J.~Fr\"ohlich \NP
{\bf B257 }, 433 (1985),
 F. David, \NP {\bf B 257},
45 (1985); V.~Kazakov, I.~Kostov and A.~Migdal, \PL {\bf 157}, 295 (1985)}
\REF\GM{D.~J.~Gross and A.~A.~Migdal, \PRL {\bf 64} (1990) 717;
M. Douglas and S.~Shenker, \NP {\bf B335} (1990) 635;
E.~Brezin and V.~Kazakov, \PL {\bf 236B} (1990) 144.}
\REF\Ising{D. J. Gross and A. Migdal, \PRL {\bf 64} (1990) 717;
E. Brezin, M.~Douglas, V.~Kazakov and S.~Shenker, \PL {\bf 237B}
(1990) 43; C.~Crnkovic, P.~Ginsparg and G.~Moore, \PL {\bf 237B}
(1990) 196.}
\REF\Doug{ M. Douglas, \PL {\bf 238B} (1990) 176.}
\REF\GMil{D. J. Gross and N. Miljkovic \journal Phys. Lett.
& 238B (1990) 217; E. Brezin, V. Kazakov, and Al. B. Zamolodchikov
\journal Nucl. Phys. &B338 (1990) 673; P. Ginsparg and J. Zinn-Justin
\journal Phys. Lett. &240B (1990) 333; G. Parisi \journal
Phys. Lett. &238B (1990) 209.}
\REF\GKl{D. J. Gross and I. R. Klebanov, \NP {\bf B344} (1990) 475.}
\REF\ind{S. Das, S. Naik and S. Wadia, \Mod {\bf A4} (1989) 1033;
J. Polchinski, \NP {\bf B324 } (1989) 123;
S.~Das, A.~Dhar and S. Wadia, \Mod {\bf A5} (1990) 799;
T. Banks and J. Lykken, \NP {\bf B331} (1990) 173;
A. Tseytlin, \IJMP {\bf A5} (1990) 1833.}
\REF\BPIZ{E. Brezin, C. Itzykson, G. Parisi and J. Zuber,
\CMP {\bf 59} (1978) 35.}
\REF\dj{ S. R. Das and A. Jevicki \journal Mod. Phys. Lett. &A5 (1990) 1639.}
\REF\ferm{D. J. Gross and I. R. Klebanov 
\journal Nucl. Phys. &B352 (1991) 671.}
\REF\sasha{A. M. Polyakov, \PL {\bf 103B} (1981) 207.}
\REF\Vort{D. J. Gross and I. R. Klebanov 
 \journal Nucl. Phys. &B354 (1991) 459.}
\REF\Comp{I. R. Klebanov and D. Lowe, Princeton preprint PUPT-1256,
to appear in Nucl. Phys. B.}
\REF\kpz{V. Knizhnik, A. Polyakov and A. Zamolodchikov,
\Mod {\bf A3} (1988) 819.}
\REF\ddk{F. David, \Mod {\bf A3} (1988) 1651;
J. Distler and H. Kawai, \NP {\bf B321} (1989) 509.}
\REF\KazMig{V.~Kazakov and A.~Migdal, \NP {\bf B311} 
(1989) 171. }
\REF\MO{P. Marchesini and E. Onofri, \JMP {\bf 21} (1980) 1103.}
\REF\Shenker{S. Shenker, Rutgers preprint RU-90-47 (1990).}
\REF\Moore{G. Moore, Yale and Rutgers preprint YCTP-P8-91, RU-91-12.}
\REF\GKN{D. J. Gross, I. R. Klebanov and M. J. Newman
\journal Nucl. Phys. &B350 (1991) 621.}
\REF\Polch{J. Polchinski \journal Nucl. Phys. &B346 (1990) 253. }
\REF\thorn{T. L. Curtright and C. B. Thorn, \PRL {\bf 48} (1982) 1309. }
\REF\ns{N. Seiberg, Rutgers preprint RU-90-29.}
\REF\pol{J. Polchinski, Proceedings of {\it Strings '90},
Texas preprint UTTG-19-90 (1990).}
\REF\gupta{S. Gupta, A. Trivedi and M. Wise, \NP {\bf B340} (1990) 475.}
\REF\Goul{M. Goulian and M. Li, Santa Barbara preprint UCSBTH-90-61.}
\REF\dif{P. Di Francesco and D. Kutasov, \PL {\bf B261} (1991) 385.}
\REF\BK{M. Bershadsky and I. R. Klebanov \journal Phys. Rev. Lett.
&65 (1990) 3088; N. Sakai and Y. Tanii, Tokyo Inst. of Tech.
preprint TIT/HEP-160 (1990).}
\REF\BKl{M. Bershadsky and I. R. Klebanov \journal Nucl. Phys.
&B360 (1991) 559.} 
\REF\trick{B. McClain and B. D. B. Roth, \CMP {\bf 111} (1987) 539;
K. H. O'Brien and C.-I. Tan, \PR {\bf D36} (1987) 1184.}
\REF\jp{J. Polchinski, \CMP {\bf 104} (1986) 37.}
\REF\AMP{A. M. Polyakov, \Mod {\bf A6} (1991) 635.}
\REF\lz{B. Lian and G. Zuckerman, Yale preprint YCTP-P18-91.}
\REF\dvv{R. Dijkgraaf, E. Verlinde and H. Verlinde, \CMP
{\bf 115} (1988) 649.}
\REF\witten{E. Witten, paper in preparation.}
\REF\GD{U. H. Danielsson and D. J. Gross, Princeton preprint PUPT-1258.}
\REF\w{J. Avan and A. Jevicki, Brown preprint BROWN-HET-801;
D. Minic, J. Polchinski and Z. Yang, Univ. of Texas 
preprint UTTG-16-91. }
\REF\moors{G. Moore and N. Seiberg, Rutgers and Yale preprint RU-91-29, 
YCTP-P19-91.} 
\REF\senwad{A. Sengupta and S. Wadia, \IJMP {\bf A6} (1991) 1961. }
\REF\boson{S. Coleman, \PR {\bf D11} (1975) 2088;
S. Mandelstam, \PR {\bf D11} (1975) 3026.}
\REF\Jev{K. Demeterfi, A. Jevicki and J. P. Rodrigues,
Brown preprints  BROWN-HET-795 and BROWN-HET-803 (1991).}
\REF\Wad{G. Mandal, A. Sengupta and S. R. Wadia,
IAS preprint IASSNS-HEP/91/8 (1991).}
\REF\GKleb{D. J. Gross and I. R. Klebanov, \NP {\bf B359} (1991) 3.}
\REF\joe{J. Polchinski, University of Texas preprint UTTG-06-91.}
\REF\dual{P. Ginsparg and C. Vafa, \NP {\bf B289} (1987) 414; 
E. Alvarez and M. Osorio\journal Phys. Rev.
&D40 (89) 1150.}
\REF\bkt{V. L. Berezinskii, \JTP {\bf 34} (1972) 610;
M. Kosterlitz and D. Thouless, \JP {\bf C6} (1973) 1181.}
\REF\ov{J. Molera and B. Ovrut\journal Phys. Rev.
&D40 (89) 1146; M. J. Duff, preprint CTP-TAMU-53/89.}
\REF\villain{J. Villain, \JP {\bf C36} (1975) 581.}
\REF\kog{Ya. I. Kogan\journal JETP Lett. & 45 (87) 709;
B. Sathiapalan\journal Phys. Rev. &D35 (87) 3277.}
\Ref\yang{Z. Yang, \PL {\bf 243B} (1990) 365.}
\REF\Ed{E. Witten, Inst. for Advanced Study preprint IASSNS-HEP-91/12.}
\REF\KS{I. Klebanov and L. Susskind, \NP {\bf B309} (1988) 175}
\REF\Par{G.~Parisi, \PL {\bf 238B} (1990) 213.}

\def\CO{{\cal O}}
\def\eqaligntwo#1{\null\,\vcenter{\openup\jot\m@th
\ialign{\strut\hfil
$\displaystyle{##}$&$\displaystyle{{}##}$&$\displaystyle{{}##}$\hfil
\crcr#1\crcr}}\,}
\catcode`\@=12
\def\EPI{Euclidean path integral}
\def\sp{\,\,\,\,}
\def\oh{{1 \over 2}}
\def\b{\beta}
\def\a{\alpha}
\def\li{\lambda_i}
\def\l{\lambda}
\def\O{\Omega}

\def\m{\mu}

\def\ap{\alpha'}

\def\NP{{\it Nucl. Phys.\ }}

\def\PL{{\it Phys. Lett.\ }}
\def\PR{{\it Phys. Rev.\ }}
\def\PRL{{\it Phys. Rev. Lett.\ }}
\def\CMP{{\it Comm. Math. Phys.\ }}
\def\JMP{{\it J. Math. Phys.\ }}
\def\JTP{{\it JETP \ }}
\def\JP{{\it J. Phys.\ }}
\def\IJMP{{\it Int. Jour. Mod. Phys.\ }}
\def\Mod{{\it Mod. Phys. Lett.\ }}

\def\rs{random surfaces}
\def\qg{quantum gravity}
\def\ap{\alpha'}
\Pubnum={PUPT-1271}
\date={July 1991}
\titlepage
\hsize=6.0in
\centerline{ \fourteenpoint STRING THEORY IN TWO DIMENSIONS\foot
{Lectures delivered at the ICTP Spring School on String Theory
and Quantum Gravity, Trieste, April 1991.}}
\vskip .3in
\centerline { \fourteenpoint Igor R. Klebanov\foot
{Supported in part by DOE grant DE-AC02-76WRO3072
and by an NSF Presidential Young Investigator Award.} }
\JHL
\bigskip
\leftskip=1.5cm \rightskip=1.5cm
{\baselineskip=14pt\centerline{\fourteenpoint ABSTRACT}
I review some of the recent progress in two-dimensional string theory,
which is formulated as a sum over surfaces embedded in one dimension.
\par}
\leftskip=0pt\rightskip=0pt
\bigskip

\chapter{INTRODUCTION.}

These notes are an expanded version of the lectures I gave at the
1991 ICTP Spring School on String Theory and Quantum Gravity. Here I have
attempted to review, from my own personal viewpoint, some of the
exciting developments in two-dimensional string theory that have taken place
over the last year and a half. Because of the multitude of new results,
and since the field is still developing rapidly, a comprehensive
review must await a later date. These notes are mainly devoted
to the matrix model approach\rk\RS~ that has truly revolutionized the
two-dimensional Euclidean \qg. Recently this approach has led
to the exact solution of \qg\ coupled to conformal matter systems with
$c\leq 1$.$^{\GM-\GKl}$ 

These lectures are about the $c=1$ model$^{\GMil,\GKl}$ that is both
the richest physically and among the most easily soluble.
This model is defined by the sum over geometries embedded in
1 dimension. The resulting string theory is, however, 2-dimensional
because the dynamical conformal factor of the world sheet geometry
acts as an extra hidden string coordinate.\rk\ind~
In fact, this is the maximal
bosonic string theory that is well-defined perturbatively. 
If the dimensionality is increased any further, then a tachyon
appears in the string spectrum and renders the theory unstable.
The matrix model maps the theory of surfaces embedded in 1 dimension
into a non-relativistic quantum mechanics of free fermions,\rk\BPIZ~ 
from which
virtually any quantity can be calculated to all orders in the
genus expansion. The second-quantized field theory of non-relativistic
fermions can be regarded here as the exact string field theory.
A transformation of this theory to a rather simple interacting
boson representation$^{\dj,\ferm}$ will also be given.
I will argue that the 2-dimensional string theory
is the kind of toy model which possesses a remarkably simple 
structure, and at the same time incorporates some of the physics 
of string theories embedded in higher dimensions. The simplicity
of the theory is apparent in the matrix model approach, but
much of it remains obscure and mysterious from the point of view of
the continuum Polyakov path integral. Until we develop
a better insight into the ``miracles'' of the matrix models,
we may be sure that our understanding of string physics is very
incomplete.

These notes are mainly a review of published papers, but I
also included a few new observations and results.
In section 2 I review the formulation of \qg\ coupled to a
scalar field as a sum over discretized \rs\ embedded in 1 dimension.
I will show that the matrix quantum mechanics generates the
necessary statistical sum. In section 3 I exhibit the reduction
to free fermions and define the double-scaling limit.\rk\GMil~
The sum over continuous surfaces is calculated to all orders in 
the genus expansion. In section 4 I show how
to calculate the exact string amplitudes
using the free fermion formalism. Section 5 is devoted
to the continuum path integral formalism\rk\sasha~ where the Liouville
field acts as an extra dimension of string theory. 
Some of the exact matrix model results will be reproduced,
but this approach still falls far short of the power of the
matrix model. In section 6 I discuss the special states 
which exist in the spectrum in addition to the massless ``tachyon''.
These states, occurring only at integer momenta, are left-overs
of the transverse excitations of string theory. Remarkably,
they generate a chiral $W_{1+\infty}$ algebra. 
In section 7 I use the matrix model
to formulate the exact string field theory both in the fermionic and
in bosonic terms. I present some manifestly finite bosonic calculations
which are in perfect agreement with the fermionic ones.
In section 8 I discuss the new physical effects that arise 
when the random surfaces are embedded in a circle of radius $R$:
the $R\to 1/R$ duality and its breaking due to \KT\ vortices.\rk{\GKl}~
This model can be regarded either as a compactified Euclidean string
theory or as a Minkowski signature string theory at finite temperature.
In section 9 I show that in the matrix quantum the effects of vortices
are implemented by the states in the non-trivial representations
of $SU(N)$.$^{\GKl,\Vort}$ Including only the $SU(N)$ singlet sector
gives the vortex-free continuum limit where the sum over
surfaces respects $R\to 1/R$ duality.\rk\GKl~ In section 10 I 
use the thermal field theory of non-relativistic fermions
to find exact amplitudes of the compactified string theory.\rk\Comp~
Finally, in section 11 I use the matrix chain model to solve
a string theory with a discretized embedding dimension.\rk\GKl~
Remarkably, when the lattice spacing is not too large, this theory is exactly
equivalent to string theory with a continuous embedding.

\chapter{DISCRETIZED RANDOM SURFACES AND MATRIX QUANTUM MECHANICS.}

In this section I will introduce the discretized approach to
summation over \rs\ embedded in one dimension 
and its matrix model implementation. If we parametrize continuous surfaces
by coordinates $(\sigma_1, \sigma_2)$, then the Euclidean geometry
is described by the metric $g_{\mu\nu}(\sigma)$, and the embedding -- by
the scalar field $X(\sigma)$. Thus, the theory of \rs\ in one embedding
dimension is  equivalent to 
2-d \qg\ coupled to a scalar field. In the Euclidean
path integral approach to such a theory, we have to sum over all
the compact connected geometries and their embeddings,
$${\cal Z}=\sum_{topologies}\int [Dg_{\mu\nu}][DX] e^{-S}
\ .\eqn\pathinteg$$
The integration measure is defined modulo reparametrizations,
\ie\ different functions describing the same geometry-embedding
in different coordinates should not be counted separately.
We assume the simplest generally covariant massless action,
$$S={1\over 4\pi}\int d^2\sigma\sqrt g({1\over\ap}
g^{\mu\nu}\partial_\mu X\partial_\nu X+\Phi R+4\lambda)\ .
\eqn\Laction$$
In 2 dimensions the Einstein term is well-known to give the
Euler characteristic, the topological invariant which depends
only on the genus of the surface $h$ (the number of handles),
$${1\over 4\pi}\int d^2\sigma\sqrt g R=2(1-h)
\ .\eqn\eq $$
Thus, the weighting factor for a surface of genus $h$ is 
$(\exp \Phi)^{2h-2}$. 
Since the sum over genus $h$
surfaces can be thought of as a diagram 
of string theory with $h$ loops, we identify the string coupling constant as
$g_0=e^\Phi$.

The main problem faced in the calculation of the \EPI\
of eq. \pathinteg\ is the generally covariant definition
of the measure for the sum over metrics
$[Dg_{\mu\nu}]$ and of the cut-off. One may
attempt to do this directly in the continuum.
Considerable success along this route has been achieved
when metrics are described in the light-cone gauge,\rk\kpz~
or in the more traditional conformal gauge.\rk\ddk~ We will
carry out some comparisons with the continuum approach in the course
of the lectures. 
\epsfxsize=4in
\topinsert\centerline{\epsfbox{fig1.ps}}
{\narrower\smallskip\singlespace  
\noindent Fig. 1.
A small section of triangulated surface. Solid lines denote the triangular
lattice $\Lambda$, and dotted lines -- the dual lattice $\tilde\Lambda$.
\smallskip} \endinsert
\noindent

The main subject of this section is 
a different approach to summing over geometries, which
has so far proven to be far more powerful than the continuum methods.
In this approach one sums over discrete approximations to
smooth surfaces, and then defines the continuum sum by taking the
lattice spacing to zero.\rk\RS~ We may, for instance, choose to approximate
surfaces by collections of equilateral triangles of side $a$.
A small section of such a triangulated surface is shown in fig. 1.
The dotted lines show the lattice of coordination number 3,
which is dual to the triangular lattice. Each face of the triangular
lattice is thought of as flat, and the curvature is entirely
concentrated at the vertices. Indeed, each vertex $I$ of the
triangular lattice has a conical singularity with deficit angle
${\pi\over 3} (6-q_I)$, where $q_I$ is the number of triangles that
meet at $I$. Thus, the vertex $I$ has a $\delta$-function of curvature
with positive, zero, or negative strength if $q_I<6,~=6,~>6$
respectively. Such a distribution of curvature may seem like a 
``poor man's version'' of geometry. However, in the continuum limit
the size of each face becomes infinitesimal, and we may define
smoothed out curvature by averaging over many triangles.
In this way, the continuum field for the metric should appear similarly
to how the quantum field description emerges in the continuum limit
of the more familiar statistical systems, such as the Ising model,
the XY model, etc. Later on, we will show some strong indications
that the discretized approach to summing over geometries is indeed
equivalent to the continuum field-theoretic approach. 

The essential assumption of the discretized approach
is that sum over genus $h$ geometries, $\int [Dg_{\mu\nu}]_h$, 
may be defined as the sum over all distinct lattices $\Lambda$,
with the lattice spacing subsequently taken to zero.
$\int [DX]$ is then defined as the integral over all possible
embeddings of the lattice $\Lambda$ in the real line.
For simplicity, the lattices $\Lambda$ may be taken to be triangular,
but admixture of higher polygons should not affect the continuum
limit. In order to complete the definition of the discretized approach,
we need to specify the weight attached to each configuration.
This can be defined by simply discretizing the action of eq. \Laction\
and counting each distinct configuration with weight $e^{-S}$.
We will find it convenient to specify the embedding coordinates 
$X$ at the centers of the triangles, \ie\ at the vertices
$i$ of the dual lattice. Then the discretized version of 
${1\over 4\pi\ap}\int d^2\sigma\sqrt g
g^{\mu\nu}\partial_\mu X\partial_\nu X $
is $\sim\sum_{<ij>} (X_i-X_j)^2 $,
where the sum runs over all the links $<ij>$ of the dual lattice.
Similarly,
$$\int d^2\sigma\sqrt g\to {\sqrt 3\over 4} a^2 V
\eqn\eq$$
where $V$ is the number of triangles, or, equivalently, the number
of vertices of the dual lattice $\tilde\Lambda$. 
Thus, the discretized version of
the path integral \pathinteg\ is
$$ {\cal Z}(g_0, \kappa)=\sum_h g_0^{2(h-1)} \sum_\Lambda 
\kappa^V\prod_{i=1}^V\int dX_i \prod_{\langle i j \rangle} G(X_i-X_j)\sp ,
\eqn\Pol$$
where $\Lambda$ are all triangular lattices of genus $h$,
$\kappa\sim \exp(-\sqrt 3\lambda a^2/4\pi)$, and
(for some choice of $\ap$) $G(X)=\exp (-\half X^2)$.
If the continuum limit of eq. \Pol\ indeed describes \qg\ coupled
to a scalar field, then there should exist a whole universality class
of link factors $G$, of which the gaussian is only one representative,
that result in the same continuum theory.

A direct evaluation of the lattice sum \Pol\ seems to be a daunting task,
which is still outside the numerical power of modern computers.
Fortunately, there exists a remarkable trick which allows us to
exactly evaluate sums over surfaces of any topology using only
analytic tools.
As was first noted by Kazakov and Migdal,\rk\KazMig~ a statistical sum
of the form \Pol\ is generated in the Feynman graph expansion
of the quantum mechanics of a $N\times N$ hermitian matrix.
\foot{
Similar tools work for other simple matter systems coupled to two-dimensional
gravity. For example, as discussed in other lectures in this volume,
in the case of pure gravity the sum over discretized surfaces
is generated simply by an integral over an $N\times N$ hermitian matrix.
}
Consider the Euclidean path integral
$$Z=\int D^{N^2}\Phi(x)\exp
\biggl [-\beta\int_{-T/2}^{T/2}dx~\Tr 
\left (\half\left ({\partial\Phi\over\partial x}\right )^2+ 
U(\Phi)\right )\biggr ]\sp .\eqn\PI
$$
where $x$ is the Euclidean time
and $U={1\over 2\ap}\Phi^2-{1\over 3!}\Phi^3$.
The parameter $\b$ enters as the inverse Planck constant.
By a rescaling of $\Phi$ eq. \PI\ can be brought to the form
$$Z\sim \int D^{N^2}\Phi(x)\exp
\biggl [-N\int_{-T/2}^{T/2}dx~\Tr  
\left(\half\left({\partial\Phi\over\partial x}\right )^2+ 
{1\over 2\ap}\Phi^2-{\kappa\over 3!}\Phi^3
\right )\biggr ]\sp ,\eqn\nPI
$$
where $\kappa=\sqrt{N/\beta}$ is the cubic coupling constant.
The connection with the statistical sum \Pol\ follows when we
develop the graph expansion of $Z$ in powers of $\kappa$.
The Feynman graphs all have coordination number 3, and are in one-to-one
correspondence with the dual lattices $\tilde\Lambda$ 
of the discretized \rs\
(fig. 1). The lattices $\Lambda$ dual to the Feynman graphs
can be thought of as the basic triangulations. One easily
obtains the sum over all connected graphs $\ln Z$, 
$$ 
\lim_{T\to\infty}\ln Z
=\sum_h N^{2-2h} \sum_{\Lambda}
\kappa^V\prod_{i=1}^{V}\int_{-\infty}^\infty dx_i 
\prod_{\langle i j \rangle} e^{-|x_i-x_j|/\ap}
\sp . \eqn\trian $$
This is precisely of the same form as eq. \Pol\ which arises
in two-dimensional \qg!
The Euclidean time $x$ assumes the role of the embedding coordinate $X$
in eq. \Pol.
We note that the role of the link factor $G$
is played by the one-dimensional Euclidean propagator.
Only for this exponential $G$ can we establish the exact equivalence
with the matrix model. This does not pose a problem, however,
as we will find plenty of evidence that the continuum
limit of the model \Pol\ with the exponential $G$ indeed 
describes quantum gravity coupled to a scalar, eq. \pathinteg.
It is evident from \trian\ that the parameter $\ap$ sets the scale of the 
embedding coordinate. In fact, we have normalized $\ap$ so that
in the continuum limit it will precisely coincide with the definition
in eq. \Laction. Whenever $\ap$ is not explicitly mentioned, its value has been
set to 1. 

Further comparing eqs. \trian\ and \Pol, we note that
the size of the matrix $N$ enters as $1/g_0$.
Let us show why. Each vertex contributes a factor $\sim N$,
each edge (propagator) $\sim 1/N$, and each face contributes $N$
because there are as many index loops as there are faces.
Thus, each graph is weighted by $N^{V-E+F}$ which, by
Euler's theorem, equals $N^{2-2h}$. The expansion
of the free energy of the matrix quantum mechanics
in powers of $1/N^2$
automatically classifies surfaces according to topology.
It would seem that, in order to define a theory with a finite
string coupling, it is necessary to consider finite $N$ which,
as we will see, is associated with severe difficulties.
Fortunately, this naive expectation is false: in the continuum
limit the ``bare'' string coupling $1/N$ becomes infinitely
multiplicatively renormalized. Thus, if $N$ is taken to $\infty$
simultaneously with the world sheet continuum limit, then
we may obtain a theory with a finite string coupling.
This remarkable phenomenon, known as the
``double-scaling limit''$^{\GM,\GMil}$, will allow us to calculate sums over
continuum surfaces of any topology.

In order to reach the continuum limit, it is necessary to
increase the cubic coupling $\kappa=\sqrt{N/\beta}\sim N^0$
until it reaches the critical
value $\kappa_c$ where 
the average number of triangles in a surface begins to
diverge. In this limit we may think of each triangle as being
of infinitesimal area $\sqrt 3 a^2/4$, 
so that the whole surface has a finite
area and is continuous. In the continuum limit $\kappa\to
\kappa_c$ we define the cosmological constant 
$$\Delta=\pi(\kappa_c^2-\kappa^2)\ .\eqn\eq$$
Recalling that  
$\kappa=\kappa_c\exp(-\sqrt 3\lambda a^2/4\pi)$, 
we establish the connection
between $\Delta$ and the physical cosmological constant $\lambda$:
$\lambda\sim \Delta/a^2$.

Above we have sketched the connection between matrix quantum mechanics
and triangulated random surfaces. It is easy to generalize
this to the case where, in addition to triangles, the surfaces consist
of other $n$-gons: we simply have to add to the matrix potential
$U(\Phi)$ monomials $\Phi^n$, $n>3$. For consistency, the
continuum limit should not be sensitive to the precise manner in which the
surfaces are discretized. In the next section we will give
the exact solution of the matrix model and show that the continuum
limit is indeed universal.

\chapter{MATRIX QUANTUM MECHANICS AND FREE FERMIONS.}

In the previous section we established the equivalence of the sum over
connected discretized surfaces to the 
logarithm of the path integral of the matrix
quantum mechanics. On the other hand, in the hamiltonian language,
$$ Z=<f| e^{-\beta H T}|i>\ .
\eqn\transition$$
Thus, as long as the initial and final states have some overlap
with the ground state, the ground state energy $E_0$ will dominate the
$T\to \infty$ limit of eq. \transition,
$$ 
\lim_{T\to\infty}{ \ln Z\over T}
=-\beta E_0
\ .\eqn\eq$$
The divergence of the sum over surfaces proportional to the length
of the embedding dimension arises due to the translation invariance.
In order to calculate the sum over surfaces embedded in 
the infinite real line $R^1$, all we need to find is the ground state energy 
of the matrix quantum mechanics.
To this end we carry out canonical quantization of the
$SU(N)$ symmetric hermitian matrix quantum mechanics.\rk{\BPIZ,\MO}~
The Minkowski time lagrangian
$$ L=\Tr\{\half\dot \Phi^2- U(\Phi)\}\eqn\lag
$$
is symmetric under time-independent $SU(N)$ rotations
$\Phi(t)\to V^\dagger\Phi(t) V$. It is convenient to decompose
$\Phi$ into $N$ eigenvalues and $N^2-N$ angular degrees of freedom,
$$\Phi (t)=\Omega^\dagger(t) \Lambda(t) \Omega(t)\eqn\eq$$ 
where
$\Omega \in SU(N)$ and $\Lambda= diag (\lambda_1, \l_2, \ldots,
\l_N)$.
The only term in the lagrangian which gives rise to dependence
on $\Omega$ is
$$\Tr\dot\Phi^2=\Tr\dot\L^2+\Tr[\L, \dot\O\O^\dagger]^2\ .\eqn\kin
$$
To identify the canonical 
angular degrees of freedom, the anti-hermitian traceless matrix
$\dot\O\O^\dagger$
can be decomposed in terms of the generators of $SU(N)$,
$$\dot\O\O^\dagger={i\over\sqrt 2}\sum_{i<j}\dot\a_{ij}
T_{ij}+\dot\b_{ij}\tilde T_{ij}+
\sum_{i=1}^{N-1}\dot\a_i H_i\eqn\dec
$$
where $H_i$ are the diagonal generators of the Cartan subalgebra,
and the other generators are defined as follows.
$T_{ij}$ is the matrix $M$ such that $M_{ij}=M_{ji}=1$, and
all other entries are 0;
$\tilde T_{ij}$ is the matrix $M$ such that $M_{ij}=-M_{ji}=-i$, and
all other entries are 0.
Substituting eq. \dec\ into eqs. \kin\ and \lag, we find 
$$L=\sum_i\left(\half\dot\li^2+U(\li)\right)
+\half\sum_{i<j} (\li-\l_j)^2(
\dot\a_{ij}^2 +\dot\b_{ij}^2)
\sp . \eqn\lagg
$$
In deriving the hamiltonian it is crucial\rk\BPIZ~
that the measure of integration
expressed in terms of $\Omega$ and $\Lambda$ assumes the form
$ {\cal D} \Phi ={\cal D} \Omega\prod_i d\l_i \Delta^2(\l)$,
where $\Delta(\l)$~is the Vandermonde determinant
$\prod_{i<j}(\l_i-\l_j)$. Because of the non-trivial
Jacobian, the kinetic term for the eigenvalues becomes
$$-{1 \over 2 \b^2 } \sum_{i=1}^N 
{1\over\Delta^2(\l)}{d\over d \lambda_i}
\Delta^2(\l){d\over d \lambda_i}\ .\eqn\eq$$
It is left as an exercise for the reader to show that this
can be rewritten as
$$-{1 \over 2 \b^2 \Delta(\lambda )} \sum_i{d^2\over {d \lambda_i}^2}
\Delta(\l)\ .\eqn\eq$$
It follows that the total hamiltonian is
$$
H= -{1 \over 2 \b^2 \Delta(\lambda )} \sum_i{d^2\over {d \lambda_i}^2}
\Delta(\lambda ) + \sum_i U(\lambda_i)
+\sum_{i < j} {\Pi_{ij}^2+\tilde\Pi_{ij}^2
 \over {(\lambda_i - \lambda_j)^2}}
\sp ,
\eqn\hamilt
$$
where $\Pi_{ij}$ and $\tilde\Pi_{ij}$ are the momenta
conjugate to $\a_{ij}$ and $\b_{ij}$ respectively, \ie,
they are the generators of {\it left} rotations on $\O$,
$\O\to A\O$. Furthermore, since $L$ is independent
of $\dot\a_i$, the wave functions must obey the constraints
$\Pi_i\Psi=0$. These constraints arise because transformations
$\O\to A\O$, where $A$ is a diagonal $SU(N)$ matrix, do not change $\Phi$.

The constraints require that only those irreducible representations
of $SU(N)$ that have a state with all weights equal to zero
are allowed in the matrix quantum mechanics.\rk\Vort~ For instance, the 
fundamental representations are excluded, and the simplest non-trivial
representation is the adjoint. 
It is not hard to classify all the irreducible representations
allowed in the matrix
quantum mechanics, \ie, the ones that have states with zero weights.
In the standard notation, the Young tableaux of
such representations consist of $i+j$ columns, with the numbers
of boxes in the columns (from left to right)
$$(N-m_1, N-m_2,\ldots, N-m_j, n_i,\ldots, n_2, n_1)
\eqn\eq$$
where $n_1\leq n_2\leq\ldots\leq n_i\leq N/2$ and
$m_1\leq\ldots\leq m_j\leq N/2$. 

Since the angular kinetic terms in \hamilt\ are positive definite, it is clear
that we have to look for the ground state among the wave functions
that are annihilated by $\Pi_{ij}$ and $\tilde\Pi_{ij}$, \ie\
in the trivial (singlet) representation of $SU(N)$. The singlet
wave functions are independent of the angles $\Omega$, and
are symmetric functions of the eigenvalues, $\chi_{sym}(\l)$. 
Because of the special form of the hamiltonian \hamilt, the eigenvalue
problem $H\chi_{sym}=E\chi_{sym}$ assumes the remarkable form\rk\BPIZ
$$
\eqalign{&\left( \sum_{i=1}^N h_i\right)
\Psi (\l)= E\Psi (\l)\ ,\cr
&h_i=-{1 \over 2 \b^2 } {d^2\over {d \lambda_i}^2}
+U(\lambda_i)\ ,\cr } 
\eqn\eq$$
where $\Psi(\l) =\Delta(\l) \chi_{sym}(\l)$ is by construction antisymmetric.
Since the operator acting on $\Psi$ is simply the sum of single-particle
non-relativistic hamiltonians, the problem has been reduced to
the physics of $N$ free non-relativistic fermions moving in the
potential $U(\lambda)$.\rk\BPIZ~ This equivalence to free fermions is at
the heart of the exact solubility of string theory in two dimensions.

The $N$-fermion ground state is obtained by filling the lowest
$N$ energy levels of $h$,
$$E_0=\sum_{i=1}^N \e_i\ .\eqn\gse$$
 The potential arising from triangulated
\rs\ is $U(\l)=\half\l^2-{1\over 3!}\l^3$, which is unbounded from below
and does not seem support any stable states. Recall, however, that
in order to find the genus expansion of the sum over surfaces,
we are only interested in the expansion about the 
classical limit $\b\to\infty~ (\hbar\to 0)$ in powers of $1/\b^2$.
Classically, there is a continuum of bounded orbits to the left of
the potential barrier. Semiclassically, these orbits are associated
with energy levels (fig. 1) whose spacing is $\CO(1/\b)$, and whose decay time
is exponentially long as $\b\to\infty$. This instability cannot be
seen in the context of the semiclassical expansion in powers of
$1/\b^2$. Our goal, therefore, is to find the expansion of
the ground state energy \gse, where $\e_i$ are the semiclassical energy levels.

To leading order, $\e_n$ are determined by the Bohr-Sommerfeld quantization
condition,
$$\oint p_n (\l) d\l={2\pi\over\b} n
\eqn\eq$$
where $p_n (\l)=\sqrt {2(\e_n-U(\l))}$, and the integral is over a closed
classical orbit. In particular, the position of the Fermi level
$\mu_F=\e_N$ is determined by
$$\int_{\l_-}^{\l_+}\sqrt { 2(\mu_F-U(\l))}=\pi {N\over\b}
\ ,\eqn\Fl$$
where $\l_-$ and $\l_+$ are the classical turning points.
Eq. \Fl\ shows that $\mu_F$ is an increasing function of $N/\b$.
Clearly, both $\mu_F$ and $E_0$ become singular 
at the point where $N/\b\to \kappa_c^2$ such that $\mu_F$ equals the height
of the barrier $\mu_c$. Recall that the cubic coupling of
the matrix quantum mechanics is $\kappa=\sqrt{N/\b}$, and that we expect
to find a singularity in the sum over graphs when $\kappa$ becomes large enough
that the average number of vertices in a Feynman graph 
(or equivalently, of triangles in a random surface) begins to diverge.
In the vicinity of this singularity the continuum limit of \qg\
can be defined.
Now we have identified the physics of this singularity in the
equivalent free fermion system: it is associated with spilling
of fermions over the barrier.

We are interested in the singularity of $E_0$ as a function
of the cosmological constant $\Delta=\pi(\kappa_c^2-{N\over\b})$.
The parts of $E(\Delta)$ analytic in $\Delta$ are not universal
and can be dropped. Indeed, if we calculate the sum over surfaces as
a function of the fixed area by inversely Laplace transforming
$E(\Delta)$, the analytic terms give contributions only for zero area.
\topinsert
\line{\hfil%
\epsfxsize=2.85in\epsfbox{fig2a.ps}\hfil\hfil%
\epsfxsize=2.85in\epsfbox{fig2b.ps}\hfil}
{\narrower\smallskip \baselineskip 10pt
\tablist{Fig. 2a) $N$ fermions in the asymmetric potential arising directly
from the triangulated surfaces.   
&
Fig. 2b) The \dsl\ magnifies the quadratic local maximum. 
\cr}
\smallskip} \endinsert \noindent

Let us introduce the density of eigenvalues
$$\rho(\e)={1\over \beta}\sum_n\delta(\e-\e_n), \eqn\dense$$
in terms of which
$$\eqalign{&\kappa^2={N\over\beta}=\int_0^{\mu_F}\rho(\e) d\e \ ;\cr
&\lim_{T\to\infty}{-\ln Z\over  T}=\b E=
\beta^2\int_0^{\mu_F}\rho(\e) \e d\e
\ .\cr}
\eqn\rels$$
Differentiating the first equation, we find
$$\partder\Delta\mu=\pi \rho(\mu_F)\sp ,\eqn\dddmu
$$
which can be integrated and inverted to obtain 
$\mu=\mu_c-\mu_F$ as a function of $\Delta$. 
Differentiating the second equation, we find 
${\partial E\over\partial\Delta}={1\over \pi}\beta (\mu-\mu_c)$.
After addition to $E$ of an irrelevant analytic term,
$E\to E+{1\over \pi}\b\m_c\Delta$, we find
$${\partial E\over\partial\Delta}={1\over \pi}\beta \mu
\eqn\dedd$$
These equations show that the continuum limit of the sum over
surfaces is determined by the 
singularity  of the single particle density of states near
the top of the barrier. Using the WKB approximation, it was found in
ref. \KazMig\ that 
$$\rho(\mu_F)={1\over \pi}\int_{\l_-}^{\l_+}
{d\l \over \sqrt{2(\mu_F-U(\l))}}\sim
-{1\over \pi}\ln\mu+\CO(1/\beta^2)\  \eqn\leading $$
It follows that $\Delta=-\mu\ln\mu+\CO(1/\b^2)$, and
$$-\b E={1\over 2\pi}(\b\m)^2\ln\mu+\ldots\approx
{1\over 2\pi}{\b^2\Delta^2\over \ln\Delta}+\ldots
\eqn\sphere$$
The critical behavior of $\rho$ comes from the part of the integral
near the quadratic maximum of the potential. 
In the continuum limit $\mu\to 0$,
the classical trajectory at $\mu_F$ spends logarithmically diverging
amount of time near the maximum. Therefore, the WKB wave functions
are peaked there. In fact, we will show that, at any order in
the WKB expansion, the leading singularity of $\rho$ is entirely
determined by the quadratic nature of the maximum. To see this,
let us rescale the coordinates $\sqrt\b (\l-\l_c)=y$, and
$e=\mu_c-\epsilon$. The single particle eigenvalue problem becomes\rk\GMil 
$$\left (\half {d^2\over dy^2}+\half y^2+\CO \bigl(1/\sqrt\b\bigr) 
y^3\right)
\psi(y)=\b e\psi(y)
\eqn\invho$$
For a finite $\b e$, all the terms beyond the quadratic are irrelevant
in the limit $\b\to\infty$ because they are suppressed by powers of $\b$.
This demonstrates the universality of the continuum limit:
any potential possessing a quadratic maximum will yield the same
sum over continuous surfaces.
In effect, our rescaling has swept all the non-universal
details of the potential out to infinity. This implies, for instance,
that, instead of working with the potential of fig. 2, we can use the double
well potential $U(\l)=-\half\l^2+\l^4$ (fig. 3). 
After rescaling $\sqrt\b\l=y$, the Schr\"odinger
equation again reduces to the motion in the inverted harmonic oscillator.
The only difference is that now fermions fill both sides of the maximum.
To all orders of the WKB expansion this simply multiplies the density
of states, and therefore the free energy, by a factor of 2.
The double well potential does not
suffer from the non-perturbative instabilities: the ground state energy
can in principle be calculated even for finite $N$ and $\b$.
We will often adopt the symmetric case because it is more convenient for
calculations, remembering that we have to divide by 2 to find
the sum over triangulated surfaces.
\topinsert
\line{\hfil%
\epsfxsize=2.85in\epsfbox{fig3a.ps}\hfil\hfil%
\epsfxsize=2.85in\epsfbox{fig3b.ps}\hfil}
{\narrower\smallskip \baselineskip 10pt
\tablist{Fig. 3a) $N$ fermions in a double well potential. 
&
Fig. 3b) The system relevant to the \dsl.
\cr}
\smallskip} \endinsert \noindent
 
From eq. \sphere\ we know the sum over all surfaces of spherical topology
embedded in $R^1$. In order to find the sums over surfaces of
any topology, we need to develop a systematic WKB expansion
of the ground state energy. According to eqs. \dddmu\ and 
\dedd, all we need to
know is the complete expansion of the single particle density
of states $\rho (\mu)$. If we consider $\mu$ such that 
$\b\m$ is $\CO(1)$, then this problem has a remarkably simple
solution,\rk\GMil~ because eq. \invho\ shows that in this
energy range the system is purely quadratic.
Let us write the density of states as
$$ \eqalign{\rho(\mu) =& \Tr\delta (h_0+\b\m)=
{1 \over \pi} {\rm Im \,Tr}
\left[ {1 \over h_0 
+\b \m - i \epsilon} \right]\,\,, \cr
h_0=&-{1\over 2} {d^2 \over dy^2} 
- {1\over 2} y^2\ .\cr} 
\eqn\rot 
$$
We will evaluate the trace in position space in order to
take advantage of the fact that the resolvent of 
the hamiltonian of a simple harmonic oscillator is well known,
$$\eqalign{&
\langle y_f|{ 1 \over  -{1\over 2} {d^2 \over dy^2} 
+{1\over 2} \omega^2 y^2  +\b \m-i\e } |y_i\rangle 
=\int_0^{\infty} dTe^{-\b \m T}
 \int_{y_i}^{y_f} {\cal D}y(t) e^{-\int_0^Tdt {1\over 2}( {\dot y}^2 +\omega^2
y^2)}\cr
&=\int_0^{\infty}dT e^{-\b \m T} \sqrt{{\omega \over 2\pi {\rm sinh} \omega T}}
e^{- \omega 
\big[(y_f^2 + y_i^2)\cosh\omega T -2y_f y_i\big]
/2\sinh\omega T} 
\ . \cr}
\eqn\path
$$
In our case, the frequency $\omega$ is actually imaginary (we
have an {\it inverted} harmonic oscillator), so we must define the resolvent by
analytic continuation. We rotate $\omega \to -i$~ while simultaneously
rotating $T \to iT$, and the relevant resolvent is 
$$
\langle y_f|{ 1 \over h_0 + \b \m-i\e}
|y_i\rangle =
i \int_0^{\infty}dT e^{-i\b \m T} 
\sqrt{{-i \over 2\pi {\rm sinh} T}}
e^{ i\big[(y_f^2 + y_i^2)\cosh T -2y_f
y_i\big]/2\sinh T} 
\sp .
\eqn\patht
$$
Now it is easy to derive the 
derivative of the density of states\rk\GMil
$$
{\partial \rho \over \partial (\b \m)}={1\over\pi}
{\partial \over \partial (\b \m)}{\rm Im} 
\int_{-\infty}^\infty  dy \langle y| {1\over h_0 +\b\m-i\e}|y
\rangle=
{1\over \pi}{\rm Im} \int _0^{\infty}dT e^{-i\b\m T} 
{T/2\over {\rm sinh}(T/2)}
\sp .
\eqn\dens
$$
Note that only for the purposes of deriving the large $\beta\mu$
expansion can we reduce the problem to pure inverted harmonic oscillator.
The integral representation \dens\ is not designed to correctly
give the non-perturbative terms $\CO(e^{-\b\mu})$. In fact,
such terms depend on the details of the potential away from the quadratic
maximum, and are not universal. While each term in the large
$\beta\mu$ expansion is related to the sum over all geometries
of a certain genus, we do not yet fully understand the geometrical
or physical meaning of the non-perturbative terms.

After expanding eq. \dens\  
and integrating once, we find the complete series of corrections to
the leading order estimate in eq. \leading,
$$\rho(\mu)={1\over \pi}\left\{-\ln\mu+\sum_{m=1}^\infty(2^{2m-1}-1)
{|B_{2m}|\over m}(2\beta\mu)^{-2m}\right\}\sp .\eqn\asymp$$

In solving eq. \dddmu\ for $\mu$ in terms of $\Delta$, it is
useful to introduce parameter $\mu_0$ defined by
$$\Delta=-\mu_0\ln\mu_0=\mu_0|\ln\mu_0|
\ .\eqn\muz$$
Then,
$$\mu=\mu_0\left\{1-
{1\over\ln\mu_0}\sum_{m=1}^\infty(2^{2m-1}-1)
{|B_{2m}|\over m(2m-1)}\left(2\beta\mu_0
\right)^{-2m}+\CO\left({1\over\ln^2\mu_0}\right)\right\}\sp .\eqn\mew$$
Integrating \dedd\ we finally arrive at the complete genus expansion
of the sum over surfaces
$$-\b E={1\over 8\pi}\left\{
(2\beta\mu_0)^2\ln\mu_0-{1\over 3}\ln\mu_0+
\sum_{m=1}^\infty {(2^{2m+1}-1) |B_{2m+2}|\over m(m+1)(2m+1)}
(2\beta\mu_0)^{-2m}\right\} \ .\eqn\genexp$$
Clearly, the sensible way to approach the
continuum limit is through ``double scaling'' $\b\to\infty$
and $\mu_0\to 0$ in such a way that $\b\mu_0=1/g_{st}$ 
is kept fixed.$^{\GMil,\GKl}$
Thus, quite remarkably, we have found closed form
expressions for sums over continuous surfaces of any topology embedded in 
one dimension. The sum over genus $h$ surfaces, 
 given by the coefficient of $g_{st}^{2h-2}$,
grows as $(2h)!$. This behavior is characteristic of closed string
theories,\rk\Shenker~ while in field theory we typically find 
that the sum over $h$-loop diagrams exhibits a slower growth
$\sim h!$. The badly divergent perturbation series \genexp\ indicates
that the theory is not well-defined non-perturbatively.

Eq. \genexp\ shows explicitly how the bare string coupling
$1/\b$ gets multiplicatively renormalized. Compared to the
$c<1$ matter coupled to gravity,\rk{\GM-\Doug}~ the new feature here is that
the renormalization is not simply by a power of the cosmological constant
$\Delta$. It appears that $\Delta$ itself is multiplicatively
renormalized: $\Delta\to \mu_0\approx \Delta/|\ln\Delta|$.
Another peculiarity of the $c=1$ matter is that the sums over spherical
and toroidal surfaces diverge in the continuum limit as $|\ln\mu_0|$.
In section 5 we will argue that all these features have a natural explanation 
in the context of Liouville theory.
Before we do that, however, we further exhibit the power of
the matrix model by calculating some correlation functions
to all orders in the topological expansion.

\chapter{CORRELATION FUNCTIONS.}

The simplest kind of matrix model operator is
$\Tr \Phi^n(x)$ where $n$ is any finite integer. This operator
inserts a vertex of order $n$ in 
all possible places on the dual graph $\tilde\Lambda$.
In terms of the basic lattice $\Lambda$, this amounts to cutting
an $n$-gon hole in the surface, pinning it at the embedding coordinate
$x$, and integrating the position of the hole all over the surface.
In the continuum limit, for any finite $n$ the size of the hole becomes
infinitesimal, and it is commonly referred to as ``a puncture''.
The continuum expression for the operator above should be roughly  
$$\int d^2\sigma\sqrt g\delta\bigl (X(\sigma)-x)
\ .\eqn\eq$$
By Fourier transforming, we derive the continuum translation
of a related operator
$$\int dx e^{iqx} \Tr \Phi^n (x)\to \int d^2\sigma\sqrt g e^{iqX}
\ , \eqn\contach$$
which resembles the basic ``tachyon'' operator of string theory.
Below we outline the procedure for calculating the correlation functions
of such operators in the matrix model. In section 5 we
show that the results can indeed be reproduced in the string theoretic
formalism.

Since the matrix model has been reduced to a system of free
fermions, it is convenient to perform calculations in
the formalism of non-relativistic second quantized field theory.
The second quantized fermion field is defined as
$$
\PH (\L,x) = \int_{-\infty}^{\infty} d\nu e^{i\nu x} a_{\E}(\nu) \PE(\nu,
\L)
\ ,\eqn\eq$$
where $\PE(\nu,\L)$ are the single fermion wave functions of energy $\nu$,
and the associated annihilation 
operators $a_{\E}(\nu)$ satisfy 
$\lbrace a_{\E}^{\dag}(\nu), a_{\E}(\nu^{\prime})
\rbrace = \D_{\E,\E^{\prime}} \D (\nu - \nu^{\prime}) $. 
Setting $\l=\sqrt 2 y$ in eq. \invho, we find that $\PE (\nu, \L)$ 
are solutions of the Whittaker equation 
$$
({d^2 \over {d \L^2}} + {\L^2 \over 4} ) \P = \nu \P .
\eqn\eq$$
and can therefore be expressed in terms of Whittaker
functions $W(\nu, \L)$.\rk\Moore~ Their spectrum is continuous
and doubly degenerate. 
Here $\E$ denotes the parity of the wavefunction $\PE(\nu,\L)$ 
and repeated $\E$ are summed over. 

The Euclidean second-quantized field theory with chemical
potential $\b\mu$ is defined by the action
$$
S = \int_{-\infty}^{\infty} d \L \int_{-\infty}^\infty dx~ \PH^{\dag} (-{d \over
{dx}} + {{d^2} \over {d \L^2}} + {{\L^2} \over 4}+\b\mu ) \PH
\eqn\eq$$
where $x$ is the Euclidean time. 
The ground state of the system (fig. 3b) 
satisfies
$$\eqalign{ 
&a_{\E}(\nu)|\b\mu\rangle=0,\qquad\qquad \nu<\b\mu\ ,\cr
&a_{\E}^\dagger (\nu)|\b\mu\rangle=0,\qquad\qquad \nu>\b\mu\ .\cr}
\eqn\eq$$

Consider correlation functions of operators of the type
$$\CO(q)=\int dx e^{iqx} \Tr f\bigl(\Phi(x)\bigr)
\ ,\eqn\op$$
where $f$ is any function. In the 
second quantized formalism, they translate into
$$\CO(q)=\int dx e^{iqx}\int d\L f(\lambda)
\PH^{\dag} \PH(x, \L)\ . \eqn\eq$$ 
Thus, to find the connected correlation functions of any set of
such operators,
we need to calculate
$$
G(q_1,\L_1;\ldots ;q_n,\L_n) = \prod_{i=1}^{n} \int
dx_i e^{iq_i
x_i} \langle \b\mu \vert  
\PH^{\dag} \PH(x_1, \L_1) \ldots 
\PH^{\dag} \PH(x_n, \L_n) \vert \b\mu \rangle_c\ .
\eqn\corr$$
Application of Wick's theorem reduces this to the sum over one-loop diagrams
with fermion bilinear insertions in all possible orders around the loop.
A convenient formula for the Euclidean Green function is\rk\Moore
$$
\eqalign{
S^E (x_1, \L_1 ; x_2, \L_2) &= \int_{-\infty}^{\infty} d\nu 
e^{-(\nu-\b\mu) \DX}
\lbrace \T (\DX) \T(\nu-\b\mu) - 
\T(-\DX) \T(\b\mu-\nu)\rbrace \times \cr \P^{\E}(\nu, \L_1)
\P^{\E} (\nu, \L_2) &=\int_{-\infty}^\infty {dp\over 2\pi} e^{-ip \DX} 
\int_{-\infty}^{\infty} d\nu  {i \over {p+ i(\nu -
\b\mu)}} \P^{\E} (\nu, \L_1) \P^{\E} (\nu, \L_2) \cr
&= i \int_{-\infty}^\infty {dp\over 2\pi} e^{-ip \DX} 
\int_0^{sgn(p) \infty} ds 
    e^{-sp +i\b\mu s} 
\langle \L_1 | e^{2i sH} | \L_2 \rangle\cr} 
\eqn\eprop$$
where $\DX = x_1 -x_2$, and we have used\rk\Moore
$$\eqalign{&
\int_{-\infty}^{\infty} d\nu e^{-i \nu s} \P^{\E} (\nu, \L_1) \P^{\E}
(\nu , \L_2) = 
\langle \L_1 | e^{2i sH} | \L_2 \rangle=\cr & 
{1 \over \sqrt{-4 \pi i \sh s}} \exp\left (-{i \over 4}
\left[{{\L_1^2+\L_2^2} 
\over {\th s}} - {{2\L_1 \L_2} \over {\sh s}} \right]\right)\cr }\eqn\eq
$$
with $ H = 1/2 p^2 - 1/8 \L^2$. 
With the Green function \eprop, 
after integration over the loop momentum, Moore
derived the representation  
$$
\eqalign{
{\partial \over {\partial \mu}} &G(q_1, \L_1;   \ldots ;q_n,\L_n) =
i^{n+1} \D({\sum q_i}) \sum_{\sigma \in \Sigma_n} ~\int_{-\infty}^{\infty}
d\xi e^{i\mu \xi}   \int_0^{\E_1
\infty} ds_1 \ldots \int_0^{\E_{n-1} \infty} ds_{n-1} \cr
 & e^{-s_1
Q_1^{\sigma} - \ldots - s_{n-1} Q_{n-1}^{\sigma} } \langle \L_{\sigma
(1)} | e^{2is_1 H} | \L_{\sigma (2)} \rangle \ldots 
\langle \L_{\sigma (n)} | e^{2i(\xi - \sum_1^{n-1} s_i )H} | \L_{\sigma
(1) } \rangle \cr }\eqn\noncomp
$$
where $Q_k^{\sigma} = q_{\sigma
(1)} + \ldots + q_{\sigma (k)} $, and $\E_k = sgn[Q_k^{\sigma}]$.  
In principle, any correlation function of operators of type
\op\ can be found by integrating $G$. A convenient way of 
calculating correlations of puncture operators is to first
introduce the operator that creates a finite boundary of length
$l$, with momentum $q$ injected into the boundary\rk\Moore~
$$O(l, q)=\int dx e^{iqx} \Tr e^{-l\Phi(x)}
\ .\eqn\fbound$$
The puncture operator should be the leading term in the small $l$
expansion of $O(l, q)$. The correlations of puncture operators
$P(q)$ can be extracted according to
$$\VEV{\prod_i O(l_i, q_i)}\sim \prod_i l_i^{|q_i|}
\VEV{\prod_j P(q_j)}
\ .\eqn\eq$$
The details of the calculations are highly technical
and can be found in ref. \Moore. Here we will simply state the 
results for the two, three, and four point functions. 
$$
\VEV{P(q) P(k)}=-\delta(q+k)
[\G(1-|q|)]^2 {\mu}^{|q|} \Bigg{[}  {1 \over {|q|}} 
- {1 \over {24  (\b\mu)^2}} (|q| -1) (q^2 - |q| -1) +\ldots\Bigg{]} 
\ .\eqn\two$$
$$
\eqalign{
&\VEV{P(q_1) P(q_2) P(q_3)}=
\delta(\sum q_i)
\prod_{i=1}^3 \left (\G(1-|q_i|) \mu^{{|q_i|\over 2}}\right)
{1\over\b\mu}\Bigg{[} 1 \cr 
& - {1 \over {24 (\b\mu)^2 } }
(|q_3|-1)(|q_3|-2) (q_1^2+q_2^2 - |q_3| -1) 
\cr
& +  
\prod_{r=1}^4 (|q_3| - r) \biggl(3(q_1^4+q_2^4)+10q_1^2 q_2^2
-10(q_1^2+q_2^2) |q_3| - 5(q_1-q_2)^2 + 12|q_3| +7\biggr) 
\cr
& \times{1 \over {5760 (\b\mu)^2}}  + \ldots \Bigg{]}\ , \cr} 
\eqn\three$$
where $q_1,~q_2>0$. 
$$\eqalign{ &\VEV{P(q_1) P(q_2) P(q_3) P(q_4)}=
-\delta(\sum q_i)
\prod_{i=1}^4 \left (\G(1-|q_i|) \mu^{{|q_i|\over 2}}\right)\times \cr
&{1\over 2(\b\mu)^2}\bigl [(|q_1+q_2|+|q_1+q_3|+|q_1+q_4|-2)+
\CO(1/(\b\mu)^2)\bigr ]\ .\cr }
\eqn\four$$
The new feature of the $c=1$ correlators, compared to
theories with $c<1$, is that they have
genuine divergences occurring 
at quantized values of the external momenta.\rk\GKN~
In this sense, the $c=1$ model is the most similar to critical string
theory where the amplitudes are well-known to have divergences
associated with the production of on-shell physical particles.
The remarkable feature of the $c=1$ amplitudes is that all the
divergences factorize into external leg factors.

Because of the divergences, eqs. \two-\four\ are strictly 
valid only if none of $q_i$ are
integer. If any of the momenta have integer values, then additional terms 
need to be taken into account, which leads to cancellation of
the infinity.\rk\GKN~ The physical meaning of this is simple:
in the matrix model there is an explicit ultraviolet and infrared cut-off,
and therefore there are no 
genuine zeroes or poles of the correlation functions.
Instead, they are regularized according to
$1/0\to |\ln\mu_0|$, and $0\to 1/|\ln\mu_0|$. 
Note that in eqs. \two-\four\ $\mu$ can be replaced
by $\mu_0$. Thus, the correlation functions scale as powers of
$\mu_0$, and not as powers of $\Delta$ found in all the $c<1$
theories coupled to gravity.\rk{\GM-\Doug}~ This provides further evidence that,
for $c=1$, a renormalization of the cosmological constant takes place.
In the next section we sketch a Liouville theory explanation 
of some of the new effects peculiar to $c=1$.

\chapter{LIOUVILLE GRAVITY COUPLED TO $c=1$ MATTER.}

In this section we outline the continuum approach to evaluating
the sum over surfaces \pathinteg. In order to integrate over all metrics,
in 2 dimensions we may pick the conformal gauge
$g_{\mu\nu}=\hat g_{\mu\nu}(\tau) e^{-\phi}$
where $\hat g_{\mu\nu}$ depends on a finite number of modular parameters
collectively denoted as $\tau$. For spherical topology, there
are no moduli, and we may choose $\hat g_{\mu\nu}=\delta_{\mu\nu}$.
Eq. \Laction\ shows that 
classically the Liouville field $\phi$ is non-dynamical.
It is well-known, however,\rk\sasha~ that the kinetic term for $\phi$ is
induced through quantum effects
arising from dependence on $\phi$ of
the path integral measure. Additional modification of the dynamics of
$\phi$ has been proposed by David and by Distler and Kawai,\rk\ddk~ 
who argued that we can simplify the integration measure 
at the expense of further renormalization of the action for 
$\phi$ such that
the ghost, $X$ and $\phi$ fields combine into a conformally
invariant theory with net conformal anomaly equal to zero.
Following their approach, we assume that the original path integral
\pathinteg\ is transformed in the conformal gauge into
$$\eqalign{&{\cal Z}=\sum_h g_0^{2h-2}
\int [d\tau][Db Dc][D\phi][DX] e^{-S_{b, c}-S}\ ,\cr
& S={1\over 4\pi}\int d^2\sigma\sqrt {\hat g}\left(
\partial_\mu X\partial^\mu X+
\partial_\mu \phi\partial^\mu \phi-4\hat R\phi+
4\Delta e^{-2\phi}\right)
\ ,\cr }\eqn\Liouv$$
and $S_{b, c}$ is the standard free action for the $b$ and $c$ ghosts.
The exponential interaction term originates from the cosmological
term in the action \Laction. \foot{Polchinski has argued\rk\Polch~ that, for 
$c=1$, the interaction term is not simply an exponential.
We will instead work with the exponential interaction,
and will show that it is also possible to explain correlation
functions within this framework. Our argument
does not substantially differ from Polchinski's, but
we prefer working with the exponential because it was shown
by Curtwright and Thorn\rk\thorn~ that with this form of interaction the
conformal invariance can be maintained exactly in the continuum limit.} 

With the action \Liouv, the simplest set of conformally invariant operators
are
$$ T(q)={1\over\pi\b}\int d^2\sigma\sqrt{\hat g} e^{iqX+\e (q)\phi}
\eqn\tachop$$
These operators inject embedding momentum $q$ into the world sheet.
They should be thought of as the  
Liouville theory implementation of the operators \contach.
In a theory with $\Delta=0$, $T(q)$ is conformally invariant
for $\e (q)=-2\pm |q|$. For $\Delta>0$, however, it was shown in ref.
\ns, \pol~ that only the operator with $\e (q)=-2+|q|$ exists.
This conclusion agrees with the scaling of correlation
functions in the matrix model. Let us now use the Liouville formalism
to calculate the spherical correlation functions.

As discovered in ref. \gupta, in performing the path integral
it is convenient to decompose $\phi=\phi_0+\tilde\phi$, 
and integrate first over the zero mode $\phi_0$.
Then, since the interaction is purely exponential, we obtain
$$\eqalign{&\VEV{\prod_{i=1}^N T(q_i)}_h= g_0^{2h-2}
\delta(\sum q_i)\half\left({\Delta\over\pi}\right )^s
\Gamma(-s)\cr &
\int [d\tau][D\tilde X][D\tilde\phi][Db Dc] e^{-S_{b, c}-S_0}
\left (\int d^2\sigma \sqrt{\hat g}e^{-2\tilde\phi}\right )^s\prod_i
{1\over\pi\b}\int d^2\sigma \sqrt{\hat g} e^{iq_i X+ (-2+|q_i|)\phi} \cr
}
\eqn\reduc$$
where 
$$S_0={1\over 4\pi}\int d^2\sigma\sqrt {\hat g}\left(
\partial_\mu \tilde X\partial^\mu \tilde X+
\partial_\mu \tilde \phi\partial^\mu \tilde \phi \right )
\eqn\eq$$ 
and $ s=2(1-h)-N+\half\sum_{i=1}^N |q_i|$.
The factor $\Gamma(-s)$ indicates that correlation functions with
non-negative integer $s$ have a divergence from the volume of $\phi_0$.
This explains the divergences in the 0-point functions on the sphere
and torus found in eq. \genexp.
Now, the difficulty is to deal with the insertion of
$T(0)^s$. Goulian and Lee\rk\Goul~ proposed to 
calculate eq. \reduc\ first for integer $s$, 
where the rules are perfectly clear, and then to define it for
non-integer $s$ by analytic continuation. This gives answers in
agreement with the matrix model.$^{\Goul,\dif}$

Now we show how the multiplicative renormalization $\Delta\to\mu_0$
takes place. Although it seems that eq. \reduc\ scales as
$\Delta^s$, the free field correlator multiplying it is, formally,
of order $0^s$. For integer $s$ this can be shown through
direct calculation. Consider, for instance, the tree level correlator
of $M$ tachyons with $q_M<0$, $q_1, \ldots, q_{M-1}>0$.
If we impose the condition $s=0$, then $q_M=2-M$. 
The integral for the coefficient of $\half \G(0)$ can be
performed explicitly\rk\dif~ and gives
$${\b^{2-M}\over (M-3)!}
\prod_{i=1}^{M-1} {\Gamma(1-|q_i|)\over\Gamma(|q_i|)}
\ ,\eqn\freecorr$$
where we have set $g_0^2=\pi^3\b^2$.
If we now send $q_1, q_2, \ldots, q_m\to 0$, 
while keeping $\sum_1^{M-1} q_i=M-2$, we obtain a tachyon
correlator with $s=m$. From eq. \freecorr\ we obtain
$$\VEV{\prod_{i=m+1}^M T(q_i)}=\half \b^{2-M+m}\G(-m)\Delta^m
0^m {1\over (M-3)!}
\prod_{i=m+1}^{M-1} {\Gamma(1-|q_i|)\over\Gamma(|q_i|)}
\eqn\rescorr$$
The physical meaning of the zero of order $m$
is that the ``tachyon'' decouples at zero momentum. This is a sign
of the fact that the ``tachyon'' is actually a massless particle
in two-dimensional string theory. We will show this more explicitly
later on. 

Now we would like to argue that in a cut-off theory 
$0^s$ should be replaced by $(1/|\ln\mu_0|)^s$. If the cut-off
is imposed, then the Liouville zero mode is enclosed in a box
of size $\sim |\ln\mu_0|$. Thus, the lowest accessible value of
the ``Liouville energy'' $2+\e (q)$ is
$\sim 1/|\ln\mu_0|$, and there is no complete decoupling
of $T(0)$. Hence, the amplitudes should scale as
$(\Delta/|\ln\mu_0|)^s=\mu_0^s$, which is the sought for renormalization
of the cosmological constant. We have identified the physical
origin of this renormalization: the decoupling of the massless particle
at zero momentum.

The continuation of eq. \rescorr\ to non-integer $s$ gives
the $N$-point tachyon amplitudes with
$q_N<0$, $q_1, \ldots, q_{N-1}>0$,\rk\dif~
$$\VEV{\prod_{i=1}^N T(q_i)}=\half \b^{2-N}
\prod_{i=1}^{N} {\Gamma(1-|q_i|)\over\Gamma(|q_i|)}
\left (d\over d\mu_0\right )^{N-3} \mu_0^{s+N-3}
\ .\eqn\eq$$
This agrees with the matrix model results \two-\four\
if the matrix model and Liouville
theory operators are related by momentum dependent normalization factor
$$ T(q)={1\over \Gamma(|q|)} P(q)
\eqn\normfact$$
With this correspondence, we can use the matrix model results to obtain
tachyon correlation functions to all orders in the genus expansion.
This is remarkable because direct Liouville theory calculations
of correlation functions have not yet been performed 
even for genus 1. For $h>3$ we do not know the region
of modular integration and could not perform a direct
Liouville calculation even in principle. From eq. \two\ we find, for instance,
$$
\VEV{T(q) T(k)}=-\delta(q+k)
\left [{\G(1-|q|)\over \G(|q|)}\right ]^2 {\mu_0}^{|q|} 
\Bigg{[}  {1 \over {|q|}} 
- {1 \over {24  (\b\mu)^2}} (|q| -1) (q^2 - |q| -1) +\ldots\Bigg{]} 
\ .\eqn\eq$$
A natural continuation of this formula to $q=0$ is to replace
$\lim_{q\to 0} |q|$ by $1/|\ln\mu_0|$. Then we obtain,
$$ 
\VEV{T(0) T(0)}\sim {\partial^2\over\partial\Delta^2} E
\ .\eqn\eq$$
This is a simple consistency check: differentiation of the path
integral with respect to $\Delta$ inserts a zero-momentum tachyon.

To conclude this section, we will describe the only $h>0$   
calculation that has been performed to date. This is the
sum over toroidal surfaces with no insertions.\rk\BK~ 
\break The reason this 
calculation is relatively straightforward 
is that $s=0$, and the complicated insertion of $T(0)^s$
is absent. Sending $s\to 0$ in eq. \reduc, and remembering
the renormalization $\Delta\to\mu_0$, we obtain the
free field path integral
$${\cal Z}_1=-\half\ln\mu_0
\int_{\cal F} d^2\tau \int [D\tilde X][D\tilde\phi][Db Dc] e^{-S_0}
\eqn\eq$$
where $\tau=\tau_1+i\tau_2$ is the modular parameter of the torus, 
and $\cal F$ denotes the fundamental region of the modular group:
$\tau_2>0, |\tau|>1,
-\half\leq \tau_1<\half$.
This path integral is standard, and we find
$${\cal Z}_1(R/\sqrt{\ap})=-\half\ln\mu_0
\int_{\cal F} d^2\tau\left({|\eta(q)|^4\over 2\tau_2}\right)
(2\pi\sqrt{\tau_2}|\eta(q)|^2)^{-1}
Z(R/\sqrt{\ap}, \tau, \bar\tau)\ ,\eqn\ef
$$
where $\eta$ is the Dedekind function and
$q= e^{2\pi i\tau} $. In the above integrand, the first term is from
the ghost determinant, the second is from the Liouville determinant, and
$Z(R/\sqrt{\ap},\tau,\bar\tau)$ is the partition function of the scalar field 
compactified on a circle, 
$$ Z(R/\sqrt{\ap},\tau,\bar\tau)=
{R\over \sqrt{\ap}}{1\over \sqrt\tau_2|\eta(q)|^2}
\sum_{n, m=-\infty}^\infty
\exp\left(-{\pi R^2|n-m\tau|^2\over\ap\tau_2}\right)\ .
\eqn\circle$$
The double sum is over the soliton winding numbers about the two non-trivial
cycles of the torus.
From eqs. \ef\ and \circle, we obtain
$$\eqalign{&{\cal Z}_1(R/\sqrt{\ap})= -\half\ln\mu_0
{R\over 4\pi\sqrt{\ap}} F(R/\sqrt{\ap})\ ,\cr &
F(R/\sqrt{\ap})=\int_{\cal F} {d^2\tau\over \tau_2^2}
\sum_{n, m}\exp\left(-{\pi R^2|n-m\tau|^2\over\ap\tau_2}\right)
\ . \cr }\eqn\part
$$ 
The $\eta$-functions cancel out, and we end up only with
the contributions of the zero modes. 
This is due to the absence of particles corresponding
to the transverse string excitations, and reflects the two-dimensional
nature of this string theory.
\foot{The isolated transverse states at discrete momenta, to be 
discussed in section 6,
do not affect the torus partition function.\rk\BKl}
Remarkably, the integral in eq. \part\ can be easily performed using
the trick of ref. \trick. The idea is to trade the $m$-sum over the winding
sectors of the string for a sum over many inequivalent fundamental
regions which together cover the strip
$-\half\leq \tau_1<\half$ in the upper half-plane.
In 26 dimensions this gives a representation of the 
string free energy in terms of 
the free energies of all the modes of the string\rk\jp.
 Similarly, eq. \part
\ becomes
$$
F(R/\sqrt{\ap})=\int_{\cal F} {d^2\tau\over \tau_2^2}
+2\int_0^\infty {d\tau_2\over \tau_2^2}
\sum_{k=1}^\infty\exp\left(-{\pi R^2k^2\over\ap\tau_2}\right)
\ .\eqn\proper
$$
The second term is simply the temperature-dependent 
one-loop free energy
for a single massless boson in $2$ dimensions, 
expressed in the proper
time representation. 
The first term is the one-loop cosmological
constant of the massless boson, which has been automatically
supplied with a `stringy' ultraviolet cut-off: 
here the $\tau$
integral is over the fundamental region, not over the strip.
Performing the integrals in eq. \proper, we find
$F(R/\sqrt{\ap})={\pi\over 3}(1+{\ap/R^2})$ and
$${\cal Z}_1(R/\sqrt{\ap})=-{1\over 24}
({R\over\sqrt{\ap}}+{\sqrt{\ap}\over R})\ln\mu_0\ .	
\eqn\rad$$
As we will show in section 9, this answer agrees with the matrix
model result. This calculation gives us hope 
that the Liouville path integral exactly describes sums over geometries
beyond the tree level.

\chapter{THE SPECIAL OPERATORS.}

The most striking feature of the tachyon correlation functions
is the occurrence of the external leg factors
${\Gamma(1-|q_i|)\over\Gamma(|q_i|)}$, which contain factorized
poles whenever any of the momenta approaches an integer value.
Although this exact factorization is not yet fully understood,
it was noted in ref. \GKN\ that the poles are related to the presence
of other physical states in the theory besides the tachyons
\tachop. These states are remnants of the transverse
excitations of the string and occur only at integer $q$. 
$^{\GKN, \AMP, \lz}$

The origin of these states can be traced to the pure
$c=1$ conformal field theory where there are special
primary fields of the form\rk\dvv
$$V_{J, m}(\partial X, \partial^2 X, \ldots) e^{2mi X(z)}\eqn\opspec$$
with conformal weight $J^2$. These states
form $SU(2)$ multiplets with spin $J$ and magnetic number $m$.
The connection with $SU(2)$ becomes apparent once we note that the
states \opspec\ are the full set of primary fields in the compact 
$c=1$ theory with the self-dual radius $R=1$, where there is a well-known
level 1 $SU(2)$ current algebra.
For each spin $J$, $V_{J, \pm J}=1$, 
which gives a tachyon operator. Other members
of the multiplet can be constructed by applying raising and lowering
operators 
${1\over 2\pi i}
\oint dz \exp (\pm 2iX(z))$. After coupling to gravity,
we can ``dress'' the special $c=1$ operators to obtain new
$(1, 0)$ operators of the form
$${\cal V}_{J, m} (z)=
V_{J, m} e^{2mi X} e^{2(J-1)\phi} (z)
\ .\eqn\newops$$
Recently, it was established that these operators are related to 
a $W_{1+\infty}$ algebra.\rk\witten\foot{A related observation
was made by I. Klebanov and A. Polyakov.} 
Undoubtedly, this algebra plays an important
role in determining the properties of the theory.

In the matrix model, the physical $(1, 1)$ operator
$$\int d^2 z{\cal V}_{J, m}(z)\bar{\cal V}_{J, m}(\bar z)\eqn\specphys$$
was, up to normalization, identified with$^{\ferm, \GD}$  
$$\int_{-\infty}^\infty dx e^{2miX} (\lambda-\lambda_c)^{2J}
\eqn\matspecial$$
where $\l_c$ is the coordinate of the quadratic maximum.
Using this identification, the correlation functions of
all operators with $m=0$ were calculated in ref. \ferm.
Recently, it was shown$^{\w, \moors}$ that the operators
\matspecial\ generate a $W_{1+\infty}$ algebra. The connection
of this algebra with the algebra of special operators 
in Liouville theory has
been established by E. Witten.\rk\witten~ 

The simplest new operator is the zero-momentum dilaton 
$${\cal V}_{1, 0}\bar{\cal V}_{1, 0}=\partial X\bar\partial X \ . \eqn\eq$$
Thus, of the full dilaton field of the critical string theory
only its zero-momentum part remains here. Similarly, the
higher $J$ excitations correspond to remnants of the higher mass
particles. The connection of the special operators to the
divergences in the correlation functions is provided
by the fusion rules. 
There are terms
$\sim 1/|z-w|^2$ in the fusion rules of the physical operators
\specphys. Integration over one operator near another can thus 
produce logarithmic divergences. 

Let us elaborate on this mechanism using, as an example,
the tachyon 2-point function.\rk\GKN~ 
The poles in it occur when $|q|$ is integer, \ie\ when tachyon
operators become members of the $SU(2)$ multiplets
of the special operators.
The operator product expansion
of two integer momentum tachyons with $n>0$ is
$$ 
e^{in X+ (-2+|n|)\phi}(z, \bar z) 
e^{-in X+ (-2+|n|)\phi}(w, \bar w)\sim \ldots +{1\over |z-w|^2}
{\cal V}_{|n|-1, 0}\bar{\cal V}_{|n|-1, 0}+\ldots
\ .\eqn\eq$$
Integrating over $z$ near $w$, we obtain a logarithmic
divergence for any integer $|q|>0$ due to the appearance of a physical
operator in the fusion rule. This explains the infinite
sequence of poles in the two-point function. 

\chapter{STRING FIELD THEORY FROM THE MATRIX MODEL.}

The success of the Liouville approach shows that the theory
of surfaces embedded in one dimension can also be viewed
as critical string theory embedded in {\it two} dimensions.$^{\ind,\Polch}$
The field $\phi$, which starts out as the conformal factor of the world
sheet metric, assumes the role of another embedding dimension similar
to $X$. Indeed, the path integral \Liouv\ can be thought of as the
sigma model for the bosonic string propagating in
two-dimensional Euclidean target space with the metric
$G_{\mu\nu}=\delta_{\mu\nu}$, the dilaton background
$D(\phi)=-4\phi$, and the tachyon background $T(\phi)=\Delta e^{-2\phi}$.
The two-dimensionality of the string theory has many 
important physical consequences. 

First of all, the mass of the particle
corresponding to motion of the string in its ground state, 
``the tachyon'' is
$$ m_T^2={2-D\over 6}
\ . \eqn\eq$$
Thus, $D=c+1=2$ is the ``critical dimension of non-critical
string theory'' where ``the tachyon'' is exactly massless.
Indeed, in the calculation of the Liouville path integral
on the torus, we found a massless two-dimensional boson
propagating around the loop.
From the decoupling
at zero momentum, observed in the study of 
the tachyon correlation functions, we deduce that the theory
is invariant under constant shifts of the tachyon field.

The second expected feature of two-dimensionality is the absence
of transverse oscillation modes of the string. Indeed, if 
we could pass to the light-cone gauge, then all transverse
oscillations would be eliminated and the entire spectrum of
the string theory would consist of one massless field.
In reality, we cannot pick the light-cone gauge because of
the lack of translation invariance in the $\phi$-coordinate.
As a result, the spectrum of the string contains the transverse
excitations constructed in section 6. 
These states are not full-fledged fields, however,
because they only occur at discrete momenta $|q|=n$. 
Indeed, the string one-loop calculation of section 5
showed that there are no massive transverse {\it fields}
in the theory. Thus, at
the minimal level, we should be able to formulate the string field
theory with a single massless field, perhaps coupled to an
infinite number of quantum mechanical degrees of freedom.

In this section we show that a string field theory with the features
anticipated above can be derived directly from the matrix quantum mechanics.
This further strengthens the connection of $c=1$ \qg\ with string
theory in $D=2$. The matrix quantum mechanics directly
reduces to the exact string field theory, which is the second-quantized
field theory of non-relativistic fermions discussed in section 4. 
In this section we will simply recast this formalism so that
some of its physical features become more transparent.
First, we will show that the non-relativistic field theory can
be expressed in the \dsl\ as a theory of free
quasi-relativistic chiral fermions.$^{\ferm, \senwad}$ These chiral fermions
have the kinetic term that is relativistic to order 
$g^0_{st}$, but receives translationally non-invariant corrections of order
$g_{st}$. Further, we will carry out the conventional bosonization\rk\boson~
of this fermionic hamiltonian, and obtain an interacting
theory of massless bosons in $D=2$ first derived by Das and Jevicki
with somewhat different methods.\rk\dj~
In conclusion, we will perform a few manifestly finite calculations
in the bosonic formalism to show that it is fully equivalent
to the original representation in terms of non-relativistic fermions.

\section{Chiral Fermions.}

In order to exhibit our general method, we first consider
fermions moving in an arbitrary potential $U(\l)$.
When we later take the double-scaling limit, as expected it will
depend only on the existence of a quadratic local maximum of $U(\l)$. 
The second quantized hamiltonian for a system of free fermions
with Planck constant $1/\b$ is 
$$
{\hat h} = \int d\l \left\{ {1 \over { 2 \b^2}} {\partial \Psi^{\dagger}  \over
\partial \l }
{\partial \Psi \over \partial\l} + U(\l)\Psi^\dagger
 \Psi - \m_F(\Psi^\dagger \Psi -N)\right\}
\,\,,
\eqn\secham
$$
where $\m_F$ is the Lagrange multiplier necessary to fix the
total number of fermions to
equal $N$. As usual, it will
be adjusted so as to equal the Fermi level of the $N$ fermion system,
$\m_F=\e_N$. The fermion field has the expansion 
$$ 
\Psi(\l,t) = \sum_i \alpha_i \psi_i(\lambda) e^{-i \b\e_i t}\,\,\,\,,
\eqn\Psy
$$
where $ \psi_i(\l) $ are the single particle
wave functions and $\alpha_i$ are the respective
annihilation operators. The time $t$ is now taken to be Minkowskian. 

All the known exact results of the $c=1$ matrix model have been derived
from this nonrelativistic
field theory.   
An important feature of this theory is that
it is two-dimensional:  
the field $\Psi(\l,t)$ depends on $t$, and also on the eigenvalue coordinate
$\l$. This is the simplest way to see how the hidden 
dimension, originating from the Liouville mode of \qg,
emerges in the matrix model. 

Note that, as $\b\to\infty$, the single-particle spectrum
is approximately linear near the Fermi surface,
$$\e_n\approx \mu_F+{1\over\b}n\omega +\CO\left({1\over\b^2}\right)
\ .\eqn\eq$$
This suggests that the behavior of excitations near the Fermi surface
is quasi-relativistic. 
The spectrum of relativistic fermions in a box
is, of course, exactly linear. 
We will therefore find non-relativistic corrections 
arising from the violations of the linearity of $\e_n$ due to  
higher orders of the semiclassical expansion.
As the Fermi level approaches the maximum,  $\m=\mu_c-\mu_F\to 0$, the spacing
of the spectrum vanishes, $\omega\approx \pi/|\ln\mu|$.
Thus, we expect that the size of the box in which the quasi-relativistic 
fermions are enclosed diverges $\sim|\ln\mu|$.

Let us introduce new fermionic variables
$\Psi_L$ and $\Psi_R$,
$$
\Psi(\l,t) ={ e^{-i\m_Ft}\over { \sqrt{2 v(\l)}}} \big[
e^{ -i \b\int^\l d \l' v(\l')  + i\pi /4   }\Psi_L(\l, t)
+e^{ i \b\int^\l d \l' v(\l')  - i\pi /4  }\Psi_R(\l, t) \big] 
\,\,\,, \eqn\rel
$$
where $v(\l)$ is the velocity of the classical trajectory
of a particle at the Fermi level
in the potential $U(\l)$, 
$$
v(\l) = {d\l \over d \tau } = \sqrt{ 2( \m_F - U(\l))}\,\,\,\,.
\eqn\vee
$$
We substitute \rel\ into \secham\ to derive the hamiltonian
in terms of $\Psi_L$ and $\Psi_R$. We drop all terms which
contain rapidly oscillating exponentials
of the form $ \exp\big[ \pm 2i \b \int^{\l}v(x)dx\big]$, since these
give exponentially small terms as $\b \sim N \to \infty$ and do not contribute
to any order of semiclassical perturbation theory.
For the same reason we can restrict the 
coordinate $\l$ to lie between the
turning points of the classical motion. For $\mu>0$, 
we will consider the case where the fermions 
are localized on one side of the barrier.
If we considered the symmetric case of fig. 3, we would have found 
two identical worlds decoupled from each other to all orders
of the semiclassical expansion. We will also discuss $\mu<0$
where the Fermi level is above the barrier, and the fermions
can be found on both sides.

After some algebra we find
$$\eqalign{ {\cal H}=
2\b{\hat h} &= \int_0^{T/2} d \tau  \biggl[ i \Psi_R^{\dagger}
\partial{\tau}\Psi_R
-i\Psi_L^{\dagger} \partial_{\tau}\Psi_L + {1 \over 2 \b v^2}
\big(\partial_{\tau}
\Psi_L^{\dagger} \partial_{\tau}\Psi_L + \partial_{\tau}
\Psi_R^{\dagger} \partial_{\tau}\Psi_R \big) \cr
&+ {1 \over 4 \b} \big(
\Psi_L^{\dagger} \Psi_L
+\Psi_R^{\dagger} \Psi_R \big)\left({ v'' \over v^3} - {5 ( v')^2 \over
2v^4} \right) \biggr]\ , \cr}
\eqn\hamt
$$
where $v' \equiv dv/d\tau$.
Here we see that the natural spatial coordinate, 
in terms of which the fermion has a standard Dirac action to 
leading order in $\b$, is
$\tau$ -- the classical time of motion at the Fermi level --
rather than $\l$. An important feature of the hamiltonian
is that it does not mix $\Psi_L$ and $\Psi_R$, 
$${\cal H}={\cal H}_L+{\cal H}_R
\ .\eqn\eq$$
The only mixing between the different chiralities is through
the boundary conditions. In order to determine the boundary
conditions, consider the leading semiclassical expression for $\Psi(\l,0)$,
$$\eqalign{
\Psi (\l,0) &= \sum_{n>0}\psi_n a_n + \sum_{m\leq 0} \psi_m b_m^\dagger
\,\,\,\,,\cr 
\psi_n &= {2 \over {\sqrt{T v_n}}} \cos\big(\b \int^{\l}d \l' v_n(\l') 
- {\pi\over 4}\big) \cr
  v_n & = \sqrt{ 2(\e_n- U(\l))}\ , \cr}
\eqn\psit
$$
where we have relabeled $\alpha_i = a_i\,\,,i>0\,\,,
\alpha_i=b_i^\dagger,\,\,\,\, i\leq 0,$ and $i$ 
is the number of the fermion energy level
starting from the Fermi level $\m_F$.
Expanding $\e_n$ about $\m_F$, we find
$$ \eqalign{
\Psi(\l, 0)&= {1 \over \sqrt{2 v}} \big[ 
e^{ i \b \int^{\l} v(\l')d\l' -i\pi/4 } \Psi_R
+ e^{ -i \b \int^{\l} v(\l')d\l' +i\pi/4  }\Psi_L \big],\cr
\Psi_R&= \sqrt{2\over T}\big( \sum_{n>0} 
e^{2\pi in \tau /T} a_n 
+\sum_{m \leq 0} 
e^{2\pi im \tau /T} b^\dagger_m\big)\,, \cr
\Psi_L&= \sqrt{2\over T}\big( \sum_{n>0} 
e^{-2\pi in \tau /T} a_n 
+\sum_{m \leq 0} 
e^{-2\pi im \tau /T} b^\dagger_m\big)
\,\,\,\,\,. \cr}
\eqn\psif
$$
Thus, $\Psi_R$ and $\Psi_L$ are expressed in terms of a single set
of fermionic oscillators, $a_n$ and $b_m$. Semiclassically, they
satisfy the boundary conditions  
$$
\Psi_R(\tau=0)=\Psi_L(\tau=0),\,\,\,\qquad
\Psi_R(\tau={T\over 2})=\Psi_L(\tau={T \over 2})
\,\,\,. \eqn\bond
$$
These ensure that the fermion number current not flow
out of the finite interval,
\ie, that 
$\bar \Psi(\tau) \gamma_1 \Psi(\tau) = \Psi_R^{\dagger}\Psi_R-
\Psi_L^{\dagger}\Psi_L$ vanish at the boundary. They also guarantee that
$\Psi_R$ and $\Psi_L$ are not independent
fields and that we are including the correct number of degrees of freedom.
The issue of dynamics at the boundary is quite subtle, however,
since the corrections to the relativistic hamiltonian in eq. \hamt\
blow up precisely at the boundary. The problem is that the semiclassical
approximation breaks down at the points where $v(\tau)=0$ even
as $\b\to 0$. One possibility of dealing with this breakdown
is to carefully regularize the physics at the boundary points.
This has lead to some success.$^{\Jev, \Wad}$ 
We will suggest another method,
which is problem-free from the beginning: to approach double-scaling
limit with $\mu<0$. As we will show, an equivalent procedure
is to work with $\mu>0$ but to construct the theory in terms of
$\Psi(p, t)$, where $p$ is the momentum conjugate to $\lambda$. 

We have succeeded in mapping the collection of $N$ nonrelativistic
fermions, which describe the eigenvalues of $\Phi$, 
onto an action which, to leading
order, is just the two-dimensional Dirac action with rather
standard bag-like boundary conditions.
However, the ${1 \over \b}$ corrections in eq. \hamt~ 
cannot be disregarded in the double
scaling limit because near the quadratic maximum
$v(\tau) =\sqrt{2\m} {\rm sinh}(\tau)$.
In the double scaling limit the surviving correction to the
relativistic hamiltonian is of order $g_{st}=1/(\b\mu)$, 
$$
{\cal H}_{g_{st}} = {1\over 4\b \m}\int_0^\infty
{d\tau\over {\rm sinh}^2(\tau)}
\biggl[ |\partial_{\tau}\Psi_i|^2+ \half |\Psi_i|^2\bigl(1 - {5\over 2}
{ \rm coth}^2(\tau)\bigr ) \biggr]
\sp ,
\eqn\gcorr$$
where $i$ runs over $L$ and $R$.
This correction does not change 
the non-interacting nature of the fermions,
but it does render the fermion propagator non-standard.

In the \dsl\ the boundary at $T/2$ recedes to infinity and is irrelevant.
In fact, the correct expression for the hamiltonian would follow had
we replaced $U(\lambda)$ by $-\half\lambda^2$.
However, the hamiltonian $\gcorr$ still suffers from the divergence
at $\tau=0$. In view of this, let us modify the problem to approach
the \dsl\ with $\mu<0$. Since string perturbation
theory is in powers of $(\b\mu)^{-2}$, analytic continuation to
positive $\mu$ should not be problematic. Now $v^2 (\tau)=
2|\mu| \ch^2 \tau$, and we find
$$\eqalign{ {\cal H}=
&= \int_{-\infty}^\infty d \tau  \biggl[ i \Psi_R^{\dagger}
\partial{\tau}\Psi_R
-i\Psi_L^{\dagger} \partial_{\tau}\Psi_L + {1 \over 4 \b |\mu|\ch^2 \tau}
\big(\partial_{\tau}
\Psi_L^{\dagger} \partial_{\tau}\Psi_L + \partial_{\tau}
\Psi_R^{\dagger} \partial_{\tau}\Psi_R \big) \cr
&+ {1 \over 8 \b |\mu|\ch^2\tau} (1-{5\over 2}\th^2 \tau)\big(
\Psi_L^{\dagger} \Psi_L
+\Psi_R^{\dagger} \Psi_R \big)
\biggr]\ . \cr}\eqn\topdouble $$
By choosing $\mu<0$ we have eliminated two problems at once.
First, now the hamiltonian has no divergence at $\tau=0$.
Second, there is now no relevant boundary at all, $\Psi_L$ and
$\Psi_R$ do not mix, and there is no need to impose boundary conditions.
The fact that the two chiralities do not mix to all orders
of the string perturbation theory is remarkable. Its impact on the 
scattering processes will be discussed after we introduce the bosonized
formalism.

\section{ The Bosonic Formalism}
 
In this section we will explicitly confirm the expectation
that the space-time picture of the $c=1$ string theory
involves the dynamics of one massless scalar field in two dimensions. To this
end we will bosonize the fields $\Psi_L$ and $\Psi_R$ following the standard
bosonization rules for Dirac fermions. \rk\boson~
 A two dimensional free massless 
Dirac fermion is equivalent to a single 
free massless scalar boson. In our case, 
although the fermions are free,
they are not truly relativistic beyond the semiclassical
limit. This will give rise to interaction terms in 
the equivalent bosonic field theory.
 
        To bosonize the system we replace the fermion fields
by\rk\boson
$$
\eqalign{
\Psi_L &={1\over\sqrt{2\pi}} 
:\exp\big[i\sqrt\pi\int^{\tau} (P - X')d\tau'\big] :\ , \cr
\Psi_R &={1\over\sqrt{2\pi}} 
:\exp\big[i\sqrt\pi\int^{\tau} (P + X')d\tau'\big] :\ , \cr}
\eqn\boz
$$
where $X$ is a massless two-dimensional periodic scalar field, 
and $P$ is its canonically conjugate
momentum. The normal ordering in eq. \boz\ is in terms of
the conventionally defined creation and annihilation operators, 
which we will utilize for the explicit calculations of section 6.3. 
        To convert eq. \hamt, 
we make use of the following easily derived expressions
$$ \eqalign{
:\Psi_L^{\dagger}\partial_{\tau}\Psi_L- \Psi_R^{\dagger}\partial_{\tau} \Psi_R:
&=
{i\over 2} : P^2 + (X')^2 : \cr
:\Psi_L^{\dagger}\Psi_L+ \Psi_R^{\dagger} \Psi_R: &=- {X'\over\sqrt\pi} \cr
:\partial_{\tau}\Psi_L^{\dagger}\partial_{\tau}\Psi_L+
\partial_{\tau}\Psi_R^{\dagger}\partial_{\tau} \Psi_R: 
&= -\sqrt\pi :P X'P + {1\over 3} (X')^3 +
{1\over 6\pi} X''':  \,\,\,\,. \cr}
\eqn\dict
$$
Substituting these into eq. \hamt, we find
$$\eqalign{
:{\cal H}: = &{1\over 2}\int_0^{T/2} d \tau :\bigg[ P^2 + ( X')^2 
- {\sqrt\pi \over \b v^2} \bigl(PX'P + {1\over 3} (X')^3 
+ {1\over 6\pi} X'''\bigr)-\cr &
{1\over  2\b\sqrt\pi} X'\left({v''\over v^3}-{5 (v')^2 \over 2 v^4}
\right)\bigg]:
\,. \cr }\eqn\hamf
$$
If we integrate by parts and discard the boundary terms,
this reduces to [\ferm]
$$
:{\cal H}: = {1\over 2}\int_0^{T/2} d \tau :\bigg[ P^2 + ( X')^2 
- {\sqrt\pi \over \b v^2} (PX'P + {1\over 3} (X')^3 )
-{1\over  2\b\sqrt\pi} X'\left({v''\over 3v^3}-{(v')^2 \over 2 v^4}
\right)\bigg]:
\,. \eqn\hamsv
$$
The boundary conditions obeyed by the field $X$~are determined by those of
$\Psi$, eq. \bond, which ensure that fermion
number not flow out of the $\tau$ box. As we have shown, the current density
$\bar \Psi(\tau) \gamma_1 \Psi(\tau) = \Psi_R^{\dagger}\Psi_R-
\Psi_L^{\dagger}\Psi_L$ vanishes at the boundary. Since this density is
proportional
to $\partial_t X(t,\tau)$, we deduce that 
$X(t, 0)$~ and $X(t, {T \over 2})$~ 
must be constant, \ie, $X$ satisfies Dirichlet boundary
conditions. The constraint on the total fermion number
requires that $X(t, 0)-X(t, {T\over 2})=0$. We are free to choose
$X(t, 0)=X(t, {T \over 2})=0$. These boundary conditions eliminate
all the winding and momentum modes of $X$. As a result,
there is no need to worry about the periodic nature of $X$:
it acts like an ordinary scalar field 
in a box with Dirichlet boundary conditions. 

The hamiltonian \hamsv\ and the boundary
condition agree with the collective field
approach.\rk{\dj,\Jev}~ Physically, the massless field $X$ describes
small fluctuations of the Fermi surface.
In the \dsl\ the hamiltonian reduces to (for $\mu>0$)
$$
:{\cal H}: = {1\over 2}\int_0^\infty d \tau :\bigg[ P^2 + ( X')^2 
- {\sqrt\pi \over 2\b \mu\sh^2 \tau} \bigl(PX'P + {1\over 3} (X')^3 \bigr )
-{1-{3\over 2}\cth^2\tau \over  12\b\mu\sqrt\pi \sh^2\tau} 
X' \bigg]:
\,. \eqn\dsh
$$
This hamiltonian, as its fermionic counterpart, suffers from
a divergence at $\tau=0$. This divergence was regularized
in ref. \Jev\ with zeta-function techniques.
Alternatively, we may take the \dsl\ with $\mu<0$. Then
$$
:{\cal H}: = {1\over 2}\int_0^\infty d \tau :\bigg[ P^2 + ( X')^2 
- {\sqrt\pi \over 2\b |\mu|\ch^2 \tau} \bigl(PX'P + {1\over 3} (X')^3 \bigr )
-{1-{3\over 2}\th^2\tau \over  12\b|\mu|\sqrt\pi \ch^2\tau} 
X' \bigg]:
\ , \eqn\dnsh
$$
and the sickness at $\tau=0$ has disappeared.
The bosonized theory of non-relativistic fermions has a remarkable
structure. In the \dsl\ there is a single cubic interaction
term of order $g_{st}$, and a tadpole of order $g_{st}$.
Therefore, we can
develop an expansion of correlation functions 
in powers of $g_{st}$ using conventional
perturbation theory. As we will show, this expansion reproduces 
the genus expansion of the string amplitudes.  

\section{Scattering Amplitudes.}

Scattering amplitudes of the $X$-quanta can be related to the
Euclidean correlation functions in the matrix model.\rk\GKleb~
We will exhibit this relation for $\mu>0$ \rk\moors~ (the
$\mu<0$ case works similarly). The finite
boundary operator $O(l, q)$ from eq. \fbound\ can be translated into 
$$
{1\over 2} \int dx e^{iqx}
\int_0^\infty d \tau e^{-l\l(\tau)} : \Psi^{\dagger}_L \Psi_L
+\Psi^{\dagger}_R \Psi_R(\tau, t): \ ,
\eqn\punct
$$
where $\lambda(\tau)$ is the classical trajectory at the Fermi level.
Upon bosonization, we find
$$O(l, q)\sim 
\int dx e^{iqx} \int d \tau e^{-l\l(\tau)}\partial_\tau X\sim
i\int_{-\infty}^\infty dk F(k, l) k \tilde X(q, k)\eqn\intop
$$ 
where
$$ \eqalign{&F(k, l)=\int_0^\infty d\tau e^{-l\l(\tau)}\cos(k\tau)\ ,\cr
&X(x, \tau)=\int dx e^{-iqx}\int dk\sin (k\tau) X(q, k)\ .\cr }
\eqn\eq$$
Evaluating $F(k, l)$ with $\l(\tau)=\sqrt{2\mu}\ch\tau$, we find\rk\moors~
$$F(k, l)=K_{ik}(l\sqrt{2\mu})={\pi\over 2\sin(ik\pi)}
(I_{-ik}(l\sqrt{2\m})-
I_{ik}(l\sqrt{2\m}))
\ .\eqn\split$$
For small $z$, we will replace
$$I_\nu(z)\to (z/2)^\nu{1\over\Gamma (\nu+1)}
\ .\eqn\eq$$
In calculating the Euclidean correlation functions, each operator
will be connected to the rest of the Feynman graph by the
propagator $1/(q^2+k^2)$. We will deform the $k$-integral of eq. \intop\
for each external leg, and pick up the residue of the propagator
pole. There is a subtlety here: each $K$-function has
to be split into a sum of two $I$-functions as in eq. \split, and the allowed
sense of deformation for the two is opposite. 
For $I_{-ik}$ we pick up the pole at $k= i|q|$, while
for $I_{ik}$ -- the pole at $k= -i|q|$. As a result, we obtain
the amputated on-shell Euclidean
amplitude times a factor for each external leg, which for small $l$ is
$$(l\sqrt{\mu/2})^{|q|}{\pi\over\sin (\pi|q|)\Gamma(1+|q|)}=
-(l\sqrt{\mu/2})^{|q|}\Gamma(-|q|)
\ .\eqn\eq$$
Therefore, the correlation function of puncture operators is 
$$\VEV{\prod_{i=1}^N P(q_i)}\sim 
\prod_{j=1}^N (-\mu^{|q_j|/2} \Gamma(-|q_j|)) A (q_1, \ldots, q_N)
\eqn\mte$$
where $A$ is the Euclidean continuation of a scattering amplitude
of $N$ $X$-quanta. The same factor for each external
leg appears in the direct fermionic calculations, \two-\four. 
The lesson from this rather technical exercise is that the 
Das-Jevicki field theory assembles the correlation functions
in a rather remarkable way. Each scattering amplitude of
the $X$-quanta has to be multiplied by the external leg factors
which arise because of the unusual form \intop\ of the external
operators.\rk\GKleb~ On the other hand, these factors containing the poles
have an important string theoretic meaning and certainly cannot
be discarded. Thus, we are still missing the precise 
interpretation of the Das-Jevicki field $X$ in the string theory
with Euclidean signature. 

After continuation of string amplitudes 
to the Minkowski signature, $X\to it$, the connection with the Das-Jevicki
field theory is more straightforward. 
The Euclidean external leg factor for the operator 
$T(q)$ is 
$-\mu^{|q|/2} \Gamma(-|q|)/\Gamma(|q|)$. Upon continuation
$|q|\to -iE$, this factor turns into a pure phase \rk\Comp
$$-\exp \left(-i{E\over 2}\ln\mu\right ){\Gamma(iE)\over
\Gamma(-iE)}
\eqn\eq$$
which does not affect any scattering cross-sections.
Since the external leg phases are unobservable, they can be
absorbed in the definition of the vertex operators.
Thus, we find that the Minkowskian scattering amplitudes of tachyons
are given by the corresponding amplitudes of the $X$-quanta.

In fact, we may consider two kinds of scattering experiments.
The usual scattering involves colliding right-moving and left-moving
wave packets. In this theory such an experiment yields trivial results
since ${\cal H}={\cal H}_L+{\cal H}_R$ \rk\GKleb, and the wave packets pass
through each other with no influence or time delay.\rk\joe~ Thus, there is
no conventional ``bulk'' scattering,
whose rate is finite per unit spatial volume when plane waves
are being scattered.

We can, however, consider another kind of scattering.\rk\joe~
For $\mu>0$ we may prepare a left-moving wave packet incident on
the boundary and wait for it to reflect. This can be interpreted
as the scattering of $n$ left-moving particles into $m$ right-moving
particles. (Recall that the number of massless particles in
one spatial dimension is a subtle concept that needs to be carefully defined.)
For $\mu<0$, a left- (right- ) moving wave packet stays left-
(right- ) moving to all
orders of perturbation theory in $g_{st}$. 
However, now there are two asymptotic 
regions, $\tau\to\pm\infty$, and the wave packet undergoes some deformation
as it passes through the interaction region near $\tau=0$.
This deformation can be interpreted as a change of state (and number) 
of particles. It is this ``non-bulk'' scattering,\rk{\joe,\Jev,\Wad}~ whose
rate is not proportional to the spatial volume, that gives, upon
Euclidean continuation, the matrix model correlators.
To demonstrate this explicitly, we will now calculate tree-level
amplitudes for $2\to 1$ and $2\to 2$ particle scattering.

We will work with the $\mu<0$ hamiltonian of eq. \dnsh\ and show that
the calculations are manifestly
finite.  Since the chiralities do not mix, we will consider
scattering of right-moving particles described 
by the hamiltonian
$${\cal H}_R={1\over 4}\int_{-\infty}^\infty
:\left [ (P-X')^2+{\sqrt\pi\over 6\b|\mu| \ch^2\tau}(P-X')^3
+{ 1-{3\over 2}\th^2\tau 
\over  12\b|\mu|\sqrt\pi \ch^2\tau} 
(P-X') 
\right ]:
\ .\eqn\rhamilt$$
Following ref. \Jev, we will perform our calculation in the hamiltonian
formalism. This is advantageous here because bosonization maps 
the fermionic hamiltonian into the normal-ordered bosonic one.
Since fermionic calculations are finite, we expect that the normal
ordering will remove all the divergences in the bosonic approach.
In other words, the hamiltonian approach provides  a definite set
of rules on how to handle the bosonized theory so that it is
perfectly equivalent to the original fermionic theory. 

We introduce the canonical oscillator basis
$$\eqalign{&X(t, \tau)=\int_{-\infty}^\infty {dk\over\sqrt {4\pi |k|}}
\left (a(k) e^{i(k\tau-|k| t)}+a^{\dag} (k) 
e^{-i(k\tau-|k| t)}\right ) ,\cr
&P(t, \tau)=-i\int_{-\infty}^\infty {dk\over\sqrt {4\pi |k|}}|k|
\left (a(k) e^{i(k\tau-|k| t)}-a^{\dag} (k) 
e^{-i(k\tau-|k| t)}\right )\ ,\cr }
\eqn\eq$$
such that $[a(k), a^{\dagger}(k')]=\delta (k-k')$.
The hamiltonian assumes the form ${\cal H}_R=H_2+H_3+H_1$.
$$\eqalign{&H_2=\int_0^\infty dk k a^\dagger (k) a(k)\ ,\cr 
&H_3=
{i\over 24\pi\b |\mu|} \int_0^\infty dk_1 dk_2 dk_3\sqrt{k_1 k_2 k_3}
\bigl [f(k_1+k_2+k_3) a(k_1) a(k_2) a(k_3)-\cr
&3f(k_1+k_2-k_3) :a(k_1) a(k_2) a^\dagger (k_3):\bigr ] 
+h. c. \ ,\cr 
&H_1=-{i\over 48\pi\b|\mu|}
\int_0^\infty dk \sqrt k g(k) \bigl(a(k)-a^{\dagger}(k)\bigr)\ ,\cr }
\eqn\newform$$
where 
$$\eqalign{&f(k)=\int_{-\infty}^\infty d\tau {1\over \ch^2\tau} e^{ik\tau}
={\pi k\over \sh (\pi k/2)}\ ,\cr
&g(k)=\int_{-\infty}^\infty
{ 1-{3\over 2}\th^2\tau 
\over  \ch^2\tau} e^{ik\tau}=
{\pi (k^3+2k)\over 4\sh (\pi k/2)}
\ .\cr} 
\eqn\eq$$
Now we calculate the $S$-matrix,
$$S=1-2\pi i\delta (E_i-E_f) T\ . \eqn\eq$$
Each right-moving massless particle has energy equal to momentum,
$E=k$. Consider the amplitude for two particles with momenta
$k_1$ and $k_2$ to scatter into a single particle of momentum $k_3$,
$$ S(k_1, k_2; k_3)=-2\pi i\delta (E_1+E_2-E_3)\sqrt{E_1 E_2 E_3}
<k_3| H_{int}|k_1 k_2>
\eqn\eq$$
where $|k_1 k_2>=a^{\dagger} (k_1) a^{\dagger} (k_2)|0>$.
Substituting the cubic interaction term, we find
$$S(E_1, E_2; E_3)= -g_{st} 
\delta (E_1+E_2-E_3) E_1 E_2 E_3
\ .\eqn\eq$$
The fact that this amplitude describes ``non-bulk'' scattering is evident
from the absence of a separate delta-function for momentum conservation.
In order to relate this to the matrix model three-puncture
correlation function, we perform Euclidean continuation 
$E_j\to i|q_j|$.
Including the external leg factors according to eq.
\mte, the result is in agreement with the tree-level contribution to eq.
\three.

Now we proceed to the more complicated case of non-forward
scattering of particles of momenta $k_1$ and $k_2$ into particles
of momenta $k_3$ and $k_4$. The amplitude is given by second-order
perturbation theory,
$$ T(k_1, k_2; k_3, k_4)=
\sqrt{E_1 E_2 E_3 E_4}
 \sum_i {<k_3 k_4|H_{int}|i><i| H_{int}|k_1 k_2>\over E_1+E_2-E_i}
\ ,\eqn\eq$$
where the sum runs over all the intermediate states $i$.
The $s$-channel contribution is easily seen to be\rk\Jev
$$ S^{(s)}(k_1, k_2; k_3, k_4)=-i 
{g^2_{st}\over 8\pi} \prod_{j=1}^4 E_j
\int_0^\infty dk \left (
{k f^2 (k_1+k_2-k)\over k_1+k_2-k+i\e}
-{k f^2 (k_1+k_2+k)\over k_1+k_2+k-i\e}\right )\ ,
\eqn\schannel$$
where the first term arises from the one-particle intermediate
state, and the second one -- from the five-particle intermediate state.
Eq. \schannel\ can be expressed as an integral over $k$ from
$-\infty$ to $\infty$. The $t$- and $u$-channel contributions assume
a similar form. Summing over the three channels, we obtain\rk\Jev 
$$ S(k_1, k_2; k_3, k_4)=-i{\pi g_{st}^2\over 8}
\delta (E_1+E_2-E_3-E_4)\prod_{j=1}^4 E_j\bigl (F(p_s)+F(p_t)
+F(p_u)\bigr )
\ ,\eqn\eq$$
where $p_s=k_1+k_2$, $p_t=|k_1-k_3|$, $p_u=|k_1-k_4|$, and
$$F(p)=\int_{-\infty}^\infty dk \left 
[{(p-k)^2\over \sh^2 (\pi (p-k)/2)}\right ] 
{k \over p-k+i\e~ {\rm sgn} (k)}
\ .\eqn\fint$$
Using 
$${1\over x-i\e}={\cal P}{1\over x}+\pi i\delta (x) \eqn\eq$$
we find
$F(p)=-i {4p\over\pi}-{8\over 3\pi}$.
The total amplitude thus becomes
$$ S(E_{1,2}; E_{3,4})=-{g_{st}^2\over 2}
\delta (E_1+E_2-E_3-E_4)\prod_{j=1}^4 E_j\bigl (
E_1+E_2+|E_1-E_3|+|E_1-E_4|-2i\bigr )
\ .\eqn\eq $$
Upon the Euclidean continuation $E_j\to i|q_j|$, 
and inclusion of the external leg factors,
this precisely agrees with the non-relativistic fermion calculation
of the 4-puncture correlator \four. 
\topinsert
\epsfxsize=4in
\centerline{\epsfbox{fig4.ps}}
{\narrower\smallskip\singlespace 
\tablist{Fig. 4a) The one-loop $\CO\left(g_{st}^2\right)$ 
contributions to the $1\to 1$ amplitude.  &
Fig. 4b) The tadpole graphs of order $g_{st}^2$. 
\cr}
\smallskip} \endinsert \noindent

Now we give an example of one-loop calculation.
Following ref. \Jev, we calculate the $\CO(g_{st}^2)$ 
correction to the two-point function ($1\to 1$ amplitude).
Like the tree-level 4-point function, this is given by
second-order perturbation theory.
The contribution to $T(k, k)$ from the one-loop graphs is
$${18 g_{st}^2k^2\over (24\pi)^2}
\int_0^\infty dk_1 dk_2 ~k_1 k_2
\left ({f^2(k_1+k_2-k)\over k-k_1-k_2+i\e}
-{f^2(k_1+k_2+k)\over k+k_1+k_2-i\e}\right )
\eqn\eq$$
where the first term is from the 2-particle intermediate state,
and the second -- from the 4-particle intermediate state (fig. 4a).
After changing variables to $s=k_1+k_2$ and $k_2$, and integrating
over $k_2$, this reduces to
$$ {3 g_{st}^2 k^2\over (24\pi)^2}
\int_{-\infty}^\infty ds s^3
{ f^2(k-s)\over k-s+i\e~ {\rm sgn} (s)}=
-{g_{st}^2\over 48\pi}
k^2 \left(ik^3+2k^2+{8\over 15}\right)
\eqn\oneloop$$
where we evaluated the integral similarly to that of eq. \fint.
This is not the complete answer because there are also the
diagrams of fig. 4b) arising 
from the tadpole term $H_1$. They contribute
equally, giving
$$ {12g_{st}^2 k^2\over (24\pi) (48\pi)}
\int_0^\infty dk_1 k_1 {f(k_1)g(k_1)\over -k_1+i\e}=
- {g_{st}^2 k^2\over 384}
\int_0^\infty dk_1 {k_1^4+2k_1^2\over \sh^2 (\pi k_1/2)}
=-{g_{st}^2\over 48\pi}\times{7k^2\over 15}\ .\eqn\eq$$ 
Adding this
to the one-loop contribution \oneloop, we find
$$T(E, E)=
-{g_{st}^2\over 48\pi}
E^2 \left(iE^3+2E^2+1\right)
\ .\eqn\eq$$
Upon the Euclidean continuation and inclusion of the external
leg factors, this agrees with eq. \two. 

The agreement
with the string one-loop calculation gives us further confidence that
the bosonic string field theory 
is finite and exact. This implies that the perturbation series
exhibits the $(2h)!$ behavior, which is unlike the $h!$ found
in the conventional field theory. This appears to be connected with
the lack of translational invariance of the hamiltonian and of the
scattering amplitudes.\rk\Jev~
In general, owing to the exponential
fall-off of the form-factors $f(k)$ and $g(k)$, the hamiltonian
\newform\ is completely free of ultraviolet divergences.
Thus, our innovation of taking the \dsl\ with $\mu<0$ 
has rendered the entire perturbative expansion manifestly finite.
On the other hand, the end results of our
calculations do not change under the continuation $\mu\to -\mu$.
This suggests that the divergences for $\mu>0$, connected with
the fact that there 
$f(k)$ and $g(k)$ did not fall off as $k\to\infty$,\rk\Jev~
are simply a result of using an inconvenient formalism. 
This is also indicated by the fact that all divergences are
correctly removed by zeta-function regularization.\rk\Jev~ In fact,
positive and negative $\mu$ are related
by the transformation that interchanges the classical coordinate
$\lambda$ with its conjugate momentum $p$. Thus,
the equation
$$(-\half {\partial^2\over \partial\l^2}-\half\l^2)\psi=-\b\mu\psi
\eqn\eq$$
can be written in terms of $p$ as
$$(-\half {\partial^2\over \partial p^2}-\half p^2)\psi=\b\mu\psi
\ .\eqn\eq$$
The lesson is that, while for $\mu<0$ it is convenient to
regard $\l$ as the coordinate, for $\mu>0$ the natural coordinate 
is $p$. If we define the original non-relativistic fermion theory
in terms of $\Psi(p, t)$, then the formalism of this section goes
through nicely for $\mu>0$: there are no boundaries, and
the chiralities do not mix. Thus, all the calculations
above can also be regarded as $\mu>0$ calculations
which are now manifestly finite. Using this approach, the 
tree-level 4-point amplitude
was calculated in the lagrangian formalism in ref. \Wad.
Another technique, which works efficiently 
for all the tree amplitudes, but is not easily generalized to
the loop corrections, 
was introduced in ref. \joe.

Despite the impressive progress in the formulation of the bosonic
string field theory of the $c=1$ \qg, there are many puzzles remaining.
In particular, how should we interpret the coordinate $\tau$
and the field $X(\tau)$? Initially, it was argued that $\tau$
is essentially the zero mode of the Liouville field $\phi$.
If so, then it is natural to think of $X$ as the tachyon field.
However, it was shown recently that $\tau$ is not locally
related to the scale factor, but is instead a conjugate variable
in a complicated integral transform.\rk\moors~ This transform is necessary
to turn a string field theory, which is highly non-local in
$\phi$-space, into a simple local hamiltonian in $\tau$-space.
Miraculously, the matrix model has automatically provided
the coordinates in which the theory looks the simplest.
However, it does not appear possible to interpret $X$ 
simply as the tachyon. Recall that, in addition to the tachyon,
the theory has a discrete infinity of other non-field degrees
of freedom. It appears that the theory has soaked up the discrete
degrees of freedom together with the tachyon into a single massless
field $X$. The form of the 
matrix model operators for the discrete observables, eq. \matspecial,  
indeed shows that they are mixed in the $X$-field together with
the tachyon. Clearly, we need to attain a much better understanding
of the precise string theoretic meaning of the bosonized field
theory of the matrix model. 

\chapter{COMPACT TARGET SPACE AND DUALITY.}

In this section we consider the discretized formulation of the sum
over surfaces embedded in a circle of radius $R$.\rk\GKl~ We will
adopt the basic definition \Pol, with the new condition
$$G(x)=G(x+2\pi R)\eqn\eq$$
that ensures the periodicity around the circle. In this model we
encounter a new set of physical issues related to the $R$-dependence
of the sum. The $c=1$ conformal field theory is symmetric
under the transformation\rk\dual~
$$R\to{\alpha^\prime\over R};\qquad\qquad
g_{st}\to g_{st} {\sqrt{\alpha^\prime}\over R}\sp .\eqn\gendual $$
The change of $g_{st}$, equivalent to a constant shift of the dilaton
background, is necessary to preserve the invariance to all orders of the
genus expansion. A symmetry of the matter system, such as \gendual,
is expected to survive the coupling to gravity. In this section we
will find, however, that in the discretized formulation the duality
is generally broken, and can only be reinstated by a careful fine-tuning.
This breaking of duality is due not to the coupling to gravity,
but to the introduction of the explicit lattice cut-off.
Indeed, a generic lattice formulation of the compact $c=1$ model
on a fixed geometry has no dual symmetry. The phase transition
that separates the small $R$ (high temperature) phase from the large $R$
(low temperature) phase is well-known in statistical mechanics
as the Kosterlitz-Thouless transition.\rk\bkt~ Physically, it is due
to condensation of vortices, configurations that are ignored 
in the ``naive'' continuum limit, but {\it are} included
in the continuum limit of a generic lattice theory.
The vortices are irrelevant for $R>R_{KT}$, but they condense
and change the behavior of the theory for $R<R_{KT}$.

In this section we will demonstrate the lack of duality of the discretized
sum over surfaces. We will also show that, if the lattice sum is
modified to exclude the vortices, then
the partition function is explicitly dual. These general results
will be of interest when, in the next section, we consider
the specific discretized sum generated by the matrix model.
We will be able to isolate the effects of vortices in the matrix model,
and will calculate the dual partition function in the vortex-free continuum
limit.

In the conformal field theory the target space duality \gendual\
is generated by the dual transformation on the world sheet\rk\ov~
$$\partial_a X\to\epsilon_{ab}\partial_b X \sp .\eqn\exdu$$
Its lattice analogue is the transformation from the lattice $\Lambda$
to the dual lattice $\tilde\Lambda$. Recall that, when we sum over
lattices $\Lambda$, we define the target space variables $x_i$ at the
centers of the faces of $\Lambda$, \ie\ at the vertices of $\tilde\L$.
We will carry out a transformation, after which the new variables, $p_I$,
are defined at the vertices of $\Lambda$.

First, to each directed link $<ij>$ of $\tilde\Lambda$ associate the 
link $<IJ>$ of $\Lambda$ which intersects it, directed
so that the cross product $\vec{ij}\times\vec{IJ}$ points
out of the surface (fig. 1). Now we define $D_{ij}=D_{IJ}=
x_i-x_j$. Let us change variables in the integral \Pol\
so that the integral is over $D_{IJ}$ instead of $x_i$.
There are $E$ links but,
apart from the zero mode, there are only $V-1$ independent $x$'s.\foot
{Here $V$, $E$ and $F$ are, respectively, the numbers of vertices,
edges and faces of $\tilde\Lambda$, and $h$ is the genus.}
Therefore, the 
$D_{IJ}$'s are not all independent but must satisfy $F-1+2h$
constraints. 
The constraint associated
with each face is that $\sum_{\langle i j \rangle}  D_{ij}=0$, where the 
sum runs over the directed boundary
of a face of $\tilde\Lambda$.
This is equivalent to the condition that the sum of the $D_{IJ}$'s 
emerging from each vertex 
of $\Lambda$ must vanish, $\sum_J D_{IJ}=0.$
Similarly, for each independent non-contractable loop on $\tilde\Lambda$,
of which there are $2h$, we find $\sum_{loop} D_{ij}=0$.
In terms of $\L$ this means that
$\sum_{\langle I J \rangle} \epsilon^a_{IJ} D_{IJ}=0$,
where the symbol $\epsilon^a_{IJ}$ is non-zero only if the link $IJ$ 
intersects the specially chosen non-contractable loop $a$
on $\tilde\Lambda$. We direct the loop $a$, and define 
$\epsilon^a_{IJ}=1$ if $\vec{IJ}\times\vec a$ 
points into the surface,
and $-1$ if it points out.

The theorem of Euler
$ V-E+F=2-2h ,$~
insures that the net number of independent variables remains unchanged.
Introducing Lagrange multipliers $p_I$ for each face of $\tilde\Lambda$
and $l_a$ for each non-contractable loop, the integral over the variables
$x_i$, for each discretization of the surface,
can be replaced with
$$\eqalign{&\int\prod_{\langle I J \rangle}  dD_{IJ}G(D_{IJ})
\int\prod_I dp_I \exp \left[ip_I\sum_J D_{IJ}\right]
\int\prod_{a=1}^{2h}dl_a
\exp\left[ il_a\sum_{\langle I J\rangle}\epsilon^a_{IJ} D_{IJ}\right ]=\cr &
\int \prod_I dp_I \prod_{a=1}^{2h}dl_a
\prod_{\langle I J \rangle}\tilde G(p_I-p_J+
l_a\epsilon^a_{IJ})
\cr }\eqn\dual$$
where $\tilde G$ is the Fourier transform of $G$.
We see that, after the transformation \dual,
the Lagrange multipliers $p_I$ assume the role of 
new integration variables residing at the vertices of $\Lambda$.
On surfaces of genus $>0$ there are 
$2h$ additional $l$-integrations. Geometrically
speaking, one 
introduces a cut on the surface along each canonically
chosen non-contractable loop, so that the values of $p$ undergo a 
discontinuity of $l_a$ across the $a$-th cut, and subsequently integrates over
$l_a$. On a sphere these additional integrations do not arise, and
one simply includes a factor $\tilde G(p_I-p_J)$
for each link of the lattice $\Lambda$. 

The discussion above literally applies to the string on a real line.
When $G$ is defined on a circle of radius $R$, then the dual variables
$l_a$ and $p_I$ assume discrete values ${n\over R}$. The integrals
over $p_I$ and $l_a$ in eq. \dual\ are replaced by discrete sums,
and the partition function, when expressed in terms of the dual variables,
bears little resemblance to the original expression.
In fact, the new variables $p_I$ describe the embedding
of the world sheet in the discretized real line with a lattice spacing
$1/R$. On a sphere, where $l_a$ do not arise, the resulting lattice
sum is precisely what is needed to describe string theory with such an 
embedding. For higher genus, the few extra variables $l_a$ 
spoil the precise equivalence, but this does not alter the essential
``bulk'' physics. 

We have found that the transformation to the dual lattice,
unlike the transformation \exdu\ in the ``naive'' continuum limit,
does not prove
$R\to\alpha^\prime/R$ duality. Instead, it establishes that string
theory embedded in a compact dimension of radius $R$
is essentially equivalent
to string theory embedded in a discretized dimension of lattice
spacing $1/R$. To conclude that $R$ is equivalent to $1/R$ is 
just as counterintuitive as to argue that theories with
very small lattice spacing in target space have the same physics as
theories with enormous lattice spacing. In section 11 we will
directly study the theory with the discrete target space. We will show 
that, if the lattice spacing is less than a critical value, then
the target space lattice is ``smoothed out'', but if it exceeds 
the critical value, then the theory drastically changes its properties.
The two phases are separated by a Kosterlitz-Thouless phase transition.
Equivalently, in the circle embedding the massless $c=1$ phase for $R>R_{KT}$
is separated by the K-T transition from the disordered (massive) $c=0$
phase for $R<R_{KT}$. The mechanism of this phase transition involves
deconfinement of vortices.\rk\bkt~ 

Consider a fixed geometry $\hat g_{\mu\nu}$,
and assume that in the continuum limit the action reduces to eq. \Laction.      
A vortex of winding number $n$ located at the origin is
described by the configuration
$$ X(\theta)=nR\theta 
\eqn\eq$$
where $\theta$ is the azimuthal angle. This configuration is singular
at the origin: the values of $X$ have a branch cut in the continuum limit.
On a lattice, however, there is no singularity, and the vortex
configurations are typically included in the statistical sum.
If we introduce lattice spacing $\sqrt\mu$, then the action of a
vortex is $S_n=n^2 R^2|\ln\mu|/4\ap$, and it seems that
the vortices are suppressed in the continuum limit. 
We will now show that this expectation is only true for large enough $R$.
Let us consider the dynamics
of elementary vortices with $n=\pm 1$. Although each one is suppressed
by the action, it has a large entropy: there are $\sim 1/\mu$ places on
the surface where it can be found. Thus, the contribution
of each vortex or antivortex to the partition function is of the order
$$ {1\over\mu} e^{-S_1}=\mu^{(R^2/4\ap)-1}
\ .\eqn\eq$$
It follows that, for $R>2\sqrt{\ap}$, the vortices are irrelevant
in the continuum limit. On the other hand, for
$R<2\sqrt{\ap}$ they dominate the partition function, invalidating 
the ``naive'' continuum limit assumed in the conformal field theory.\rk\bkt~
 In fact, in this phase the proliferation of vortices 
disorders $X$ and makes its correlations short range. 
The critical properties of this phase are those of pure
gravity. Is there any way to
salvage the ``naive'' continuum limit? The answer is yes, but only
at the expense of fine tuning the model, so that
the vortices do not appear even on the lattice. Below we find 
an explicit example of a model without vortices, and show that
it {\it does} exhibit the $R\to 1/R$ duality.

First, we have to give a clear definition of a vortex on a lattice.
Loosely speaking, a face $I$ of lattice $\tilde\Lambda$ contains 
$v$ units of vortex number
if, as we follow the boundary of $I$, the coordinate $x$ wraps around
the target space circle $v$ times. This does not quite define the
vortex number because $x$ is only known at a few discrete points along the
boundary and cannot be followed continuously. In order to define vortex   
number on a lattice, it is convenient to adopt the Villain link factor
\rk\villain~
$$G(x)=\sum_{m=-\infty}^\infty 
e^{-\oh (x+2 \pi mR)^2}
\eqn\lf $$
where the sum over $m$ renders $G$ periodic under $x\to x+2 \pi R$. Now, eq.
\Pol\ becomes
$$ {\cal Z}(g_0, \kappa)=\sum_h g_0^{2(h-1)} \sum_\Lambda
\kappa^{\rm Area}\prod_{i=1}^V\int_0^{2\pi R} {dx_i\over\sqrt {2 \pi}} 
\prod_{\langle i j \rangle} 
\sum_{m_{ij}=-\infty}^\infty
e^{-\half (x_i-x_j+2 \pi m_{ij}R)^2}.
\eqn\newpol$$
In this model the number of vortices inside a face $I$
can be defined as
$$v_I=\sum_{\partial I} m_{ij}\ , \eqn\vnum $$
where ${\partial I}$~is the directed boundary of $I$.
After the dual transformation, one obtains a discrete 
sum with 
$$\eqalign{&\tilde G(p_I-p_J + l_a \epsilon^a_{IJ})
=e^{-\oh (p_I-p_J+l_a \epsilon^a_{IJ})^2}\cr
& p_I={n_I\over R} , \ l_a={n_a\over R} \cr}\sp . \eqn\eq $$
This sum is closely connected  
to the sum over all possible vortex numbers inside each face of 
$\tilde\Lambda$.\rk\villain~

We shall now change the definition 
of the partition function \newpol\ to exclude the vortex configurations.
It is helpful to think of $m_{ij}$ as link gauge fields.
The vortex numbers $v_I$ defined in eq. \vnum\ are then analogous
to field strengths. If there are no vortices, the field strength is
zero everywhere. However, we still need to sum over all possible windings
around non-contractable loops on the 
lattice $\tilde\Lambda$, $l_A=\sum_{loop~a} m_{ij}$.
Thus, the space of $m_{ij}$ we need to sum over is
$$m_{ij}=\epsilon^A_{ij} l_A+m_i-m_j, \eqn\newm$$ 
where $m_i$ range over all integers and play the role of gauge transformations.
$\epsilon^A_{ij}$ is defined above and is
non-zero only for the links $\langle ij \rangle$
intersecting the specially chosen non-trivial loop $A$ 
on the lattice $\Lambda$. 
Thus, $l_A$ is the winding number for the non-trivial cycle
$a$ on $\tilde\Lambda$ which intersects the loop $A$.
Summing over $m_{ij}$ from eq. \newm\ only, the partition function
becomes
$$ {\cal Z}(g_0, \kappa)=2 \pi R \sum_h g_0^{2(h-1)} \sum_\Lambda
\kappa^V\prod_{i=1}^{V-1}
\int_{-\infty}^\infty {dx_i\over \sqrt{2 \pi}} 
\sum_{l_A=-\infty}^\infty\prod_{\langle i j \rangle} 
e^{-\oh (x_i-x_j+ 2 \pi l_A\epsilon^A_{ij}R)^2},
\eqn\newpol$$
where the factor of $2 \pi R $~arises from integration 
over the zero mode of $x$.
After transforming to the dual lattice we obtain
$$ {\cal Z}(g_0, \kappa)= {1\over R}\sum_h 
\left ({g_0\over\sqrt{2 \pi} R}\right )^{2h-2} 
\sum_\Lambda
\kappa^V\prod_{I=1}^{F-1}
\int_{-\infty}^\infty {dp_I\over \sqrt{2 \pi}} 
\sum_{l_a=-\infty}^\infty\prod_{\langle I J \rangle} 
e^{-\oh (p_I-p_J+ l_a\epsilon^a_{IJ}/ R)^2}.
\eqn\nnewpol$$
Thus, after elimination of the vortices, the transformation to the dual lattice
clearly exhibits duality under 
$$   \sqrt{2 \pi} R \rightarrow {1 \over \sqrt{2 \pi} R },\
\sp g_0 \rightarrow {g_0\over \sqrt{2 \pi}R} \sp . \eqn\duality $$  
Remarkably, the duality is manifest even before the continuum limit is taken.
In the next section we consider the matrix model for random surfaces
embedded in a circle. We will find a clear separation between the
vortex contributions, and the ``naive'' vortex-free continuum sum. This
will allow us to calculate explicitly the dual partition function of
the vortex-free model.

\chapter{MATRIX QUANTUM MECHANICS AND THE CIRCLE EMBEDDING.}

The Euclidean matrix quantum mechanics is easily
modified to simulate the sum
over discretized random surfaces embedded in a circle
of radius $R$: we simply define the matrix variable
$\Phi(x)$ on a circle of radius $R$, \ie\ $\Phi(x+2\pi R)=\Phi(x)$.
The path integral becomes\rk\GKl
$$Z=\int D^{N^2}\Phi(x)\exp
\biggl [-\beta\int_0^{2\pi R}dx~\Tr 
\left (\half\left ({\partial\Phi\over\partial x}\right )^2+ 
U(\Phi)\right )\biggr ]\sp .\eqn\CPI
$$
The periodic one-dimensional propagator
$$G(x_i-x_j)=\sum_{m=-\infty}^\infty e^
{-|x_i-x_j+2 \pi mR|}\sp \eqn\odp$$
gives the weight for each link in the random surface interpretation.
In terms of the hamiltonian \hamilt,
eq. \CPI\ is simply a path integral representation for the partition function
$$Z_R=\Tr e^{-2 \pi R\b H}\sp ,
\eqn\eq$$
so that $2 \pi R$ plays the role of the inverse temperature.
The problem of calculating the finite temperature partition function
seems to be drastically more complicated than the zero temperature
problem, where only the ground state energy was relevant. 
Now we have to know all the energies and degeneracies of 
states in arbitrarily high representations of $SU(N)$.
We also have to explain the sudden enormous jump in the number
of degrees of freedom as we slightly increase the temperature
from zero. In view of the discussion in the last section, the reader
should not be surprised if we claim that the new degrees of freedom,
incorporated in the non-trivial representations of $SU(N)$,
are due to the K-T vortices. Since the vortices are dynamically
suppressed for $R>R_{KT}$, the problem of calculating $Z_R$
is not as complicated as it may seem.

Let us begin by calculating the contribution to $Z_R$ from the
wave functions in the trivial representation of $SU(N)$.
Since the singlet spectrum is that of $N$ non-interacting
fermions moving in the potential $U$, its contribution 
to the partition function can be calculated with the standard methods of
statistical mechanics. 

Instead of working with a fixed number of fermions $N$, we will take the 
well-known route of introducing a chemical potential $\mu_F$ adjusted so
that
$$\kappa^2={N\over \beta}=\int_0^\infty\rho(\e)~
{1\over 1+e^{2 \pi R\beta(\e-\mu_F)}}d\e \sp .\eqn\chem
$$
In the thermodynamic limit, $N\to\infty$, the free energy satisfies
$${\partial F\over\partial N}=\mu_F\sp .
\eqn\eq$$
Shifting the variables, $\mu=\mu_c-\mu_F$, $e=\mu_c-\e$, we can write
the equations for the singular part of $F$ as
$$\eqalign{&{\partial\Delta\over\partial\mu}=\pi\tilde\rho(\mu)=
\pi\int_{-\infty}^\infty de \rho(e){\partial\over\partial\mu}
{1\over 1+e^{2 \pi R\beta(\mu-e)}}\cr
&{\partial F\over\partial\Delta}={1\over \pi}\beta \mu\sp ,
\cr}\eqn\newrel$$
to emphasize resemblance with eqs. \dddmu\ and \dedd\ in the $R=\infty$ case.
Indeed, we will calculate $F(\Delta)$ using the same method.
The change introduced by a finite $R$ only affects the first of the
equations, determining $\mu(\Delta)$. We can think of $\tilde\rho(\mu)$
as the temperature-modified density of states. 
Differentiating it, we find
$${1\over\beta}{\partial\tilde\rho\over\partial\mu}=
\int_{-\infty}^\infty de {1\over\beta}\partder\rho e~
{\pi R\beta\over 2\cosh^2 [\pi R\beta (\mu-e)]}\sp .
\eqn\nddd$$
Now, substituting the integral representation \dens\ and
performing the integral over $e$,
we find
$${1\over\beta}{\partial\tilde\rho\over\partial\mu}=
{1\over \pi}{\rm Im} \int _0^{\infty}dT e^{-i\b\m T} 
{T/2\over {\rm sinh}(T/2)}
{T/2R\over\sinh (T/2R)}\sp ,\eqn\newborel
$$
up to terms 
$\CO(e^{-\beta\mu})$ and
$\CO(e^{-\beta\mu 2 \pi R})$ 
which are invisible in the large-$\beta$ asymptotic expansion.
Thus, this integral representation should only be regarded as the 
generator of the correct expansion in powers of 
$g_{st}$.
Integrating this equation and fixing the integration constant to agree
with the WKB expansion, we find 
$${\partial\Delta\over\partial\mu}
={\rm Re}
\int_\mu^\infty {dt\over t} e^{-it}~
{t/2\beta\mu\over\sinh t/2\beta\mu}~
{t/ 2 \beta\mu R\over\sinh (t/2 \beta\mu R)}\sp .\eqn\newdense
$$
This relation has a remarkable duality symmetry under
$$R\to{1\over  R},\qquad\qquad\beta\to{R \beta } \sp .
\eqn\dsym$$ 
This is precisely the kind of duality expected in the vortex-free
continuum limit. This strongly suggests that, in discarding
the contributions of the non-singlet states to $Z_R$, we have
suppressed the vortices. Later on we will offer additional arguments
to support this claim.

To demonstrate the duality in the genus expansion, consider the
asymptotic expansion of eq. \newdense
$$\eqalign{&{\partial\Delta\over\partial\mu}=
\left [-\ln\mu+\sum_{m=1}^\infty\left ( 2\beta\mu\sqrt R\right )^{-2m}
f_m(R)\right ],\cr
&f_m(R)=(2m-1)! \sum_{k=0}^m |2^{2k}-2|~|2^{2(m-k)}-2|
{|B_{2k}|~|B_{2(m-k)}|\over (2k)! [2(m-k)]!} R^{m-2k}\sp .
\cr}\eqn\db $$
Note that the functions $f_h(R)$ are manifestly dual: 
$f_1={1\over 6} (R+1/R)$, $f_2={1\over 60}(7 R^2 +10+ 
7 R^{-2})$, and so forth. 
Solving for $\mu(\Delta)$ and integrating eq. 
\newrel\ we find the sum over connected surfaces 
${\cal Z}=-2 \pi R \b F$:
$${\cal Z}=
{1\over 4}\left\{
(2\beta\mu_0\sqrt R)^2\ln\mu_0-2f_1(R)\ln\mu_0+
\sum_{m=1}^\infty {f_{m+1}(R)\over m(2m+1)}
(2\beta\mu_0\sqrt R)^{-2m}\right\}\ . \eqn\radexp$$
Comparing this with eq. \genexp, valid in the case of infinite radius,
 we find that the coupling constant $1/(\b\mu_0)$ has been replaced by
$$ g_{eff} (R)=
 {1\over \b\mu_0\sqrt R}\ ,\eqn\eq$$ 
and that the contribution of each genus
$h$ contains the function $f_h(R)$. 
Since both $g_{eff}(R)$ and $f_h (R)$ are invariant under the dual
transformation \dsym, we have confirmed that
the genus expansion
of the sum over surfaces is manifestly dual.
As in the critical string theory, the effective coupling
constant depends on the radius. Indeed, the transformation $R\to 1/R$
keeping $\beta\mu_0$ fixed is not a symmetry, as it
 interchanges a weakly and a strongly
coupled theory.

From eq. \radexp\ we find that the sum over toroidal surfaces is
$-{1\over 12}(R+1/R)\ln\mu_0$. This agrees with the direct
calculation in the continuum, eq. \rad. 
The extra factor of 2 in the matrix model
is due to the doubling of free energy for models with symmetric potentials.
To get the basic sum over triangulated surfaces, we have to divide eq.
\radexp\ by 2, which then gives perfect agreement with Liouville theory.
This provides additional evidence that the singlet free energy
in the matrix model evaluates the sum over surfaces in the
vortex-free continuum limit. To see this more explicitly, 
we have to consider the non-singlet corrections to the free energy
and show that, like the vortices, they are irrelevant for
$R>R_{KT}$. 

This problem was discussed in ref. \Vort, where 
it was argued that the total partition
function can be factorized as
$$\Tr e^{-2\pi R\b H}=
\Tr_{singlet} e^{-2\pi R\b H}
\left (\sum_n D_n
e^{-2\pi R\b\delta E_n}\right)
\eqn\nsfact$$
where $\delta E_n$ is the energy gap between the ground state
and the lowest state in the $n$th representation, and $D_n$
is the degeneracy factor. Let us discuss the leading correction
to the free energy coming from the adjoint representation.
Although the degeneracy factor 
$D_{adj}=N^2-1$  diverges as $N\to\infty$, 
an estimate of ref. \Vort\ shows that the energy
gap also diverges in the continuum limit,
$$\b\delta E_{adj}=c|\ln\mu|\ .\eqn\eq$$
As a result,  the correction 
$$D_{adj}
e^{-2\pi R\b\delta E_{adj}}\sim 
N^2 \mu^{2\pi R c}
\eqn\eq$$
is negligible for $R>{1\over \pi c}$  in the continuum limit 
where $N\mu$ is kept constant. Taking the logarithm of eq. \nsfact,
we find that the total free energy is
$$ F=F_{singlet}+
\CO\left(N^2 \mu^{2\pi R c}\right)+\ldots
\eqn\nss$$
Thus, on the one hand, the higher representations are enhanced by
enormous degeneracy factors; on the other hand, they are
suppressed by energy gaps which diverge logarithmically 
in the continuum limit.
This struggle of entropy with energy is 
precisely of the same type as occurs in the physics of the 
\KT\ vortices, as discussed above.
In fact, we will now argue that the leading non-singlet correction in
eq. \nss\ is associated with a single vortex-antivortex pair.
We perturb the action of the continuum theory with the
operator which creates elementary vortices and anti-vortices,
$$O_v=\int d^2\sigma \sqrt{\hat g}\cos\bigl[{R\over\alpha'} (X_L-X_R)\bigr]
\eqn\eq
$$
where $X_{L, R}$ are the chiral components of the scalar field $X$.
In string theory language, this is the sum of vertex
operators which create states of winding number 1 and $-1$.
The insertion of such an operator on a surface creates
an endpoint of a cut in the values of $X$.
The conformal weight of $O_v$ is $h=\bar h={R^2\over 4\a'}$;
therefore, $O_v$ becomes relevant for $R<2\sqrt{\a'}$, 
causing a phase transition at $R_{KT}=2\sqrt{\a'}$\rk\kog.
The phase transition is connected with the instability towards
creation of cuts in the values of $X$, \ie, with vortex condensation.

This picture of the \KT\ phase transition can be easily adapted
for coupling to two-dimensional quantum gravity. The sum
over genus zero surfaces coupled to the periodic scalar reduces to the path
integral
$${\cal Z}=N^2\int [DX] [D\phi]\exp\biggl(-S_L-S_0-k
\int d^2\sigma \sqrt{\hat g}e^{\beta\phi}
\cos\bigl[{R\over\alpha'} (X_L-X_R)\bigr]\biggr)
\eqn\eq $$
where $S_L$ is the Liouville gravity action for $c=1$,
and $\beta=- 2+{R\over\sqrt{\a'}}$ so as to make the
conformal weight of the perturbing operator $(1, 1)$.
The gravitational dimension of the perturbation is then
${R\over 2\sqrt{\a'}}-1$. For small values of the cut-off
$\mu$, and for $R>2\sqrt{\a'}$,
$${{\cal Z}\over N^2}=-\mu^2 R\ln\mu+ \CO\left(k^2\mu^{R/\sqrt{\a'}}\right)+
\CO\left(k^4\mu^{(2R/\sqrt{\a'})-2}\right )+\ldots \ ,
\eqn\eq
$$
which demonstrates that the \KT\ transition occurs at
$R_c=2\sqrt{\a'}$, the same value as in a fixed gravitational background.
The term $\CO(k^2)$ in ${\cal Z}$ originates from the 2-point function
of dressed winding mode vertex operators, \ie\ from surfaces
with one cut. This leading correction has the same form as eq.
\nss\ found in the context of matrix quantum mechanics.
The further corrections, coming from additional vortex-antivortex
pairs,
are even more suppressed for $R>R_{KT}$ in the
continuum limit.

For $R<R_{KT}$, however, the expansion in $k$ diverges
badly, indicating an instability with respect to
condensation of vortices. In this phase we expect the field
$X$ to become massive, and the continuum limit of the theory
should behave as pure gravity. In the matrix model this is easy to
show for very small $R$. The argument is standard: at very
high temperature a $d+1$-dimensional theory reduces to $d$-dimensional
theory because the compact Euclidean dimension of length
$2\pi R=1/T$ becomes so small that variation of fields along it can be
ignored. Thus, for small $R$, the matrix quantum mechanics \CPI\
reduces to the integral over a matrix (zero-dimensional theory)\rk\yang 
$$\int D^{N^2}\Phi\exp
[-2\pi R\beta \Tr U(\Phi)]\sp ,
\eqn\eq$$
well-known to describe pure gravity.

We have accumulated a considerable amount of evidence that
the non-singlet wave functions of the matrix quantum mechanics
implement the physical effects of the \KT\ vortices.
For $R>R_{KT}$ their corrections to the partition function are
negligible, and the singlet free energy \radexp\ gives the sum over
surfaces. For $R<R_{KT}$ the non-singlets dominate the partition
function, and the matrix model no longer describes
$c=1$ conformal field theory coupled to gravity. In this phase 
the singlet free energy gives the sum over surfaces in the
vortex-free ``naive'' continuum limit. As we showed
in section 8, this continuum limit can be achieved at the expense
of a careful fine tuning of the lattice model. 
In the matrix model the equivalent operation turns out to be simple:
just discard the non-singlet states.
We conclude, therefore, that for any $R$
the singlet sector of the matrix quantum mechanics describes
$c=1$ conformal field theory coupled to gravity.
This remarkable fact means that, like the $R=\infty$ theory,
the compact theory is exactly soluble in terms of free fermions.
In the next section, we will use this to calculate compact
correlation functions exactly to all orders in the genus expansion.

\chapter{CORRELATION FUNCTIONS FOR FINITE RADIUS.}

In this section we evaluate the correlation functions of the
compact $c=1$ model coupled to gravity using the formalism
of free non-relativistic fermions at finite temperature.\rk\Comp~
We have already used this approach to find the sum over surfaces
with no insertions, eq. \radexp. The calculation of correlation
functions has a very simple relation to the zero-temperature
($R=\infty$) calculation described in section 4.
We again consider general operators of type \op, which reduce to
fermion bilinears. Thus, the goal is to calculate the generating
function $G(q_1, \lambda_1;\ldots; q_n, \lambda_n)$ of eq.
\corr\ in the compact case. 

The free fermions are now described by the thermal Euclidean
second-quantized action
$$
S = \int_{-\infty}^{\infty} d \L \int_0^{2\pi R} dx \PH^{\dag} (-{d \over
{dx}} + {{d^2} \over {d \L^2}} + {{\L^2} \over 4}+\b\mu ) \PH
\ ,\eqn\eq$$
and the fermion field satisfies the antiperiodic boundary conditions
$\PH(\lambda, x+2\pi R)=-\PH(\lambda, x)$. Compactifying the Euclidean
time is a standard device for describing the theory at a finite temperature
$T=1/(2\pi R)$. 
The thermal vacuum satisfies
$$
\langle \b\mu | a_{\E}^{\dag} (\nu) a_{\E^{\prime}}(\nu) \vert \b\mu \rangle = 
 \D_{\E \E^{\prime}} f(\nu) \equiv \D_{\E \E^{\prime}} {1 \over
{e^{(\b\mu - \nu)/T} +1}}
\eqn\eq$$
where $f(\nu)$ is the Fermi function and $\mu$ is the Fermi level.
The thermal Euclidean Green function may be obtained by replacing
$\theta(\nu-\b\mu)$ by $f(\nu)$ in the $T=0$ Green function of eq. \eprop,
$$
\eqalign{
S^E (x_1, \L_1 ; x_2, \L_2) &= 
\int_{-\infty}^{\infty} d\nu e^{-(\nu-\b\mu) \DX}
\lbrace \T (\DX) f(\nu) - \T(-\DX) (1-f(\nu)) \rbrace \times \cr
\P^{\E}(\nu, \L_1) \P^{\E} (\nu, \L_2) 
&= {i\over 2\pi R} \sum_{\W_n} e^{-i\W_n \DX} 
\int_0^{sgn(\W_n) \infty} ds 
   e^{-s\W_n +i\b\mu s} 
\langle \L_1 | e^{2i sH} | \L_2 \rangle \ .\cr} 
\eqn\eq$$
The only change compared to the $R=\infty$ formula
\eprop\ is that the allowed frequencies
become discrete, $\omega_n=(n+\half)/R$, 
where $n$ are integers. This ensures the antiperiodicity
of $S_E$ around the compact direction.

As in the zero-temperature case, the calculation of eq. \corr\ reduces
to a sum over one-loop diagrams. Since the frequencies are now
discrete, the single integration over the loop momentum is replaced
by a sum. This is the only change introduced by the finite temperature.
As we show below, its effects are easy to take into account.

First of all, each external momentum is quantized, $q_i=n_i/R$.
As a result, the momentum conserving delta function 
$\D(\sum q_i)$ occurring
in the non-compact case
is replaced by
$R$ times the Kroenecker function $\D_{\sum q_i}$. 
Consider a diagram which corresponds to a given ordering of the
external momentum insertions, which we denote as
$q_1, q_2, \ldots, q_n$. 
Factoring out the delta-function, the contribution of this diagram is
$$\eqalign{&{i^n \over R}\sum_{\W_n}  
\int_0^{sgn(\W_n) \infty} d\alpha_1 
   e^{-\alpha_1\W_n +i\b\mu \alpha_1} 
\langle \L_n | e^{2i \alpha_1 H} | \L_1 \rangle
\int_0^{sgn(\W_n+q_1) \infty} d\alpha_2 
   e^{-\alpha_2(\W_n+q_1) +i\b\mu \alpha_2}\times 
\cr &\langle \L_1 | e^{2i \alpha_2 H} | \L_2 \rangle
\ldots \int_0^{sgn(\W_n+\sum_1^{n-1}q_i) \infty} d\alpha_n 
   e^{-\alpha_n(\W_n+\sum_1^{n-1}q_i) +i\b\mu \alpha_n} 
\langle \L_{n-1} | e^{2i \alpha_n H} | \L_n \rangle\cr 
\ .}\eqn\eq$$
It is necessary to break the sum over $\omega_n$ into parts
where the $sgn$ functions in the exponents are constant.
Thus, we form $p_k=-\sum_1^k q_k$ and put the $p$'s in the increasing
order. If $p$ and $p'$ are two consecutive $p$'s, then the summation
that needs to be performed is, defining $\xi=\sum_1^n \alpha_i$, 
$${1\over R}\sum_
{\omega_n=p+{1\over 2R}}^
{p'-{1\over 2R}} ~e^{-\omega_n\xi}=
{e^{-p\xi}-e^{-p'\xi}\over\xi}
{\xi /2R\over \sh (\xi /2R)}\ .\eqn\eq$$ 
This factor 
multiplies the integrand of an $n$-fold $\alpha_i$-integral.
Comparing this with the $R=\infty$ case, we find that there the loop
momentum integration between $p$ and $p'$ gives the factor 
$(e^{-p\xi}-e^{-p'\xi})/\xi$ multiplying the same $n$-fold integrand. 
It follows that, in every 
single contribution to the final answer, the compactification
of the target space simply inserts the extra factor
${\xi /2R\over \sh (\xi /2R)}$ into the integrand. 
Thus, from eq. \noncomp\ we arrive at
$$
\eqalign{
&{1\over\b}{\partial \over {\partial \mu}} G_R(q_1, \L_1;   \ldots ;q_n,\L_n) =
i^{n+1} R\D_{\sum q_i} 
\sum_{\sigma \in \Sigma_n} ~\int_{-\infty}^{\infty}
d\xi e^{i\b\mu \xi} 
{\xi /2R\over \sh (\xi /2R)} \int_0^{\E_1
\infty} ds_1 \ldots \cr &\int_0^{\E_{n-1} \infty} ds_{n-1} 
 e^{-s_1
Q_1^{\sigma} - \ldots - s_{n-1} Q_{n-1}^{\sigma} } \langle \L_{\sigma
(1)} | e^{2is_1 H} | \L_{\sigma (2)} \rangle \ldots 
\langle \L_{\sigma (n)} | e^{2i(\xi - \sum_1^{n-1} s_i )H} | \L_{\sigma
(1) } \rangle \cr }
\eqn\compint$$
This formula allows us to calculate correlation functions
of any set of operators of type \op\ in the compact $c=1$ model.
The integral representation of any such correlator is obtained from
that in the non-compact case by insertion of the factor
${\xi /2R\over \sh (\xi /2R)}$ into the integrand,
and by replacement $\D(\sum q_i)\to R\D_{\sum q_i}$.
A special case
of this rule is evident in passing from the $R=\infty$ equation
\dense\ to the finite $R$ equation \newborel.  

In general, comparing eqs. \compint\ and \noncomp, we find that, if
$$\VEV{\CO_1(q_1)\ldots\CO_n(q_n)}_{R=\infty}=
\delta(\sum q_i) F(q_1, \ldots, q_n; \mu, \b)
$$
then
$$\eqalign{&\VEV{\CO_1(q_1)\ldots\CO_n(q_n)}_R
=R \delta_{\sum q_i} F_R(q_1, \ldots, q_n; \mu, \b)\ ,\cr
&F_R(q_1, \ldots, q_n; \mu, \b)=
{{{1 \over 2 R\b} \DM } \over {\sin( {1 \over 2 R\b} \DM )}}
F(q_1, \ldots, q_n; \mu, \b)\ .\cr }
\eqn\relation$$
This remarkable relation exists because the entire dependence on
$\mu$ is through the factor
$e^{i\b\mu \xi}$ in the integrands of eqs. 
\noncomp\ and \compint. Therefore, a function $f(\xi)$ can 
be introduced into the integrand simply by acting on the integral
with operator $f(-{i\over\b}\DM)$. 

Since $\mu$ has the interpretation of the renormalized cosmological constant,
in each sum over surfaces of fixed area the compactification simply
introduces the non-perturbative factor 
${A/2R\b \over \sin( A/2R\b)}$. The remarkable feature of this modification,
and, equivalently, of the operator in eq. \relation, is that they are entirely
independent of which operators are inserted into the surface.
To demonstrate the power of eq. \relation, we will use the non-compact
correlators of eqs. \two-\four\ to develop the corresponding genus
expansions in the compact case with virtually no extra work.
To this end, we expand
the operator in eq. \relation\ as
$$
{{{1 \over 2R\b} \DM } \over {\sin( {1 \over 2R\b} \DM )}}
= 1+ \sum_{k=1}^{\infty} {{(1-2^{-2k+1}) |B_{2k}|  } \over
{(\b R)^{2k} (2k)!}} \left(\DM\right)^{2k}\ .
\eqn\eq$$
Substituting this expansion into eq. \relation,
we determine the finite $R$ genus $g$ correlators in terms
of the $R=\infty$ genus $\leq g$ correlators,
$$\eqalign{& F_R^{g=0}(q_i; \mu)=F^{g=0}(q_i; \mu)\ ,\cr
& F_R^{g=1}(q_i; \mu)=F^{g=1}(q_i; \mu)+
{1\over 24 R^2}{\partial^2\over\partial\mu^2}F^{g=0}(q_i; \mu)\ ,\cr
& F_R^{g=2}(q_i; \mu)=F^{g=2}(q_i; \mu)+
{1\over 24 R^2}{\partial^2\over\partial\mu^2}F^{g=1}(q_i; \mu)
+{7\over 5760 R^4} {\partial^4\over\partial\mu^4}F^{g=0}(q_i; \mu)\ ,\ etc.\cr
}
\eqn\recursion$$
 In particular, 
apart from the discretization of momenta and the
change in the overall $\delta$-function,
all spherical correlation functions are not affected by the compactification.
This was, of course, expected in the Liouville approach.
The advantage of the matrix model formalism is that the entire 
genus expansion of any correlator can be calculated.
Surprisingly, as we have demonstrated, the compact case is hardly
any more difficult than the non-compact case. From the point of view of
the Liouville approach, this conclusion is quite remarkable. There 
the finite $R$ calculations have the additional complication of having
to perform $2h$ sums over the winding numbers around the non-trivial cycles 
of a genus $h$ surface. Thus,
eqs. \relation\ and \recursion, which are
completely general, pose a new challenge for the Liouville
approach.

Below we give genus expansions for some correlation functions of puncture
operators $P(q)$. 
For the two-point function up to genus three we get
$$
\eqalign{&\VEV{P(q) P(-q)}=
-R[\G(1-|q|)]^2 {\mu}^{|q|} \Bigg{[}  {1 \over {|q|}} 
- {(\sqrt R \b\mu)^{-2} \over 24} (|q| -1)\times
\cr & {\left\{ R(q^2 - |q| -1) - 
{1 \over R}\right \} } 
+ {(\sqrt R \b\mu)^{-4} \over 5760 }
\prod_{r=1}^3 (|q|-r) \times \cr &\left\{R^2(3q^4 - 10|q|^3 - 5q^2 + 12 |q| +7)
- 10(q^2 - |q| -1) +  {7 \over R^2}\right\} \cr & 
- {(\sqrt R\b\mu)^{-6} \over 2903040 }
\prod_{r=1}^5 (|q|-r) \biggl\{R^3( 9q^6 - 63 |q|^5 + 42q^4 +217 |q|^3 -
205|q| -93)  \cr
&- 21 R(3q^4 - 10|q|^3 - 5q^2 + 12|q| +7) +
{{147} \over R}(q^2 - |q| -1) - {93 \over {R^3}} \biggr\} 
 + \ldots \Bigg{]} \cr }
\eqn\rtwo$$
We see that the external leg factors are the same as in the non-compact
case. The relation between the tachyon operators $T(q)$
and $P(q)$ is also the same, eq. \normfact. For the same reason
as in the non-compact case, 
eq. \rtwo\ is, strictly speaking, valid only for non-integer
$q$, where  we may
replace $\mu$ by $\mu_0$.
As in the sum over surfaces, we find that the effective 
string coupling constant is 
$g_{eff}\sim {1 \over {\sqrt{R} \b\mu_0}}$. The $R$-dependence
of the correlator at each genus is not dual: the duality is broken
by insertions of momentum into the world sheet.
In the limit $q_i\to 0$ the duality is restored, and the 
tachyon correlators reduce
to derivatives of the sum over surfaces ${\cal Z}(\mu_0, R)$,
$$\VEV{ T(0) T(0)}\sim {\partial^2\over\partial\Delta^2}
{\cal Z}(\mu_0, R)\ .\eqn\eq$$

The three-point function for the case $q_1, q_2 > 0$ is, up to genus two,
$$
\eqalign{
&\VEV{T(q_1) T(q_2) T(q_3)}=
R \delta_{\sum q_i}
\prod_{i=1}^3 
\left ({\G(1-|q_i|)\over\Gamma(|q_i|)} \mu^{{|q_i|\over 2}}\right)
{1\over\b\mu}\Bigg{[} 1 \cr 
& - {(\sqrt R\b\mu)^{-2} \over 24}
(|q_3|-1)(|q_3|-2) \left\{ R(q_1^2+q_2^2 - |q_3| -1) - {1 \over R}
\right\} +  
\prod_{r=1}^4 (|q_3| - r)\times \cr 
&\biggl\{ R^2 \biggl(3(q_1^4+q_2^4)+10q_1^2 q_2^2
-10(q_1^2+q_2^2) |q_3| - 5(q_1-q_2)^2 + 12|q_3| +7\biggr) \cr
& - 10(q_1^2+q_2^2 - |q_3| -1) + {7 \over {R^2}}\biggr\}   
{(\sqrt R\b\mu)^{-4} \over 5760 }  + \ldots \Bigg{]} \cr}
\eqn\rthree$$
In the case of the four-point function we must consider two independent
kinematic configurations which will give amplitudes that are not related
by analytic continuation.\rk\Moore~ The simplest case is when $q_1,q_2,q_3>0$,
$$
\eqalign{
&\VEV{T(q_1) T(q_2) T(q_3) T(q_4)}=
- R \delta_{\sum q_i}\prod_{i=1}^4 
\left ({\G(1-|q_i|)\over\Gamma(|q_i|)} \mu^{{|q_i|\over 2}}\right)
(\b\mu)^{-2} \Bigg{[} (|q_4|-1) \cr
& - {(\sqrt R\b\mu)^{-2} \over 24}
\prod_{r=1}^3 (|q_4|-r) \left\{ R(q_1^2+q_2^2+q_3^2 - |q_4| -1) -
{1 \over R} \right\}  \cr
& + {(\sqrt R\b\mu)^{-4} \over 5760 } 
\prod_{r=1}^5 (|q_4|-r) \biggl\{ R^2 \bigg{(} 3(q_1^4+q_2^4+q_3^4)
+ 10(q_1^2 q_2^2 + q_1^2 q_3^2 + q_2^2 q_3^2) \cr
& - 10|q_4|(q_1^2+q_2^2+q_3^2) - 5(q_1^2+q_2^2+q_3^2) 
+ 10(q_1 q_2+q_1 q_3+q_2 q_3) + 12 |q_4| +7 \bigg{)} \cr
& - 10(q_1^2+q_2^2+q_3^2 -|q_4| -1) + {7 \over {R^2}} \biggr\} 
 +\ldots \Bigg{]} \cr }
\eqn\rfouro$$
The other kinematic configuration is 
$
q_1, q_2>0$, $q_3, q_4<0$ 
with 
$
min(|q_i|) = q_1$, $max(|q_i|) = q_2$.
This leads to a much more complicated formula, so we just give the
answer out to genus one
$$
\eqalign{
&\VEV{T(q_1) T(q_2) T(q_3) T(q_4)}=
- R \delta_{\sum q_i}\prod_{i=1}^4 
\left ({\G(1-|q_i|)\over\Gamma(|q_i|)} \mu^{{|q_i|\over 2}}\right)
{q_2-1\over (\b\mu)^2} \Bigg{[} 1 - {(\sqrt R\b\mu)^{-2} \over 24}\cr
& \times \biggl\{ R \bigg{(} S^3(S-2q_2-6) + q_2^2(S^2+q_3^2+q_4^2)
+13 S^2 q_2 - 8S q_2^2 - 5q_2(q_3^2+q_4^2) \cr
& + 6S(S-2q_2) + 6(q_3^2 + q_4^2)
+10q_2^2 +S -2q_2 -6 \bigg{)} - {1 \over R}(S-2)(S-3) 
\biggr\}
+\ldots \Bigg{]} \cr }
\eqn\rfourt$$
where we have defined $S= {1 \over 2} \sum |q_i|$.

Using the matrix model, we have directly calculated the correlation
functions of the tachyon operators for the momentum modes,
$$ T(q)={1\over\b}\int d^2\sigma\sqrt{\hat g}e^{iq(X+\bar X)} e^{(-2+|q|)\phi}
\ ,\eqn\eq$$
where $X$ and $\bar X$ are the holomorphic and anti-holomorphic parts of the
$X$-field. In string theory we also have the winding operators
$$ \tT(\tq)=
{1\over\b}\int d^2\sigma\sqrt{\hat g}e^{i\tq(X-\bar X)} e^{(-2+|\tq|)\phi}
\ ,\eqn\eq$$
where $\tq$ is quantized as $nR$. Although these operators are
hard to introduce in the matrix model directly, their correlators
are obtained from those of the momentum operators
through the dual transformation
$$ X+\bar X\to X-\bar X\ ,\qquad\qquad R\to {1\over R}\ ,\qquad\qquad
\beta\to\beta R .
\eqn\dual$$
Carrying out the dual transformation \dual\ on eqns. \rtwo-\rfourt, we find
$$\eqalign{
\VEV{\tT(\tq) \tT(-\tq)}=&
R\left [{\G(1-|\tq|)\over \G(|\tq|)}\right]^2 
\mu^{|\tq|} \Bigg{[}  {1 \over {|\tq|}}\cr & 
- {(\sqrt R \b\mu)^{-2} \over 24} (|\tq| -1) 
\left\{{1\over R}(\tq^2 - |\tq| -1) - R\right\} 
+\ldots \Bigg{]}\ ,\cr } 
\eqn\eq$$
\etc\ In general, if 
$$\eqalign{&\VEV{T(q_1)\ldots T(q_n)}_{R=\infty}=
\delta(\sum q_i) F(q_1, \ldots, q_n; \mu, \b)\ ,\qquad {\rm then}\cr
&\VEV{\tT(\tq_1)\ldots\tT(\tq_n)}_R= R^{n-1} 
\delta_{\sum \tq_i} 
{{{1\over 2\b} \DM } \over {\sin({1\over 2\b} \DM )}}~
F(\tq_1, \ldots, \tq_n; \mu, \b R)\ .\cr}
\eqn\dualrelation$$
This formula was obtained by carrying out the dual transformation on
the similar formula \relation\ for the momentum states. Since we do not have
a direct matrix model calculation of the winding operator correlations,
we do not yet know the expressions for correlators where both the
momentum and the winding operators are present.

Is there any real time interpretation of the exact correlation
functions we have calculated? In the $R=\infty$ case we found that,
after the continuation to Minkowski signature, the external leg
factors turn into pure phases, and the correlation functions become
scattering amplitudes of the $X$-quanta of the Das-Jevicki field theory.
We can perform the same continuation, $|q_j|\to -iE_j$, for the
finite $R$ correlation functions, expecting to obtain the same
field theory at a finite temperature $T=1/(2\pi R)$. 
The external leg factors again
turn into phases and can be absorbed in the definition of the
vertex operators, and we find the scattering 
amplitudes at temperature $T$. For instance,
$$\eqalign{&S(E_1, E_2; E_3)=-g_{str}\D(E_1+E_2-E_3)~E_1E_2E_3
\Biggl{[} 1+ \cr
& {g_{str}^2\over 24}(1+iE_3)(2+iE_3)
\bigl(1-iE_3+E_1^2+E_2^2+(2\pi T)^2\bigr)
+\ldots \Biggr{]}\cr}
\eqn\tempthree$$
Is the concept of finite-temperature $S$-matrix well-defined?
In the usual theories the answer is negative, because interaction
with the heat bath stops a particle before it reaches the interaction
region. However, in the hamiltonian \dnsh\ the interaction is exponentially
localized in a finite region of space near $\tau=0$. Thus, the interaction
with the heat bath is negligible at infinity, and the concept
of $S$-matrix still appears to be meaningful. It would 
be interesting to compare the fermionic 
results like eq. \tempthree\ with bosonic 
finite temperature calculations with the hamiltonian \dnsh.

In conclusion, we should also mention the fascinating speculation due to
Witten that the $c=1$ matrix model
describes physics in the background of a two-dimensional
black hole.\rk\Ed\foot{For a detailed discussion, see
H. Verlinde's lectures in this volume.}
The black hole has finite temperature, and its Euclidean description 
 is, perhaps, related to the compact
$c=1$ model with a definite radius $R$. It would be remarkable if
the formulae reported in this section 
could be interpreted in terms of massless particle scattering
off the black hole and in terms of Hawking radiation.

\chapter{DISCRETIZED TARGET SPACE.}

In this section we will study another interesting model:
the theory of surfaces embedded in a discretized real line
with lattice spacing $\e$. As shown in section 8, the lattice
duality transformation relates it to the model of surfaces
embedded in a circle of radius $1/\epsilon$. Although for
genus $>0$ this transformation generates $2h$ extra variables
which violate the precise equivalence of the two theories,
the basic physical effect -- the \KT\ phase transition --
is of the same nature.

In fact, the model with discretized target space is very interesting
in its own right. Here we can test how string theory
responds to introduction of a small lattice spacing in space-time.
It has been argued\rk\KS~ that in this respect string theory
is very different from any field theory: if the lattice
spacing is smaller than some critical value, then the theory
on a lattice is {\it precisely} the same as the theory in the continuum.
This is to be contrasted with any known field theory,
where the continuum behavior may only be recovered as the lattice
spacing is sent to zero. The exactly soluble $c=1$ string theory
is an ideal ground for testing this remarkable stringy effect.
Below we give the exact solution of the model with the discretized
target space and confirm that, for $\e<\e_c$, the lattice is smoothed
out so that the theory is identical to the embedding in $R^1$. 

The matrix model representation of the partition function is now in terms
of an integral over a chain of $M$ matrices with nearest neighbor
couplings\rk\Par~
$$Z(\epsilon)=\prod_{i=1}^M\int D^{N^2}\Phi_i\exp\left [-\beta\sum_i
\left({1\over 2\epsilon}\Tr (\Phi_{i+1}-\Phi_i)^2+
\epsilon\Tr W(\Phi_i)\right )\right ]\sp .
\eqn\mc $$
It is easy to check that the perturbative expansion of this integral
gives the statistical sum of the form
$$\lim_{M\to\infty}{\ln Z(\epsilon)\over M}=
 \sum_h g_0^{2(h-1)} \sum_\Lambda
\kappa^V\prod_{i=1}^{V-1}\sum_{n_i} 
\prod_{\langle i j \rangle} e^{-\e |n_i-n_j|}\ , 
\eqn\discpol$$
which is precisely what is needed to describe the embedding
in a discretized real line with lattice spacing $\epsilon$.
As in the matrix quantum mechanics \PI, the integral \mc\ 
can be expressed in terms of the eigenvalues
of the matrices $\Phi_i$. The only modification here is that, instead
of quantum mechanics of $N$ identical non-interacting fermions, we now
find quantum mechanics with a discrete time step $\epsilon$. Thus,
$$\lim_{M\to\infty}{\ln Z(\epsilon)\over M}=
\sum_{i=1}^N \ln\mu_i \sp ,
\eqn\eq$$
where $\mu_i$ are the $N$ largest eigenvalues of the transfer matrix
$$\eqalign{&\mu_i f_i(x)=\int_{-\infty}^\infty dy K(x, y)f_i(y),\cr
& K(x, y)=\left({\beta\over 2\pi\epsilon}\right)^{1/2}\exp\left [-{\beta\over 2}
\left \{{(x-y)^2\over\epsilon}+\epsilon\bigl(W(x)+W(y)\bigr)
\right\}\right ]\ .\cr}
\eqn\transfer $$
Although it is hard to solve this problem exactly, as usual
things simplify in the continuum limit where we expect to
find universal behavior controlled by the quadratic maximum of $W(x)$.
Indeed, if we take
$$W(x)=-\half x^2+\CO( x^3)
\eqn\eq $$
and rescale $x\sqrt\beta=z$, we find
$$ K(z, w)={1\over \sqrt{ 2\pi\epsilon}}\exp\left [-
{(z-w)^2\over 2\epsilon}+{1\over 4}\epsilon\left
(z^2+w^2+\CO\left({1\over\sqrt\beta}\right)(z^3+w^3)\right)
\right ]\ .\eqn\parab $$
Thus, the terms beyond the quadratic order are suppressed by powers of $\beta$
and are, therefore, irrelevant. The quadratic problem can be solved exactly
through finding the equivalent quantum mechanical problem with 
Planck constant $1/\b'$ and hamiltonian $H(\e)$ such that  
$$ K(x, y)=\bra{x} e^{-\epsilon\beta' H(\epsilon)}\ket{y}\ .
\eqn\he $$
Then, $\mu_i=\exp (-\epsilon\beta' e_i)$ where $e_i$ are the $N$ lowest
eigenvalues of $H(\epsilon)$. For the quadratic transfer
matrix,  
$$K(x, y)=\left({\beta\over 2\pi\epsilon}\right)^{1/2}\exp\left [-{\beta\over 2}
\left ({(x-y)^2\over\epsilon}-
\half\epsilon(x^2+y^2)\right)\right ]\sp ,\eqn\tran $$
the hamiltonian is also quadratic
$H={p^2\over 2}-{\omega^2 x^2\over 2}. $ 
Comparing 
$$ \bra{x} e^{-\epsilon\beta' H}\ket{y}
=\left({m\omega\beta'\over 2\pi\sin\omega\epsilon}\right)^{1/2}
\exp\left [-{m\omega\beta'\over 2}
\left ((x^2+y^2)\cot\omega\epsilon-2xy\sin^{-1}\omega
\epsilon\right)\right ]\eqn\osc $$
with eq. \tran, we find
$$ \cos \epsilon\omega=1-\half\epsilon^2\ ,
\qquad\qquad\qquad \beta'=\beta {\sin (\omega\e)\over\omega\e}
\sp . \eqn\trans $$
Thus, we have reduced the problem of random surfaces embedded in discretized
target space to quantum mechanics of $N$ fermions moving in an
inverted harmonic oscillator potential, \ie\ to random surfaces
embedded in $R^1$! Eq. \trans\ dictates how the curvature of the potential
and $\beta'$ depend on $\epsilon$. Recalling that $\ap={1/\omega^2}$
and $g_{st}={1\over \b'\mu_0}$, we find that the target space
scale and the coupling constant of the equivalent continuum 
string theory depend
on $\e$. Thus, changing the lattice spacing in target space
simply changes the parameters of the equivalent string theory
embedded in continuous space. This effect was originally demonstrated
in higher-dimensional string theory 
using completely different techniques.\rk\KS~ 
It appears to be a general property of string theory.

Intuitively, we do not expect the equivalence with the continuum string
theory to be there for arbitrarily large $\e$. Indeed, eq. \trans\
shows that, if $\e$ exceeds $\e_c=2$ then the solution for 
$\omega$ becomes complex, which is a sign of instability.
Since large lattice spacing corresponds, in the dual language, to small
radius, it is clear that the instability is with respect to
condensation of vortices. Thus, for $\e>\e_c$, we expect 
the vortices to eliminate the massless field on the world
sheet corresponding to the embedding dimension, so that the
pure gravity results. This is easy to check for very large 
$\e$ where the sites of the matrix chain become
decoupled,
$$\lim_{M \to\infty}{\ln Z(\epsilon)\over M}\to
\int D^{N^2}\Phi e^{-\b\e\Tr W(\Phi)}, \eqn\eq$$
and we obtain the one-matrix model well-known to describe the pure gravity.

We will now give a continuum explanation of the \KT\ transition
from the small $\e$ phase, where the lattice is irrelevant, to
the large $\e$ phase, where it has the dominant effect.
We will replace
the discretized real line by a continuous variable $X$
with a periodic potential $V(X)=\sum_{n>0} a_n\cos({2\pi n\over\e} X)$,
which implements the effects of the lattice. The conformal
weight of the operator $\int d^2\sigma\cos({2\pi n\over\e} X)$ is 
$h={\pi^2 n^2 \ap\over \e^2}$. Thus, for small $\e$ all the
operators perturbing the action have $h>1$ and are irrelevant. 
This is the essential reason why the string does not feel
a lattice in target space with spacing $<\CO(\sqrt{\ap})$.
The phase transition to lattice-dominated phase takes place when
the perturbation with $n=1$ becomes relevant, \ie, for
$$\e_c=\pi\sqrt{\a'}\eqn\position$$ 
We can compare this result 
with the position of the K-T transition in the matrix chain model,
where we found $\e_c=2$ and $\sqrt{\a'}(\e_c)=2/\pi$.
The agreement of these values with the continuum relation \position\ 
strengthens our explanation of the phase transition in the matrix model. 
It is satisfying that we have found the matrix model confirmation
of a general effect, smoothing out of a target space lattice,
which applies to string theories in any dimension.

\chapter{CONCLUSION.}

I hope that I have convinced the reader that the 2-dimensional
string theory is a highly non-trivial toy model for string theory
in higher dimensions. Its matrix model formulation as a sum
over surfaces embedded in 1 dimension is an example
of a perfectly regularized generally covariant definition of the
Polyakov path integral. It also turns out to be remarkably
powerful, giving us the exact solution of a non-trivial string
theory. Many of its physical features, such as the $R\to 1/R$
duality (and its breaking due to the vortices), smoothing
out of the target space discreteness, presence of poles in
the correlation functions, \etc, carry over to string theories
in higher dimensions. Also, even though this string theory is
two-dimensional, it has some interesting remnants of transverse
excitations, manifested in the isolated states at integer momenta 
which generate the $W_{1+\infty}$ algebra. 
Thanks to the matrix models, we have
acquired a wealth of exact information on the perturbative
properties of this string theory. Unfortunately, we still do not
have a detailed understanding or interpretation of most of it
in the conventional continuum formalism. Hopefully, this gap
will be closed in the not too distant future. 

\ack
I am grateful to 
D. Gross, D. Lowe, M. Bershadsky and M. Newman, in collaboration
with whom much of this material was developed. 
I am indebted to A. Polyakov for many helpful discussions and
comments. I also thank D. Lowe for reading the manuscript,
and R. Wilkinson for his help in making figures.
I am grateful to ICTP and the organizers of the Spring School for
their hospitality.

\refout
\bye

It is plausible that the energy gap
is proportional to the quadratic Casimir invariant,
$$\delta E_n={n\over \beta }\delta (\mu)+\CO(1/N^3)\ ,\eqn\de
$$
where $n$ is half the number of boxes in the Young tableux
of the irreducible representation of $SU(N)$ (because of 
matrix model constraints, only the self-conjugate representations are
allowed). The relevant degeneracy factor is the sum of the 
dimensions of the self-conjugate representations with $2n$ boxes,
which to leading order in $N$ is
$$D_n={N^{2n}\over n!}
$$


\bye

\def\m{{\mu}~}

\def\refmark#1{[#1]}                        


\def\ln{{\rm ln}}
\def\sp{\,\,\,\,}

\def\D#1{\partial_{#1}}                     

\def\p{{\phi}}                     

\def\E{{\cal E}}
\def\P{{\cal P}}
\def\g2{g^2_{\rm string}}

\def\b{{\beta}}

\def\l{{\lambda}~}



\def\op{{\hat \p}}



\def\ie{{\it i.e.}~}

\def\oh{{\textstyle {1\over 2}}}

\def\text{\textstyle}
\def\th{^{\rm th}}

\def\CO{{\cal O}}
\def\NP{{\it Nucl. Phys.\ }}
\def\PL{{\it Phys. Lett.\ }}
\def\PR{{\it Phys. Rev.\ }}
\def\PRL{{\it Phys. Rev. Lett.\ }}
\def\CMP{{\it Comm. Math. Phys.\ }}
\def\JMP{{\it J. Math. Phys.\ }}
\def\JTP{{\it JETP \ }}
\def\JP{{\it J. Phys.\ }}
\def\IJMP{{\it Int. Jour. Mod. Phys.\ }}
\def\Mod{{\it Mod. Phys. Lett.\ }}

\REF\GM{D.~J.~Gross and A.~A.~Migdal, \PRL {\bf 64 } (1990);
M.Douglas and S.Shenker, Rutgers preprint RU-89-34, Oct. 1989;
E.~Brezin and V.~Kazakov, Ecole Normale preprint Oct. 1989}
\REF\Joe{J. Polchinski, \NP {\bf B324 }, (1989) 123}
\REF\ind{S. Das, S. Nair and S. Wadia, \Mod { A4} (1989) 1033}
\REF\one{E.~Brezin, V.~A.~Kazakov and Al.~B.~Zamolodchikov,
Ecole Normale preprint, LPS-ENS 89-182;
P.~Ginsparg and Z.~Jinn-Justin, Harvard preprint.}
\REF\P{G.~Parisi, 
Roma Tor Vergata preprint, ROM2F-89/30} 
\REF\Brezin{ E.~Brezin, C.~Itzykson, G.~Parisi and J.~Zuber,
\CMP {\bf 59}, (1978) 35 }
\REF\KT{M. Kosterlitz and D. Thouless, \JP {\bf C6} (1973) 1181; 
V. L. Berezinski, \JTP {\bf 34}  (1972) 610; J. Villain
\JP {\bf C36} (1975) 581}
\REF\Par{
G.~Parisi,  
Roma Tor Vergata preprint, ROM2F-90/2 }
\REF\FS{ W. Fischler and L. Susskind\journal Phys. Lett. &171B (86) 383}
\REF\pri{E.~Brezin, V.~A.~Kazakov and Al.~B.~Zamolodchikov, 
private communication}
\REF\Kostov{I. Kostov\journal Phys. Lett. &215B (88) 499}
\REF\kog{Ya. I. Kogan\journal JETP Lett. & 45 (87) 709}
\REF\sat{B. Sathiapalan\journal Phys. Rev. &D35 (87) 3277}
\date={March, 1990}
\Pubnum{PUPT-1172}
\titlepage
\title{\bf ONE DIMENSIONAL STRING THEORY ON A CIRCLE}
\author{DAVID J. GROSS
\foot{Research supported in part by NSF grant PHY80-19754}} 
\andauthor{IGOR KLEBANOV
\foot{Research supported in part by DOE grant DE-AC02-76WRO3072}} 
\address{\JHL}
\abstract
	We discuss random matrix model representations of
$D=1$ string theory, with particular emphasis on 
the case in which the target space is a 
circle of finite radius. The duality properties
of discretized strings are analyzed and shown to depend on the dynamics
of vortices. In the representation in terms of a continuous circle of matrices 
we find an exact expression for the partition function, 
neglecting non-singlet states, as a function of 
the string coupling and the radius which exhibits
exact duality. In a second version, based on a discrete chain of matrices, 
we find that vortices induce, for a finite radius, a Kosterlitz-Thouless
phase transition that takes us to a $c=0$ theory.

\endpage
\chapter{\bf Introduction}

\chapter{Discretized Random Surfaces and Duality.}

The Polyakov path integral for the $c=1$ string can be defined in
discretized form as
$$ {\cal Z}(g_{st}, \kappa)=\sum_G g_{st}^{2(G-1)} \sum_\Lambda 
\kappa^{\rm Area}\prod_i\int dt_i \prod_{\langle i j \rangle} G(t_i-t_j)\sp ,
\eqn\Pol$$
where the sum is over discrete random surfaces of genus G, denoted by
$\Lambda$, and 
the product is over all the nearest neighbor vertices $\langle ij
\rangle$ of $\Lambda$. 
This is the partition function of a string whose target space, $t$,
is one dimensional.
For a string on the infinite real line $R_1$, the
$t$-integrals range from $-\infty$ to $+\infty$, and the factor
$G=e^{-E}$ for each link is usually taken to be Gaussian
$$ G(t_i-t_j)=e^{-\oh (t_i-t_j)^2}\sp ,
\eqn\gauss $$
yielding a discrete version of the standard continuum path integral.
In this paper we will be primarily concerned with the string theory
whose target space is a circle of radius $R$. In this case, the
$t$-integrals in eq. \Pol\ range from $0$ to $2\pi R$, and the link
factor $G(t)$ must be periodic under $t\to t+ 2\pi n R$, \ie 
$$ G(t_i-t_j)=e^{-R^2 E(\theta_i-\theta_j)}\sp ,\eqn\rform $$
where $\theta=t/R$ is an angular variable.
 
String theory compactified on a circle is more complicated and, in many
ways, more interesting than the string on a real line. 
In particular, we can study 
the dependence of string partition function on the free parameter $R$,
which has the dimension of length in the target space. 
A hint about what to expect from this dependence is provided to us by
critical string theory and conformal field theory. In those cases, which
are characterized by the decoupling of 2-d geometry, it was found
that the theory is invariant under the transformation 
\refmark\dual~
$$R\to{\alpha^\prime\over R};\qquad\qquad
g_{st}\to g_{st} {\sqrt{\alpha^\prime}\over R}\sp .\eqn\gendual $$
In non-critical string theory it is reasonable to expect this symmetry
to survive the interaction with the fluctuations of the 2-d geometry. 

The symmetry \gendual, which might appear miraculous from the target space
point of view, 
is generated by
the dual transformation on the world sheet 
\refmark\ov~
$$\partial_a X\to\epsilon_{ab}\partial_b X \sp .\eqn\exdu$$
In this section 
we will discuss duality in the context of discretized world sheets.
We will show that, when eq. \gendual\ is a symmetry of the theory,
it is generated by a standard in statistical mechanics duality transformation
which is a discrete analogue of eq. \exdu. We will also find that
the target space duality is a much more subtle and intricate phenomenon
in regularized string theory than in conformal field theory.
In fact, with the conventional definition \Pol, the string partition function
is not symmetric under \gendual. For large $R$, the continuum limit of
eq. \Pol\ describes a massless field $t$ on 2-dimensional surfaces.
On the other hand, for a small $R$ eq. \rform\ can be expanded in powers
of $R^2$ which shows that the variable $t$ is disordered, \ie,
has only short range correlations.
Thus, as the radius is decreased from
$R=\infty$, there is a radius $R_c$ below which the critical properties
of the compactified string are those of the $d=0$ rather than the
$d=1$ target space. The phase transition at $R_c$ is the Kosterlitz-Thouless
transition on a random surface. This phase transition is well-known to occur
in the $O(2)$ theory on a regular lattice. We will find that, as expected,
the fluctuations of geometry do not remove the phase transition.

Clearly, the Kosterlitz-Thouless transition destroys the duality symmetry, 
eq. \gendual. This transition is caused by the liberation of
vortices which disorder the variable $t$. 
At the end of this section we will show how to 
alter the definition of the string partition
function \Pol\ to forbid all vortex configurations. Not surprisingly,
this eliminates the K-T transition and renders the
theory explicitly dual.

To show how all of this works in practice, let us carry out the 
transformation of the partition function \Pol\ to the dual lattice.   
We shall replace the variables $t_i$, which
are defined on the vertices of the original lattice $\Lambda$, by the variables
$T_I$, defined on the dual lattice $\tilde \Lambda$, whose vertices 
correspond to the faces of $\Lambda$.
Associate with each pair of vertices on $\Lambda$, $\{t_i,~ t_j\}$, 
connected by the  link 
$\langle ij\rangle$,
a pair $\{T_i,~T_j\}$ on the dual lattice $\tilde \Lambda$, 
connected by the link $\langle IJ\rangle$
that intersects $\langle ij\rangle$.\foot{
The cross product of
$\langle ij\rangle$ with
$\langle IJ\rangle$ has to point, say, out of the surface. 
For a triangulation of
an orientable surface there is a consistent orientation of the surface which
can be described by the normal to each face of the triangulation that 
is preserved under the transformation to the dual triangulation.}
We shall determine the values of the $T_I$'s by demanding that
$T_I-T_J= D_{IJ}=D_{ij}=t_i-t_j$.
Let us first change variables from the $t_i$ to the link variables, 
$D_{ij}= D_{IJ}$.

and are coupled by exponential interactions. Thus we shall
be discussing the discretized models listed in Table 1.
\smallskip

\centerline{ Table 1. The original and the dual lattice representations }
\centerline{ of the partition functions defined by matrix models.} 
\smallskip

\begintable
Original lattice $\Lambda$\| Dual lattice $\tilde\Lambda$\crthick
$\int \prod_i dt_i\prod_{\langle i j\rangle} e^{-\mu \vert t_i-t_j\vert}$|
$\int \prod_I dT_I \prod_{a=1}^{2G} dl_a\prod_{\langle I J\rangle}
\left [ \mu^2+
(p_I-p_J+l_i\epsilon^a_{IJ})^2\right ]^{-1}$\cr
$\int_0^{2\pi R}\prod_i dt_i\prod_{\langle i j\rangle}
 \sum_{m_{ij}}e^{-\mu \vert t_i-t_j+ 2 \pi m_{ij}R \vert} $|
$\sum_{N_I, l_a}\prod_{\langle I J\rangle}
\left [\mu^2+( R^{-2}(N_I-N_J+l_a\epsilon^a_{IJ})^2
\right ]^{-1}$\cr
$\sum_{n_i}\prod_{\langle i j\rangle}e^{-
 \mu\vert n_i-n_j\vert /R}$|
$\int_0^{2\pi R} \prod_I dp_I \prod_{i=1}^{2G} dl_a\prod_{\langle I J\rangle}
\left [\mu^2+(2R)^2\sin^2 \left (
{p_I-p_J+l_a\epsilon^a_{IJ}\over 2R}\right
)\right ]^{-1}$
\endtable

Lattice duality enables us to define discretized string on a circle in
2 different ways. The first option, as in line 2 of table 1, is to
define the original variables $t_i$ on the lattice $\Lambda$ on a circle
of radius $R$. Then the dual variables $p_I$ describe strings hopping
on discretized real line with lattice spacing $1/R$. The $2G$
extra summations destroy the precise equivalence of the path integral
on the dual lattice with the correct definition
of a string on a discretized line. We will follow this option
in section 4. There, within a truncated treatment, we will
find explicit duality to all orders in the genus expansion.
However, using a high temperature (small $R$) expansion, it is easy
to see that duality cannot survive in a complete treatment, and the 
theory makes a transition to a small $R$ $c=0$ phase.

Alternatively, as in line 3 of table 1, the original variables can be
defined on a discretized real line with lattice spacing $1/R$.
Then the dual variables $p_I$ live on a circle of radius $R$.
Once again, the correctness of the dual lattice path integral is violated
by the $2G$ integrals over $l_a$. However, on a sphere the definition
is precisely correct and allows us to study explicitly the free energy       
as a function of $R$.
 Furthermore, a finite number of extra variables for $G>0$
should not affect any of the ``surface phenomena'', such as phase transitions.
In this representation we will find a Kosterlitz-Thouless transition,
at a genus independent radius, to a theory
with $c=0$. 

As in critical string theory,
the free energy is a function of $R$ and $g_{st}^2/2R$. 

We have reached a surprising conclusion that, if all the
$SU(N)/H$ non-singlet states are neglected, the partition function
nevertheless respects the $R\to\alpha^\prime/R$ duality. 
Unfortunately, an exact treatment of all non-singlet states seems impossible.
At the very least, we need to know the order of magnitude of the corrections
introduced to the partition function by the non-singlet states.
We will now show that these corrections are exponentially small for
large $R$. 

One of the interesting properties of the matrix quantum mechanics
is that the excitation energies among the singlet states vanish
in the continuum limit as $\sim 1/|\ln\Delta|$. 
We shall now show that, as $\Delta\to 0$, there exists a lower bound
on the energy splitting between the lowest non-singlet state and the ground
state. This splitting is produced by the angular part of the hamiltonian
$$H_U=\sum_{i<j}{1\over\beta^2(\lambda_i-\lambda_j)^2}
\Pi_{ij}^2.\eqn\ang$$
The motion of all fermions is confined, up to tunneling corrections,
to the region between the maxima of the potential located at
$\lambda_+$ and $\lambda_-$. Therefore,
${1\over(\lambda_i-\lambda_j)^2}
\geq {1\over(\lambda_+-\lambda_-)^2},$
and
$$E_r-E_0\geq
{1\over\beta^2(\lambda_+-\lambda_-)^2}
\sum_{i<j}\Pi_{ij}^2
={1\over\beta^2(\lambda_+-\lambda_-)^2}
C^{(2)}(r)\sp , \eqn\qqq 
$$
where $r$ is an $SU(N)$ representation and $C^{(2)}(r)$ is its
quadratic Casimir invariant. Since 
$C^{(2)}(r)\sim N$, we find $\beta (E_r-E_0)\geq \CO(1)$.
Thus, the splitting between the lowest states in different representations
is of the same order in $N$ as the splittings among the singlet states,
but bounded from below by a constant. Thus, as $\Delta\to 0$, there is
an infinite number of singlet excitations lower in energy than the
lowest non-singlet excitation. Nevertheless, the non-singlets introduce
$\CO(e^{-cR})$ corrections to the partition function, 
which become important at small $R$. In fact, for small $R$ we expect
major modifications to the behavior exhibited by the singlet states.
The singlet states alone exhibit duality, \ie, for all $R$ the theory
has $c=1$. However, it is easy to see from the lattice definition
in line 2 of table 1 that, in some neighborhood of $R=0$ the
matter field $t$ has short range correlations. In fact, for any lattice 
$\Lambda$, standard high temperature (small $R$) expansion indicates
that, for a small enough $R$, $t$ is disordered, \ie, has $c=0$.
\foot{The disordered high-temperature phase is universal, \ie,
it exists for any continuous non-vanishing link factor $G$.
We thank D. Fisher and E. Lieb for emphasizing this to us.}
Thus, after averaging over random lattices $\Lambda$, we find the
critical behavior of pure 2-d gravity. In section 5 we will
explicitly study the $c=1$ theory down to $R=R_c$, the Kosterlitz-Thouless 
point, and prove the validity of these arguments.
It is remarkable that, without the non-singlet contributions, the theory
exhibits duality, and the phase transition disappears.
In view of the arguments in section 2, this suggests that discarding
the non-singlets somehow forbids vortices, but we have not been able to show
this directly.

\chapter{ Kosterlitz-Thouless phase transition on random
surfaces. }

In this section we discuss the model defined in line 3 of table 1.
On the original lattice $\Lambda$ this model describes string theory
on a discretized real line with lattice spacing $\epsilon=1/R$.
In this context this model was discussed by Parisi. \refmark\Par~
However, as shown in table 1, the dual variables now live on a circle.
On spherical topology the dual lattice partition function defines
string theory on a circle of radius $R$, but with link factor
$$\tilde G(p_I-p_J)=\left [2+(2R)^2\sin^2\left ( 
(p_I-p_J)/2 R\right
)\right ]^{-1},\eqn\nlf$$
which is quite different from the link factor in section 4, eq. \odp.
On $G>0$ surfaces the precise equivalence of the dual lattice partition
function with string theory is spoiled by the $2G$ $l_a$ integrations.
However, we do not expect them to affect the ``surface properties''
of the theory, such as phase transitions.

The matrix model representation of the partition function is now in terms
of an integral over a chain of $M$ matrices with nearest neighbor
couplings
$$Z(\epsilon)=\prod_{i=1}^M\int d\Phi_i\exp\left [-\beta\sum_i
\left({1\over 2\epsilon}\Tr (\Phi_{i+1}-\Phi_i)^2+
\epsilon\Tr W(\Phi_i)\right )\right ]\sp .
\eqn\mc $$
As in eq. \PI, this integral can be expressed in terms of the eigenvalues
of the matrices $\Phi_i$. The only modification here is that, instead
of quantum mechanics of $N$ identical non-interacting fermions, we now
find quantum mechanics with a discrete time step $\epsilon$. Thus,
$$\lim_{M\to\infty}{\ln Z(\epsilon)\over M}=
\sum_{i=1}^N \ln\mu_i \sp ,
\eqn\eq$$
where $\mu_i$ are the $N$ largest eigenvalues of the transfer matrix
$$\eqalign{&\mu_i f_i(x)=\int_{-\infty}^\infty dy K(x, y)f_i(y),\cr
& K(x, y)=\left({\beta\over 2\pi\epsilon}\right)^{1/2}\exp\left [-{\beta\over 2}
\left \{{(x-y)^2\over\epsilon}+\epsilon\bigl(W(x)+W(y)\bigr)
\right\}\right ]\ .\cr}
\eqn\transfer $$
A method for finding $\mu_i$ was proposed in \refmark\Par. 
It consists of finding
a quantum mechanical hamiltonian $H(\epsilon)$ such that
$$ K(x, y)=\bra{x} e^{-\epsilon\beta H(\epsilon)}\ket{y}\ .
\eqn\he $$
Then, $\mu_i=\exp (-\epsilon\beta e_i)$ where $e_i$ are the $N$ lowest
eigenvalues of $H(\epsilon)$. We can now apply the methods of section 3 
to the quantum mechanics defined by the hamiltonian $H(\epsilon)$
in order to find the critical behavior of $Z(\epsilon)$. 
In \refmark\Par~it was shown how to determine 
$H(\epsilon)$ order by order in perturbation
theory about $\epsilon=0$. However, the exact determination of $H(\epsilon)$
appears to be impossible. Fortunately, it turns out that we do not need
to know the exact form of $H(\epsilon)$ to find $Z(\epsilon)$ to
all orders in $1/\beta^2$. In fact, as in section 3, the only term
in $W(x)$ that affects this expansion is the quadratic term
about its maximum.
To see this, let us expand $W(x)=x^2-\lambda x^4$ about its
maximum at $x=1$:
$$W(\tilde x)=\half-2\tilde x^2+\CO(\tilde x^3)
\eqn\eq $$
where $\tilde x=x-1$. After a rescaling $\tilde x\sqrt\beta=z$ 
and shifting $W$ by a constant, we find
$$ K(z, w)={1\over \sqrt{ 2\pi\epsilon}}\exp\left [-
{(z-w)^2\over 2\epsilon}+\epsilon\left
(z^2+w^2+\CO\left({1\over\sqrt\beta}\right)(z^3+w^3)\right)
\right ]\ .\eqn\parab $$
Thus, the terms beyond the quadratic order are suppressed by powers of $\beta$
and, therefore, are irrelevant in the calculation of $H(\epsilon)$ 
to $\CO(\beta^0)$ in powers
of $\epsilon$. The resulting $\CO(\beta^0)$ terms in $H(\epsilon)$
are again quadratic and can be easily found to all orders
in $\epsilon$. Indeed, we have reduced the problem to
determination of $H(\epsilon)$ for
$$K(x, y)=\left({\beta\over 2\pi\epsilon}\right)^{1/2}\exp\left [-{\beta\over 2}
\left ({(x-y)^2\over\epsilon}-2\epsilon(x^2+y^2)\right)\right ]\sp .\eqn\tran $$
Let us recall that, for an upside down harmonic oscillator with hamiltonian
$H={p^2\over 2m}-{m\omega^2 x^2\over 2}, $ the propagator is
$$ \bra{x} e^{-\epsilon\beta H}\ket{y}
=\left({m\omega\beta\over 2\pi\sin\omega\epsilon}\right)^{1/2}
\exp\left [-{m\omega\beta\over 2}
\left ((x^2+y^2)\cot\omega\epsilon-2xy\sin^{-1}\omega
\epsilon\right)\right ].\eqn\osc $$
Comparing eqs. \tran\ and \osc, we find
$$\cos \epsilon\omega(\epsilon)=1-2\epsilon^2\sp , \eqn\trans $$
so that the energy levels of $H(\epsilon)$ are 
$i(n+1/2)\omega(\epsilon)/\beta$. Thus, for small $\epsilon$, changing
the $\epsilon$ simply amounts to changing the energy scale of the 
quantum mechanics problem. As a result, the free energy,
$E(\epsilon, \Delta)={\omega(\epsilon)\over 2} E(\Delta) $,
where $E(\Delta)$ is the $N$-fermion ground state energy of the matrix
quantum mechanics of section 3, eq. \energy. 
As $\epsilon\to 0$, $\omega(\epsilon)\to 2$ and we recover
matrix quantum mechanics from the infinite chain of matrices.

It is interesting that, for this treatment of the
string on a circle of finite radius, the vacuum energy
is independent of $R$, except for a trivial multiplicative factor.
Presumably, this is due to the presence of the extra $l_a$ integrations
in line 3 of Table 1. 
These $2G$ extra variables violate the precise identification
of this model as a string on a circle for world sheets with genus $G>0$.

Alternatively, this model can be viewed as a representation of the
string on a discretized real line where the identification with
the Polyakov path integral is exact.
We have shown explicitly that introducing a small lattice spacing 
into the target space, which here is a real line, does not change the critical
properties of string theory. This result was found using quite different
methods in \refmark\KS~. However this breaks down for large enough
lattice spacing, which corresponds in the dual prescription to small enough
radius. We find that, for $\epsilon> 1$,
$\omega(\epsilon)$ and
$H(\epsilon)$ become complex, which is a sign of instability of 
the $c=1$ phase of string theory. 
This is the K-T transition in a theory with averaging over random
lattices $\tilde\Lambda$.
We have determined the precise
location of the K-T transition on a random surface to lie at
$$R_c={1\over\epsilon_c}=1 \sp . \eqn\rcrit$$

What is the nature of the phase for $R<R_c$? Standard arguments suggest
that the matter field acquires a mass and no longer affects the
critical properties. Therefore, we expect the partition function to
describe pure 2-d gravity, $c=0$. This is in fact the case.
In the limit $\epsilon\to\infty$ the matrix chain reduces to
a collection of decoupled sites, each one described by the
one-matrix model, well known to simulate $c=0$ gravity.\refmark{\Par}

\chapter{Discussion.}

In the introduction we remarked that, instead of thinking of the string
coordinate $t$ as a spatial dimension, it may be easier to interpret it
as the Euclidean time. With this interpretation, when $t$ is compactified
on a circle of radius $R$, the path integral describes $d=1$ string
theory at temperature ${1\over 2\pi R}.$
As in section 4, we can then think of the string partition function
as the free energy of matrix quantum mechanics. There we found that,
in the asymptotic large-$N$ expansion, the free energy of the singlet
states exhibits duality under the inversion of temperature.
Such a symmetry was also found in critical string theory. However,
this symmetry is almost absurd from the thermodynamical point of view.
Indeed, the high-temperature expansion, as well as the explicit example
of section 5, indicate that the duality is broken at each order of
string perturbation theory by
the Kosterlitz-Thouless phase transition.
Although we discussed the phase transition on a random surface,
it also occurs in critical string theory where a regular lattice suffices.
As observed in \refmark{\kog, \sat}, in critical strings 
the K-T transition temperature
$T_{KT}$ coincides with the Hagedorn temperature and also with 
appearance of extra tachyons coming from
winding modes about the temperature direction. 
Could these 3 phenomena be related in general?
Here the $d=1$ string provides a new soluble example, and the answer 
appears to be no. In the $d=1$ string the ground state is not tachyonic,
and therefore no new tachyons can appear. Also, the Hagedorn temperature
is infinite. Nevertheless, the $K-T$ transition occurs, and its role
is, as usual, to destroy the criticality of the string coordinate $t$
on random surface. Above $T_{KT}$  this coordinate 
becomes disordered, decouples, and we find the $d=0$ string. 

If we imagine that extra spacelike dimensions are added,
the K-T transition always seems to remove 
just one dimension. 
This behavior
is reminiscent of conventional field theory: the high $T$ limit
of $d+1$-dimensional field theory is described by $d$-dimensional
Euclidean path integral. The unusual stringy effect is that
the $(d+1)$-st dimension disappears abruptly at a specific temperature.
This  argument is not safe for bosonic string
theories because for $d>1$ the theory contains
tachyons. Even in the case $d=1$ we need to improve our understanding
of the space-time picture of the string theory before thermodynamics
can be convincingly discussed.

We have found that a straightforward discretized definition of $d=1$
string theory does not exhibit $R\to\alpha'/R$ duality.
What would happen if critical string theory was defined on discretized
world sheets? There, it is sufficient to study  
a regular lattice, and the theory is well-known to exhibit the
K-T phase transition which breaks duality. Therefore, the standard continuum
rules for string amplitude calculations are not consistent with the lattice
regularization. 

Now we have a dillema. One option would be to take
the discretized definition as primary, and to modify the continuum
calculations, through inclusion of singular world sheet vortices, 
to exhibit the K-T transition and break the duality. 

The second option is to modify the lattice definition to explicitly
suppress vortices, as we did in section 2. Then, at least
in string perturbation theory, duality
emerges on a lattice. It seems that discarding the non-singlet
states in the matrix quantum mechanics of section 4 effectively suppresses
vortices and gives a partition function consistent with the second option.

The majority of critical string literature is based on the second option,
and takes duality seriously. It would be interesting,
however, following refs. \refmark{\kog, \sat}, to also explore
the consequences of breaking of duality by the K-T phase transition. 

\centerline{\bf ACKNOWLEDGMENTS}

We would like to thank C.~Callan, 
A.~Migdal, M.~Newman, A.~Polyakov, and A.~Zamolodchikov 
for discussions.

\refout

\bye